\documentclass[acmtog,screen]{acmart}

\usepackage{booktabs} 

\usepackage[linesnumbered,ruled,noline]{algorithm2e} 

\SetKwComment{Comment}{$\triangleright$\ }{}
\SetAlFnt{\small}
\SetAlCapFnt{\small}
\SetAlCapNameFnt{\small}
\SetAlCapHSkip{0pt}

\SetCommentSty{mycommfont}

\usepackage{mathtools}
\usepackage{bm}
\usepackage{cases}
\usepackage{comment}
\usepackage{threeparttable}
\usepackage{subcaption}
\usepackage{wrapfig}
\usepackage{enumitem}
\usepackage{overpic}
\usepackage{siunitx}
\usepackage{makecell}
\usepackage{multirow}
\usepackage{colortbl}
\usepackage{float}
\usepackage{lipsum}
\usepackage{physics}

\usepackage[dvipsnames]{xcolor}
\usepackage{tikz}
\usetikzlibrary{3d}
\usepackage{tikz-3dplot}
\usepackage{tikz-dimline}
\usepackage{pgfplots}
\pgfplotsset{compat=1.18}

\newcommand{\rv}[1]{#1}

\newlength{\imagelength}

\setcopyright{acmlicensed}
\acmJournal{TOG}
\acmYear{2025}
\acmVolume{44}
\acmNumber{6}
\acmArticle{265}
\acmMonth{12}
\acmDOI{10.1145/3763279}

\begin{document}

\title{The Granule-In-Cell Method for Simulating Sand--Water Mixtures}

\author{Yizao Tang}
\email{skytang0205@outlook.com}
\orcid{0009-0006-6732-7296}
\affiliation{
\institution{School of Electronics Engineering and Computer Science, Peking University}
\city{Beijing}
\country{China}
}
\authornote{joint first authors}

\author{Yuechen Zhu}
\email{yuechen.zhu@stcatz.ox.ac.uk}
\orcid{0009-0008-9399-5970}
\affiliation{
\institution{Mathematical Institute, University of Oxford}
\city{Oxford}
\country{United Kingdom}
}
\authornotemark[1]

\author{Xingyu Ni}
\email{xingyuni.cs@gmail.com}
\orcid{0000-0003-1127-2848}
\affiliation{
\institution{The University of Hong Kong}
\city{Hong Kong}
\country{Hong Kong SAR}
}
\authornote{corresponding authors}

\author{Baoquan Chen}
\email{baoquan@pku.edu.cn}
\orcid{0000-0003-4702-036X}
\affiliation{
\institution{State Key Laboratory of General Artificial Intelligence, Peking University}
\city{Beijing}
\country{China}
}
\authornotemark[2]

\begin{abstract}
The simulation of sand--water mixtures requires capturing the stochastic behavior of individual sand particles within a uniform, continuous fluid medium.
However, most existing approaches, which only treat sand particles as markers within fluid solvers, fail to account for both the forces acting on individual sand particles and the collective feedback of the particle assemblies on the fluid.
This prevents faithful reproduction of characteristic phenomena including transport, deposition, and clogging.
Building upon kinetic ensemble averaging technique, we propose a physically consistent coupling strategy and introduce a novel Granule-In-Cell (GIC) method for modeling such sand--water interactions.
We employ the Discrete Element Method (DEM) to capture fine-scale granule dynamics and the Particle-In-Cell (PIC) method for continuous spatial representation and density projection.
To bridge these two frameworks, we treat granules as macroscopic transport flow rather than solid boundaries within the fluid domain. This bidirectional coupling allows our model to incorporate a range of interphase forces using different discretization schemes, resulting in more realistic simulations that strictly adhere to the mass conservation law.
Experimental results demonstrate the effectiveness of our method in simulating complex sand--water interactions, uniquely capturing intricate physical phenomena and ensuring exact volume preservation compared to existing approaches.
\end{abstract}

%
%

%
%



\maketitle
\begin{figure}
  \centering
  \includegraphics[width=\linewidth]{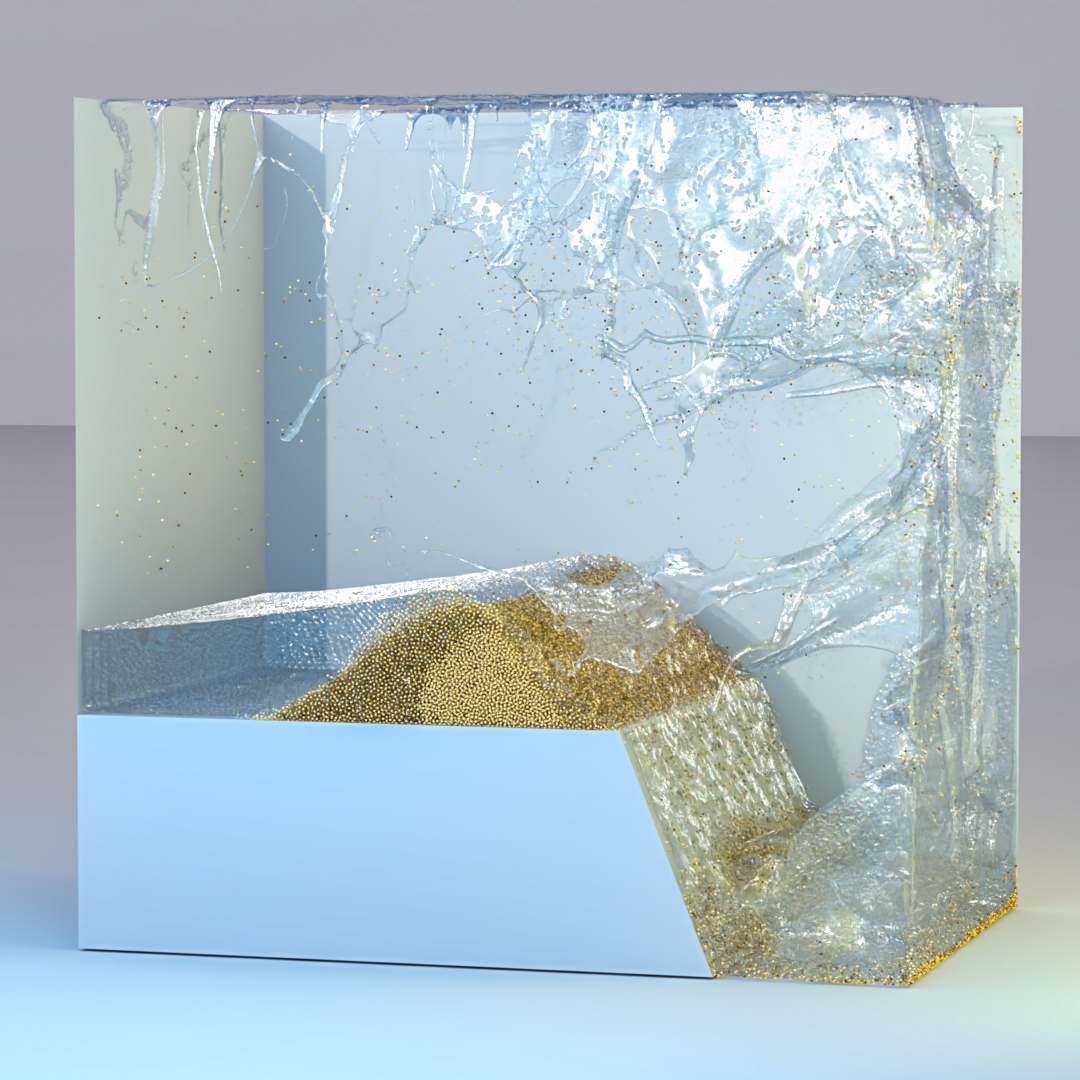}
  \vspace{-2em}
  \caption{
    Dam breaking simulated using out method (GIC). We implemented bidirectional coupling of sand and water, maintaining the conservation of the mixture's volume, and enabled the transition of sand particles from dry to wet.
  }
  \Description{a figure of dam breaking}
  \label{fig:lifting}
\end{figure}

\section{Introduction}


The simulation of sand--water mixtures is not only a core issue in hydraulic engineering \cite{white2003fluid, yang1996sediment, simons1992sediment} but also has attracted significant attention in the field of computer graphics due to its rich visual details and complex movement patterns \cite{rungjiratananon2008real, yan2016multiphase, su2023real, qu2023power}. The current challenge lies in the necessity to illustrate both the stochastic micro-movement of single granules and the statistical macro-transport processes of granular cluster ensemble, which demands a unified, self-consistent, and comprehensive dynamical description linking the mesoscopic and the macroscopic scales.

In previous graphics researches, two predominant frameworks have emerged: one from the Lagrangian perspective and the other from the Eulerian perspective. 
The former, including works based on Smoothed Particle Hydrodynamics (SPH) \cite{lenaert2009mixing, ren2014multiple, yan2016multiphase, ren2021unified} and the work combining the SPH method and the discrete element method (SPH-DEM) \cite{wang2021visual}, represents both the granules and the fluid by Lagrangian particles.
Since this framework focuses on analyzing the forces and motions of individual particles, the simulated scale of physical effects is limited to the grain size. 
When significant interactions between granule and fluid arise, the particles become inadequate to accurately reflect the collective influence of the granule ensemble in the mixture.
For example, the water absorption in sand cannot only be regarded as a microscopic birth-and-death process of particles.
The latter, represented by the material point method (MPM) \cite{tampubolon2017multi, gao2018animating}, evolves the governing equations with Eulerian approach, treating both the granule and the fluid as continuous media.
This framework analyzes the movement patterns of particle groups on a larger scale and successfully reflects the statistical properties of granule ensembles.
However, it will encounter difficulties when capturing finer details of granule movement near the phase interface.

Fundamentally, the most critical issue in previous methods is that the description of sand granules follows the fluid framework, resulting in the sand movement depending on the fluid dynamics.
Specifically, in MPM-based approaches, analyzing the sand--water mixture from the same continuum perspective not only results in the loss of distinct discrete boundaries for granules, but also reduces the constitutive model of the sand--water mixture to smooth transitions between dry and soaked states, thereby neglecting its essential physical characteristics as a porous medium.
In the SPH-DEM method, although DEM is utilized to model rigid, ball-like sand granules, the SPH component represents sand as a traversable density cloud.
\rv{Near the sand--water interface, DEM particles experience capillary forces, but fail to hold water. While within the fluid interior, although no actual liquid bridges form between submerged particles, their interactions are still calculated using this capillary model.}
This unnatural treatment explains why, in their simulation, the fluid is neither absorbed nor expelled by the granules.
\rv{As demonstrated across scenarios, funnel experiments (Fig.~\ref{fig:funnel}, \ref{fig:funnel2}) shows that sand particles, with any radius, would flow through the neck without forming aggregations or obstructing water passage.}
Similarly in dam-break simulations (Fig.~\ref{fig:dam}), sand's obstruction effect lags behind water's infiltration process, leading sand deposits to prefer permitting water permeation rather than providing hydrodynamic resistance.

In this paper, we develop a novel \emph{Granule-In-Cell (GIC)} method that captures not only the general characteristic of mixture but also the discrete granule movement in fluid. 
In GIC, sands are considered as identical mesoscopic entities, and through ensemble averaging, their macroscopic transport equations are consistent with mesoscopic Newton's laws of motion, allowing us to obtain the collective behavior of sand.
To be specific, we use DEM to describe sand and Particle-In-Cell (PIC) \cite{harlow1964particle} to describe fluid. 
In the mesoscopic view, sand is driven by fluid, while from the macroscopic perspective, inspired by implicit density projection (IDP) \cite{kugelstadt2019implicit}, sand constrains the volume of fluid, thereby influencing fluid's movement.
Utilizing this bidirectional coupling, we can depict the characteristics of the sand--water mixture system at both levels, achieving a set of realistic visual simulation results.

Our technical contributions are summarized as follows:
\begin{itemize}[nolistsep]
    \setlength{\parskip}{0pt}
    \item A novel consistent sand--water coupling strategy reflecting the discrete characteristic of the granules,

    \item An extended IDP algorithm respecting the volume constrain that conserves the total volume, and

    \item Discretization schemes for the systematically derived forces from the ensemble averaging. 
    
\end{itemize}

\section{Related Work}
\begin{figure*}
    \centering
    \setlength{\imagelength}{.245\linewidth}
    \includegraphics[width=\imagelength]{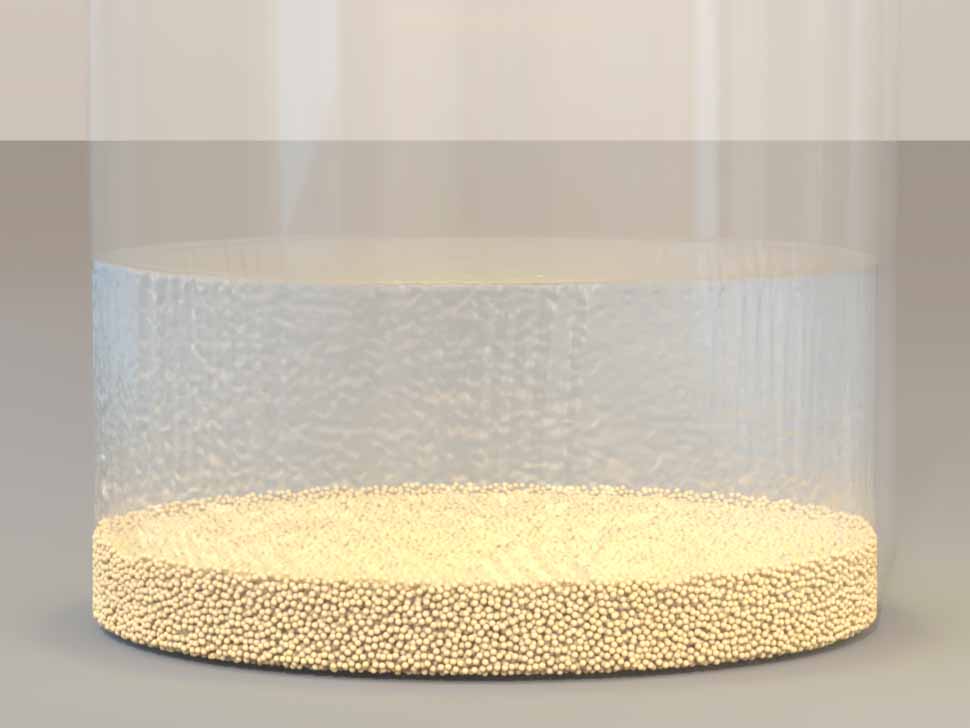}
    \hfill
    \includegraphics[width=\imagelength]{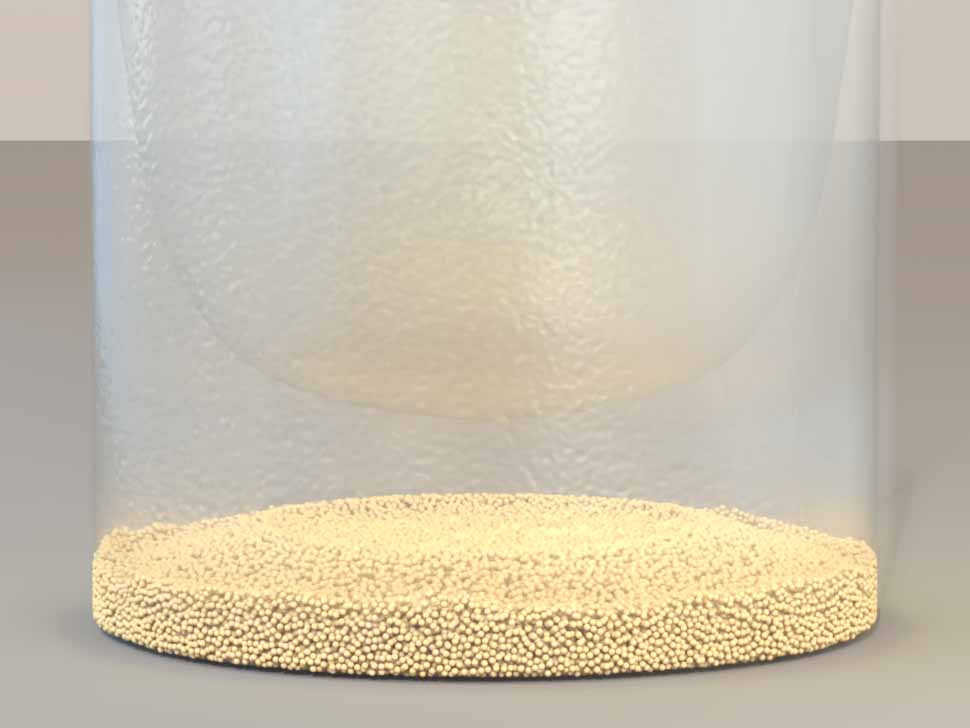}
    \hfill
    \includegraphics[width=\imagelength]{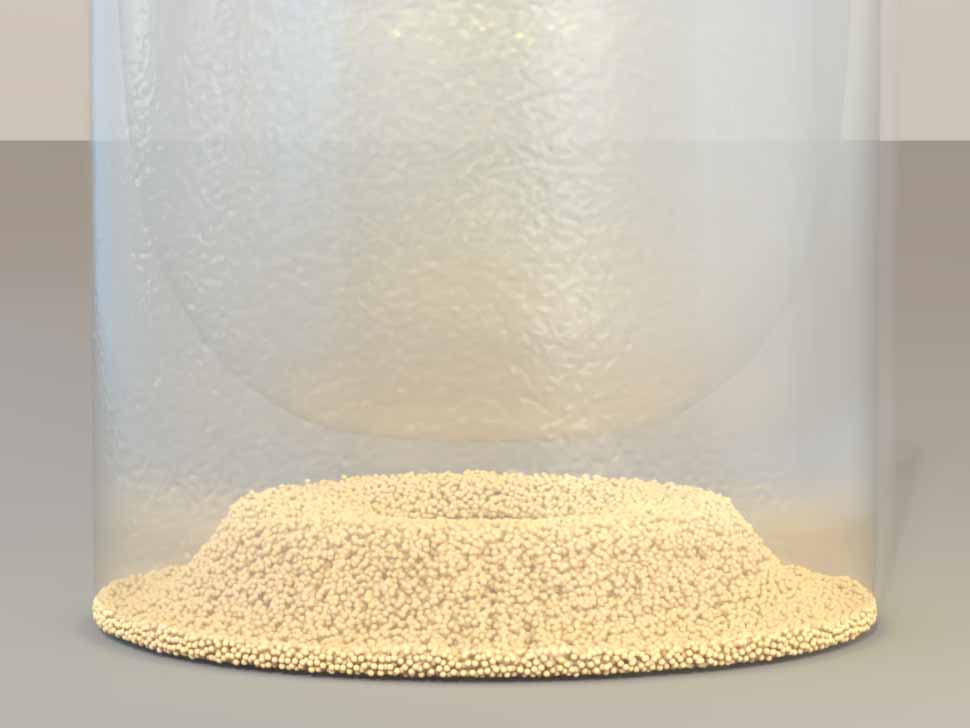}
    \hfill
    \includegraphics[width=\imagelength]{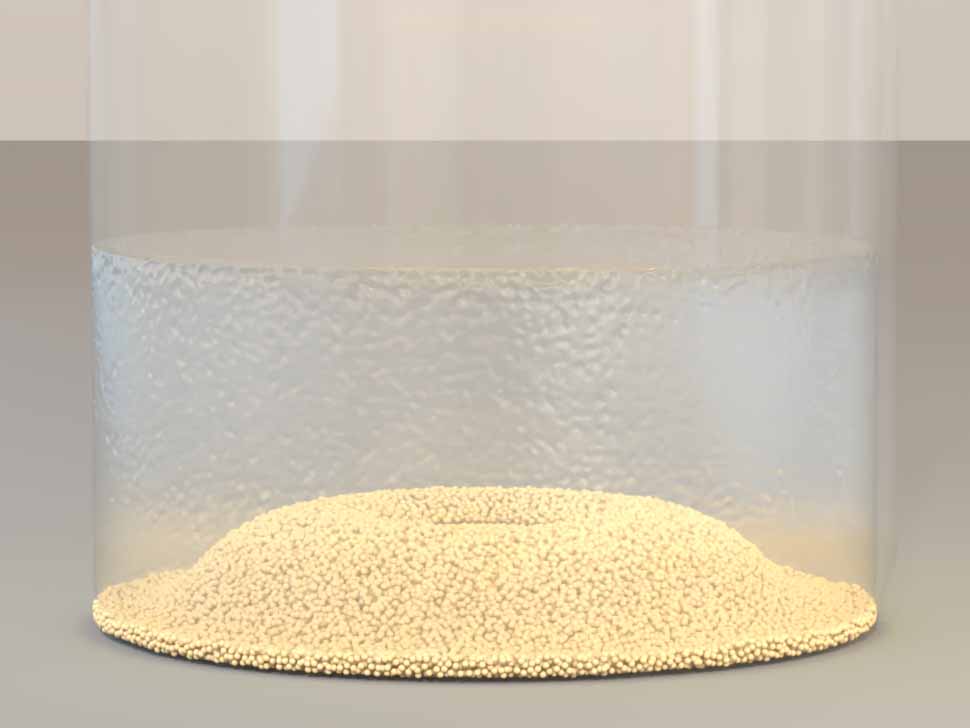}
    \\ \vspace{0.00667\linewidth}
    \includegraphics[width=\imagelength]{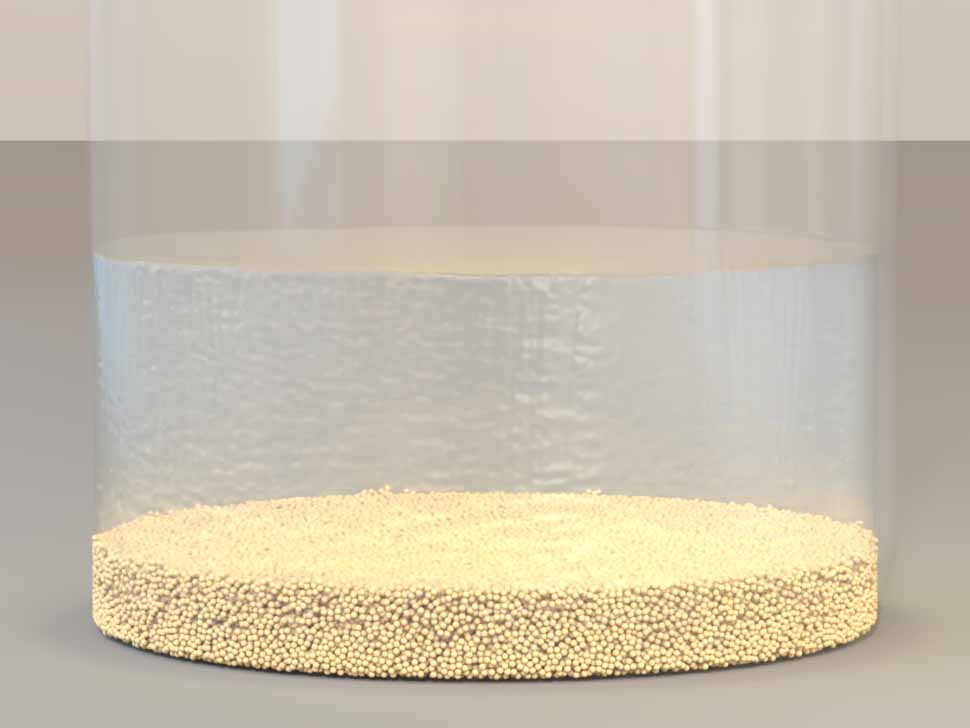}
    \hfill
    \includegraphics[width=\imagelength]{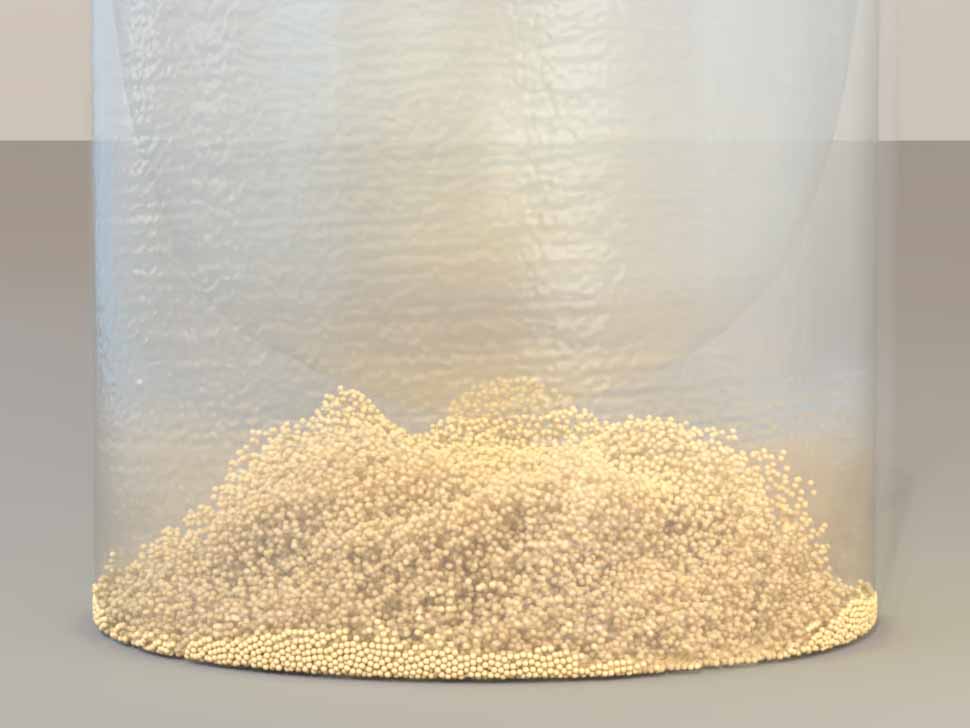}
    \hfill
    \includegraphics[width=\imagelength]{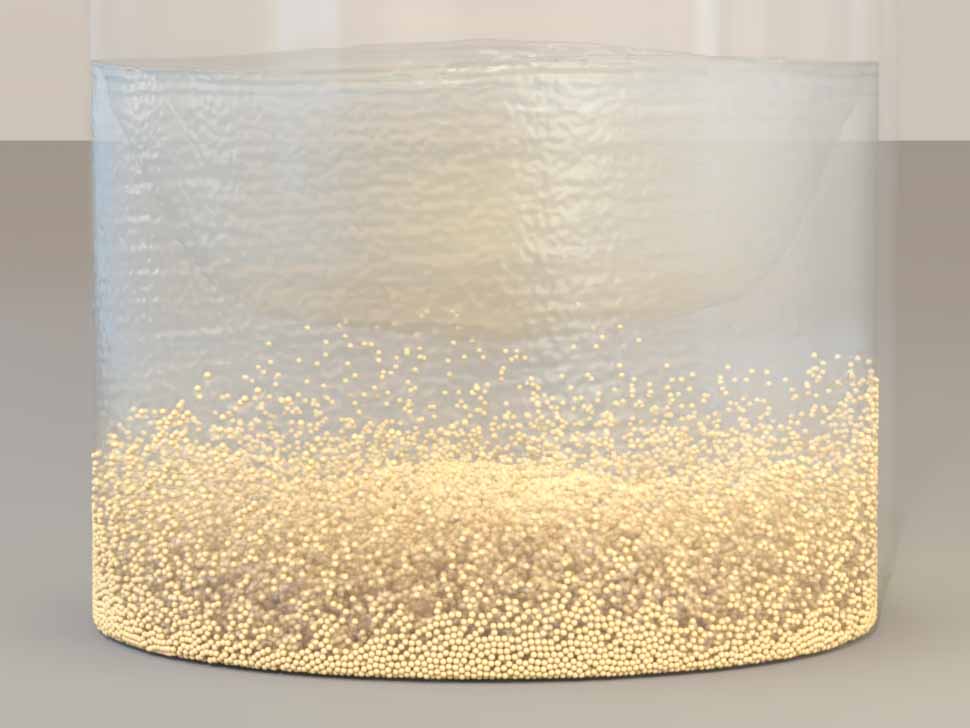}
    \hfill
    \includegraphics[width=\imagelength]{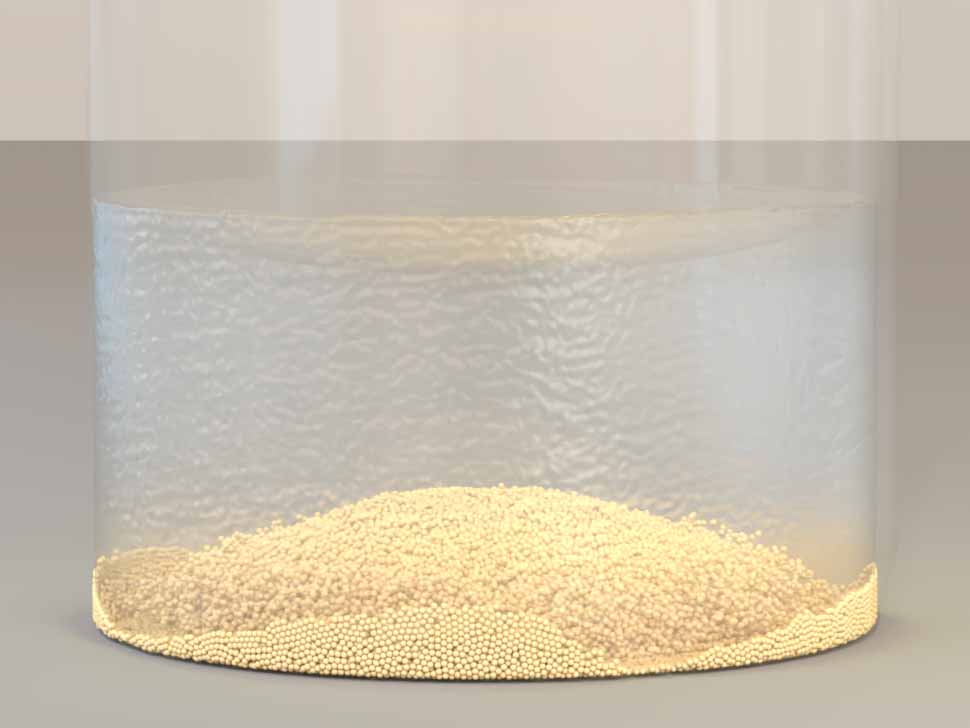}
    \\ \vspace{0.00667\linewidth}
    \includegraphics[width=\imagelength]{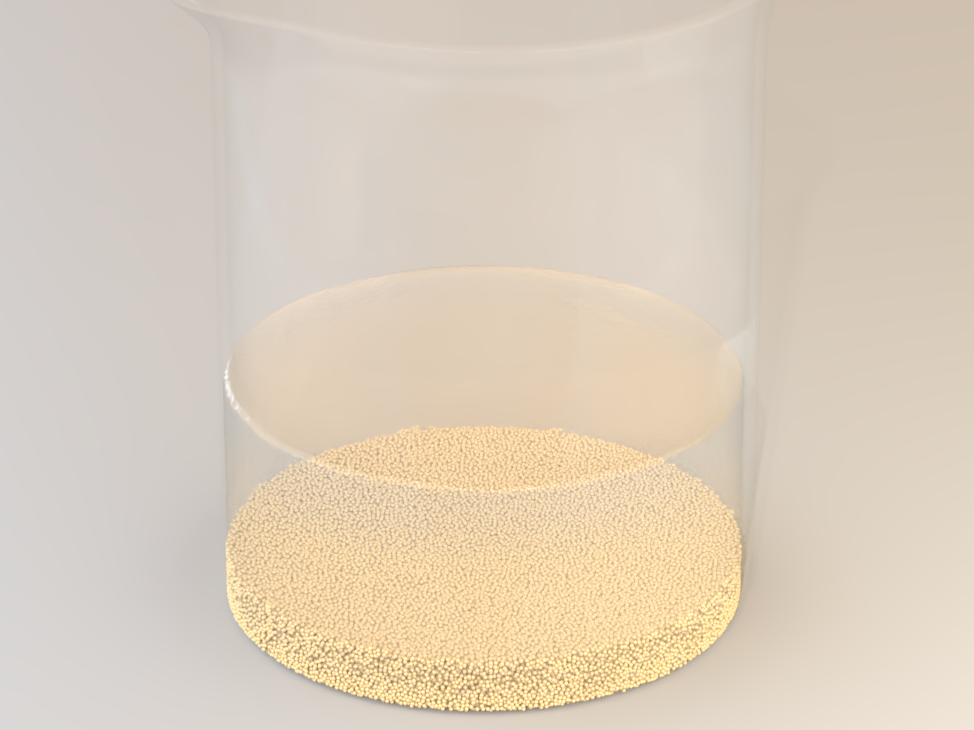}
    \hfill
    \includegraphics[width=\imagelength]{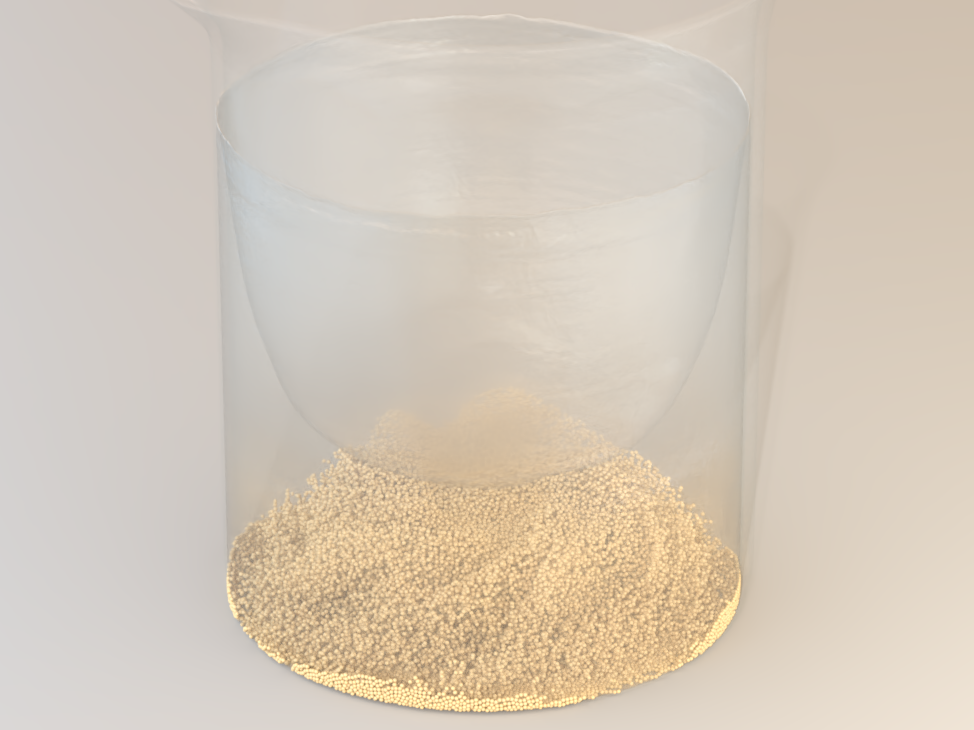}
    \hfill
    \includegraphics[width=\imagelength]{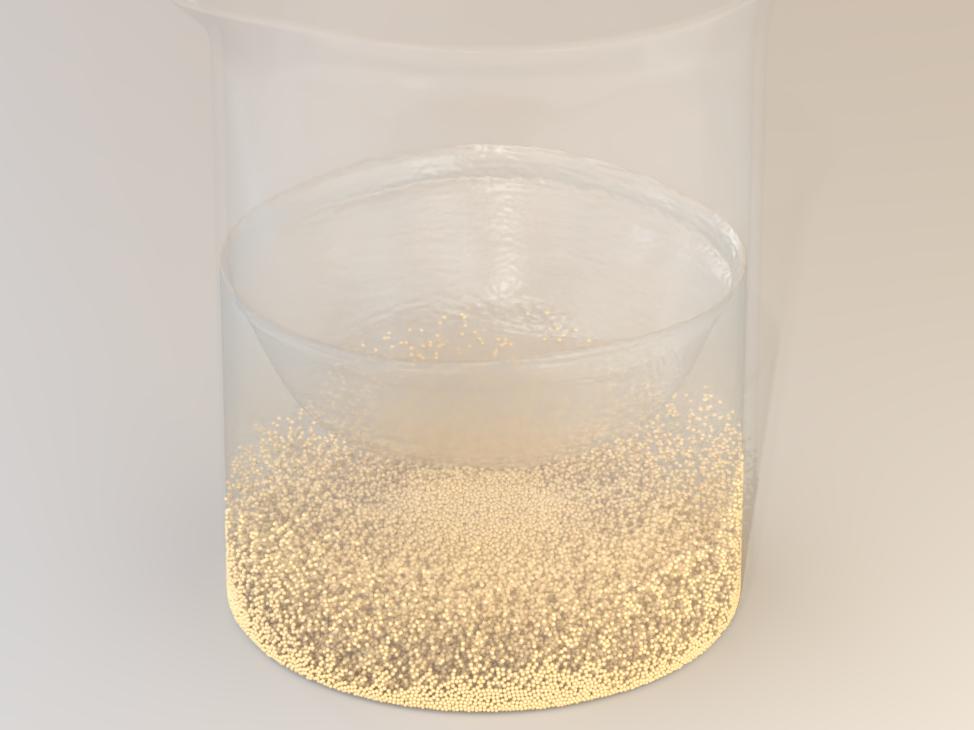}
    \hfill
    \includegraphics[width=\imagelength]{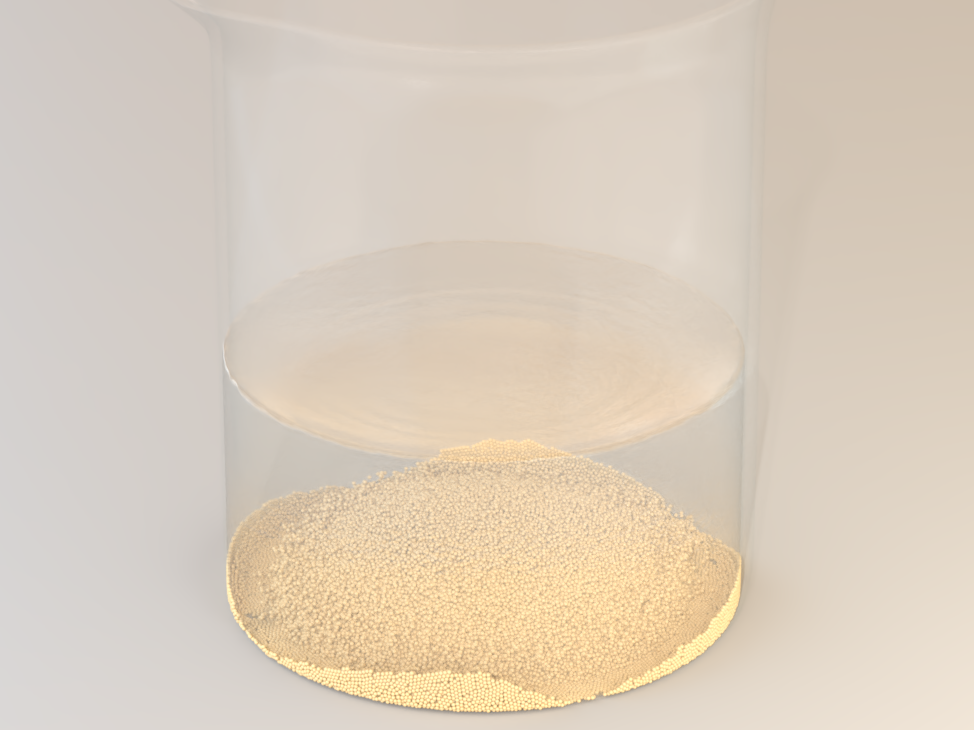}
    \\ \vspace{0.00667\linewidth}
    \includegraphics[width=\imagelength]{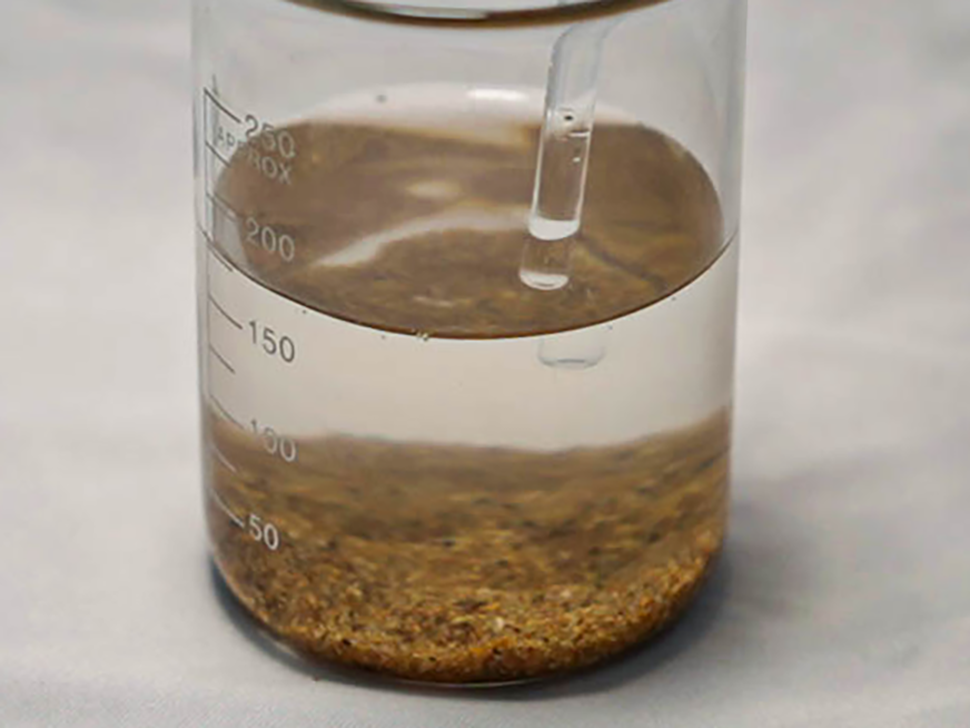}
    \hfill
    \includegraphics[width=\imagelength]{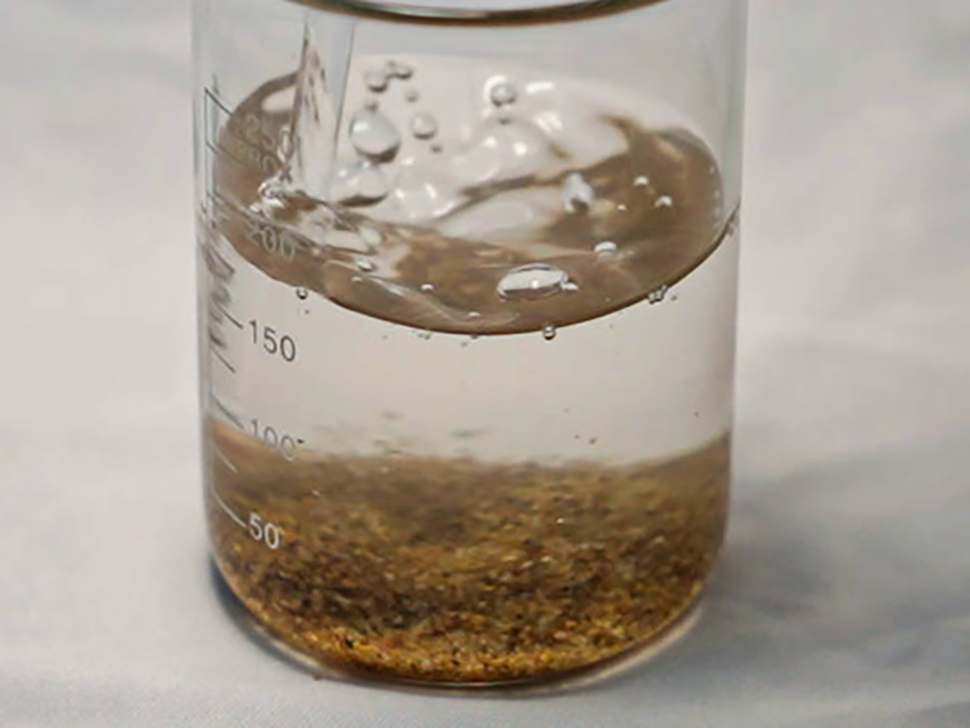}
    \hfill
    \includegraphics[width=\imagelength]{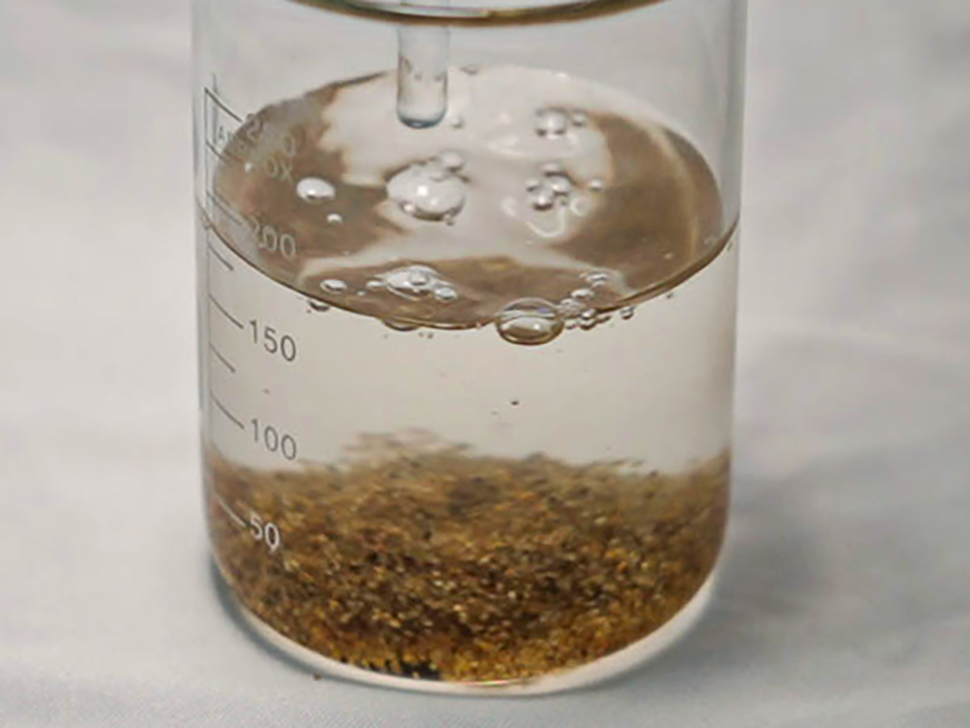}
    \hfill
    \includegraphics[width=\imagelength]{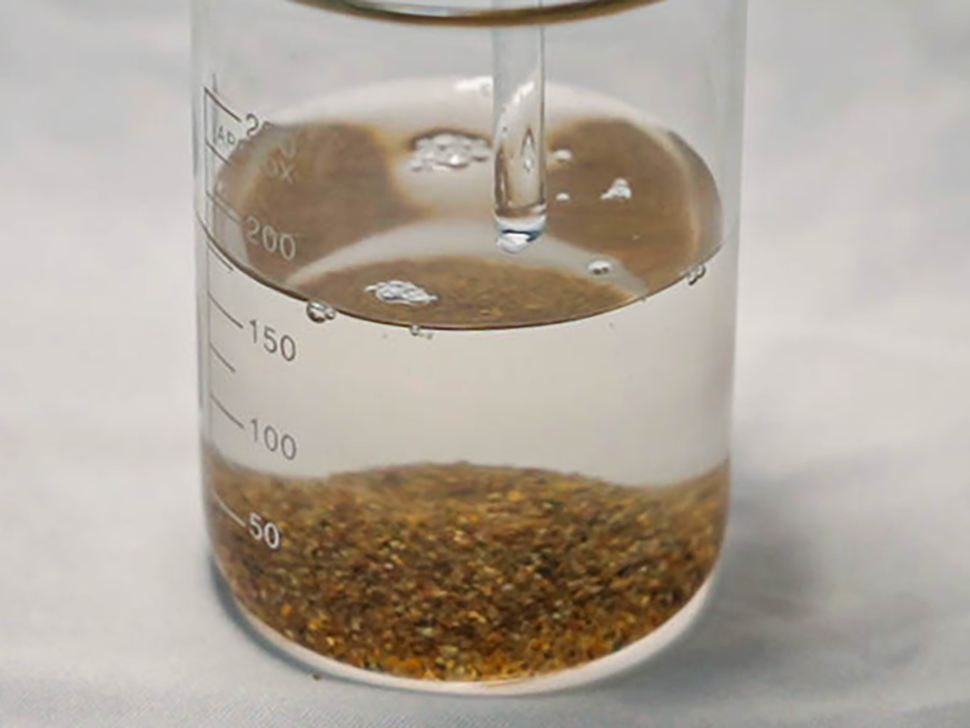}
\caption{Stir. In these figures, we show the comparison between simulation results and real experiments of sand motion in rotating water. \rv{The first rows shows the simulation result of MPM \cite{gao2018animating}. The second and third rows show the results of our method, with the second row displaying the side view and the third row the birds' eye view, illustrating the state of the sand at $t=0, 1, 2, 6\ \mathrm{s}$ after the tangential force is applied to the water. The fourth row shows the state of the sand during the real experiment, from the beginning of stirring to the final formation of a sand pile.}}
\label{fig:stir}    
\end{figure*}

We begin by reviewing the work closely related to our approach, with subsections focusing on strategies for PIC/FLIP solvers, sand simulation, and sand--water coupling, respectively.


\subsection{PIC/FLIP Methods}
Since its introduction by Harlow et al.\ \shortcite{harlow1955machine, harlow1964particle}, the PIC method has been applied to simulate a wide range of materials, including fluids and solids. 
To mitigate dissipation during information transfer between particles and the grid, the Fluid Implicit Particle (FLIP) method retrieves the increment of particle momentum from the grid \cite{brackbill1986flip, zhu2005animating}. The Affine Particle-In-Cell \cite{jiang2015affine} and Polynomial Particle-In-Cell \cite{fu2017polynomial} methods use higher-order shape functions in space to enhance the grid's representation capabilities and reduce back-transfer errors \cite{hu2018moving}.
These transfer schemes typically assume a sufficiently dense and uniform particle distribution but lack robust strategies to consistently ensure this condition. In practice, particles often exhibit clumping and void formation due to suboptimal interpolation schemes and accumulated numerical errors.
To address this, the Power PIC method \cite{qu2022power} defines the kernel function on particles rather than grids. Using particle volume as the source and grid volume as the target, it employs the optimal transport equation to determine interpolation weights. Particle positions are then determined by the centroid of this weight function. This method ensures an incompressible fluid representation with uniformly distributed particles in divergence-free fields.

In the simulation of sand--water mixtures, it is crucial to 
ensure the conservation of the total volume. When multiple components coexist, the divergence of a single phase is no longer zero; instead, the mixture must satisfy the mass transport equation.
Our approach builds on the concept of IDP\cite{kugelstadt2019implicit}, which addresses density variations during fluid simulation. By projecting particle positions to achieve consistent density on the grid, this method ensures uniform particle distribution, maintains incompressibility, and preserves visual fidelity, even with large time steps.
We extended IDP by ensuring that the fluid and sand particles complement the spaces of each other, using projection to maintain local volume consistency of the mixture.

\subsection{Sand Simulation}
For sand simulations, the static friction complicates the macroscopic continuous description. \citet{zhu2005animating} were the first to apply the PIC/FLIP method to sand simulation, decomposing the motion of sand particles into rigid movement and incompressible shearing flow based on the yield surface. Recently, \citet{klar2016drucker} introduced the Drucker-Prager elastoplastic model into the MPM constitutive model. While projecting onto the Drucker-Prager yield surface can eliminate relative motion between particles, it cannot guarantee interlocking within the friction cone. \citet{yue2018hybrid} used MPM to save computational costs in DEM, employing DEM particles externally and MPM quadratic points internally, while ensuring both to reflect the same motion in reconciliation zone. This method interlocks sand particles within the friction cone near phase boundaries, effectively capturing clogging effects related to particle size.

Another method for simulating sand involves treating each granule as an individual entity and modeling their motion and interactions separately. Each granule is modeled as a soft ball with interaction forces determined by penalty functions based on relative displacement with its neighbors
\cite{cundall1979development,kruggel2007review}. In graphics, \citet{bell2005particle} applied Hertz contact force to set the normal part and Coulomb law to obtain the tangential part. The interactions in DEM are simple and flexible, effectively capturing the granularity and high-frequency details. When handling complex scenes, the key is on developing suitable interaction models. SPH further soften the boundaries by emulating the Dirac $\delta$ function \cite{monaghan1992smoothed, muller2003particle}. The values carried by the particles are treated as marks in the corresponding spatial locations. To emphasize fluid incompressibility, iterative prediction-correction schemes are employed to maintain constant density \cite{he2012local, solenthaler2009predictive}, and pressure projection is applied to ensure divergence-free conditions at particle positions \cite{bender2015divergence, ihmsen2013implicit, takahashi2018efficient, liu2024dual}.

\subsection{Sand--Water Coupling}
\paragraph{Eulerian Viewpoint} When coupling continuum sand with a continuum fluid, Power Plastics \cite{qu2023power} used weakly compressible fluid \cite{xie2023contact} for the water phase and the Herschel-Bulkley model \cite{yue2015continuum} for the sand phase. The bidirectional coupling was achieved through interaction forces on the same MPM grid. However, this single grid requires identical velocities for both medium, leading the homogenized mixture moving under the summation of each force based on their own physical properties rather than that of the mixture.

Both \citet{tampubolon2017multi}  and \citet{gao2018animating} used two sets of grids to discretize the governing equations of the two phases. The former considered the transition of sand from dry to wet, including mass and momentum exchange between sand and water. And the latter derived equations for the mixture, coupling the two phases through inter-phase forces while maintaining the constraint of incompressibility. When simulating weakly compressible fluids modeled using MPM, explicit time integrators necessitate a very small time step due to the large bulk modulus. In the case of sand particles using MPM, the delayed stress response to strain results in motion dominated by fluid behavior, which makes it difficult to capture the rigid response of the sand particle system to external forces accurately.


\paragraph{Lagrangian Viewpoint} \rv{A series of SPH based methods represent materials as particles, where spatial information is derived via particle interpolation.
To model sand using SPH, \citet{yan2016multiphase} employed the Drucker--Prager model and developed a unified multiphase SPH framework to handle phase changes \cite{yang2017unified, jiang2020divergence}. \citet{ren2021unified} further developed this approach by recording the outer and inner volume fractions on fluid particles to distinguish their spatial relationship with sand grains acting as porous materials. While the fluid phase parameters passively advect with a mixture model, these parameters are also transported between different SPH particles based on a drift velocity \cite{ren2014multiple}.}

\rv{For sand particles modeled in DEM, they continuously absorb SPH particles, increasing their moisture ratio while being influenced by external forces from the water \cite{rungjiratananon2008real}.} \citet{wang2021visual} extended this approach by employing a seepage model to simulate the infiltration impact from water on soil structure. They introduced a saturation-based capillary model to account for cohesive forces, complementing momentum exchange, buoyancy, and drag forces.
However, as the moisture ratio of soil depends on the volume of surrounding water particles, accurately calculating it near the center of ``spherical clumps'' becomes challenging. Additionally, using the Rayleigh timestep for both fluid and sand simulations, which is much shorter than the fluid's  Courant–Friedrichs–Lewy (CFL) timestep, results in significant computational inefficiency.

\paragraph{PIC-DEM Coupling} Compared to continuous schemes, DEM independently handles each particle and resolve their complete dynamic processes without additional equations for conservation, thereby providing higher spatial resolution and capable of capturing high-frequency details. In Computational Fluid Dynamics (CFD), when coupled with DEM, it's crucial to maintain smooth fractions over time, even in small cells containing large granules. The mapping from particle to grid in CFD is also referred to as coarse graining, averaging, or aggregation \cite{xiao2011algorithms, zhu2002averaging}. In earlier studies, coupling was performed on the same set of grids, requiring a sufficiently large grid size to accurately represent the particle volume fraction \cite{anderson1967fluid}. Otherwise, if particles cross cell boundaries, sharp changes will occur, leading to significant spatial and temporal fluctuations and resulting in numerical integration instability \cite{link2005flow, link2008pept}. In the two-layer mesh approach, a coarser grid is used to average the particles and assumes that the fluid cells have the same average voidage of sand grains as the coarser grid \cite{deb2013novel, su2015two, jing2016extended}, or that the particle volume fractions can be expanded within the support domain \cite{zhang2023calculation, sun2015diffusion, zhao2022euler, zhang2024simple}.

\newcommand{\sm}[1]{{\scriptscriptstyle#1}}
\newcommand{\tf}{\sm{\mathrm{f}}}
\newcommand{\ts}{\mathrm{s}}
\section{Background}
In this section, we present the governing equations in sand--water mixture systems. We use DEM to characterize the sand granules and the PIC/FLIP method to characterize the motion of the water. After a brief review of background in \S\ref{sec:cag} and \S\ref{sec:gel}, \S\ref{sec:couple} describes the specific form of coupling force.

\subsection{Discrete Viewpoint of Granular Material}
\label{sec:cag}


DEM treats each granule as an independent entity and simulates the entire system in a Lagrangian framework. Following the soft ball model, each sand particle is simplified to an identical sphere with radius $r$, undergoing translational motion governed by Newton's laws \cite{cundall1979development}.
The model calculates the elastic and frictional forces between particles based on their relative displacements, adopting the same formulation as \citet{hentz2004discrete} for the $i$-th and $j$-th particles within a distance $d_{ij} = \norm*{\vb*{x}_j - \vb*{x}_i} < 2r$, the force on the $i$-th particle is simplified as
\begin{subequations}
    \begin{align}
    \vb*{F}_{i, \mathrm{n}} &= k_{\mathrm{n}}\qty(2r - d_{ij}) \hat{\vb*{x}}_{ij},\\
    \label{eq:k_n}
    \vb*{F}_{i, \mathrm{t}} &= \min\qty{\norm*{\vb*{f}_{i, \mathrm{t}}}, \norm*{\vb*{F}_{i, \mathrm{n}}} \tan{\varphi}}\hat{\vb*{f}}_{i, \mathrm{t}},\\
    \vb*{f}_{i, \mathrm{t}} &= -k_{\mathrm{t}}\qty(\vb*{v}_{ij} - \vb*{v}_{ij}\vdot \hat{\vb*{x}}_{ij}\hat{\vb*{x}}_{ij}),
    \end{align}
\end{subequations}
%
where $k_{\mathrm{n}}, k_{\mathrm{t}}$ are the normal and tangential stiffness coefficients respectively, related to Young's modulus and Poisson's ratio, $\varphi$ is the friction angle, $\hat{.}$ denotes the unit vector in the direction of the vector, i.e. $\hat{\vb*{x}}=\vb*{x}/\norm*{\vb*{x}}$ and $\vb*{x}_{ij} \coloneqq \vb*{x}_i - \vb*{x}_j, \vb*{v}_{ij} \coloneqq \vb*{v}_i - \vb*{v}_j$.

\subsection{Continuous Viewpoint of Mixture}
\label{sec:gel}

\begin{figure*}[h]
    \centering
\includegraphics[width=.9\linewidth]{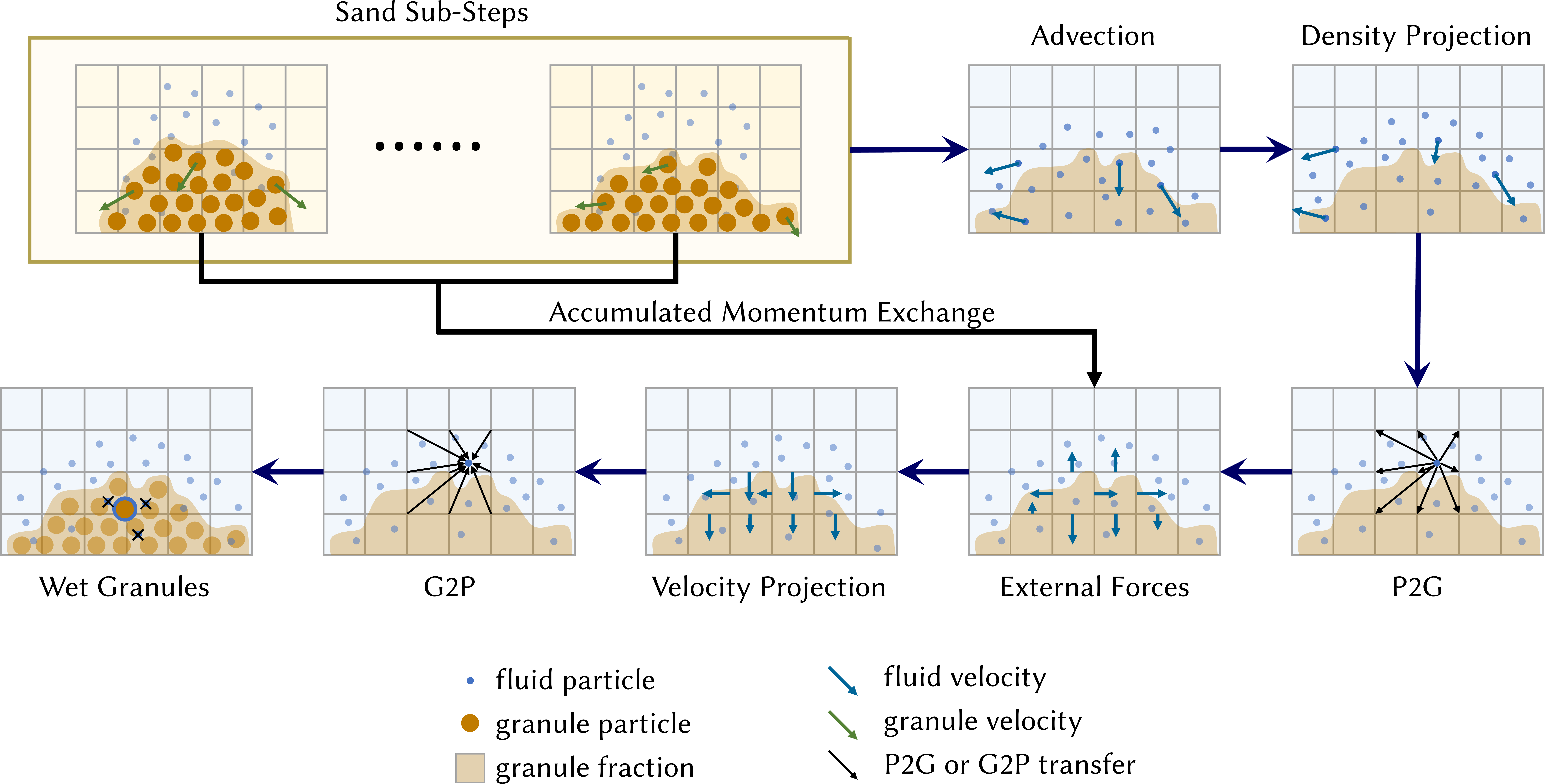}
\vspace{-.5em}
\caption{Algorithm steps of the Granule-In-Cell method. The frame on the top left portion illustrate the sub-steps for sand, and the following scenes depict the steps for calculating fluid. Note that for one step in our simulation, there are multiple sand sub-steps. And the coupling forces calculated in each sub-steps are accumulated and transferred to the water as external forces like the black arrow illustrates. }
\label{fig:alg}    
\end{figure*}

In sand--water mixtures, the space is occupied by both granules and fluid, necessitating a combined treatment of the two phases in the analysis. Within this framework, the fluid phase is treated as one component of a multiphase flow system, and its mass conservation equation is given by
\begin{equation}
    \label{eq:mass}
    \pdv{\alpha_\tf}{t} + \div\qty(\alpha_\tf\vb*{v}_\tf) = 0,
\end{equation}
where $\vb*{v}_\tf$ denotes the velocity field of the fluid, $\alpha_\tf$ represents the average volume fraction in the vicinity of a point, and the density is assumed constant, $\rho = \rho_0$; rigorous definition is provided in \S\ref{sec:kinetic}.
Applying Newton's second law, the fluid dynamics equation is derived as
\begin{equation}
    \label{eq:momentum}
    \pdv{\qty(\alpha_\tf \rho \vb*{v}_\tf)}{t} + \div\qty(\alpha_\tf\rho\vb*{v}_\tf\vb*{v}_\tf) = \alpha_\tf \rho \vb*{b}_\tf - \alpha_\tf\grad p + \div(\alpha_\tf \vb*{T}) + \vb*{F}_\ts,
\end{equation}
where $\vb*{b}_\tf$ represents the body force, $p$ is the internal pressure, and $\vb*{F}_\ts$ is the force exerted by sand particles on the fluid. The deviatoric stress tensor $\vb*{T}$ for a Newtonian fluid is given by $\vb*{T} = \mu(\grad \vb*{v}_\tf + \grad \vb*{v}_\tf^{\top})$, where $\mu$ denotes the fluid's viscosity. For detailed derivations, refer to \S\ref{sec:kinetic}.

In the mixture, the sum of the volumes occupied by sand and water in any neighborhood of a point must not exceed the total volume of that space. 
Let $\alpha_\ts$ denote the average volume fraction of granules. Above constraint is formulated as
\begin{equation}
    \label{eq:constraint_0}
    \alpha_\ts + \alpha_\tf \leq 1,
\end{equation}
where the equality holds in fluid regions, since the fluid could fill the interstitial space between the granules.

\subsection{Forces on Granule in Liquid}
\label{sec:couple}
For granules moving in fluid, the influence of fluid motion must be considered in addition to inter-granule collisions. From a microscopic perspective, the non-uniform flow of the fluid results in asymmetric motion of fluid molecules near the surface of the spherical granule. This asymmetry manifests macroscopically as a pressure force \cite{stokes1851effect, basset1888treatise, boussinesq1885resistance, oseen1927hydrodynamik}. By integrating over the surface of the granule, the total force can be decomposed into three components: the pressure gradient force corresponds to the fluid pressure gradient at the granule's position, evaluated in its absence, $\vb*{F}_{\mathrm{p}}$; the drag and lift forces caused by the velocity difference between the two medium, $\vb*{F}_{\mathrm{d}}$; and the virtual mass force arising from the changes in the velocity of fluid induced by the granule's motion, $\vb*{F}_{\mathrm{v}}$. \citet{maxey1983equation} pointed out that these forces have the following forms:
\begin{subequations}
    \begin{align}
        \vb*{F}_{\mathrm{p}} &= -\frac{4}{3}\pi r^3 \grad p_\tf, \label{eq:gradient-force}\\
        \vb*{F}_{\mathrm{d}} &= 12\pi r^2 \rho_\tf \mu \norm*{\vb*{v}_\tf - \vb*{v}_\ts} \qty(\vb*{v}_\tf - \vb*{v}_\ts),\\
        \vb*{F}_{\mathrm{v}} &=\frac{2}{3}\pi r^3 \rho_\tf\qty(\dv{\vb*{v}_\tf}{t} - \dv{\vb*{v}_\ts}{t}),
    \end{align}
\end{subequations}
where $\vb*{v}_\ts$ is the velocity of the granule. For our model, since sand particles are limited to translational motion, the Magnus force, arising from rotational effects, is not considered.

The presence of a fluid not only generates the three types of forces mentioned above but also influences the interactions between granules. 
One granule perturbs the velocity distribution of the fluid, thereby inducing indirect interactions on the other granules.
However, current research is limited to two-particle models, with no reliable results available for multi-particle systems.
\citet{jeffrey1984calculation} analytically calculated this interaction under the infinite boundary condition. This interaction depends on the relative velocity of the two granules and holds the similar form as the tangential interaction force in DEM collisions, thus requiring no additional treatment. However, when performing ensemble averaging on mesoscopic, the volume fraction gradient of sand $\grad \alpha_\ts$ introduces an additional macroscopic force known as the concentration gradient force \cite{hsu2003two, zhong2014drift}:
\begin{equation*}
    \vb*{F}_{\alpha} = -\frac{\rho_\ts \vb*{D}_\ts}{\alpha_{\tf} \tau_\ts} \vdot \grad \alpha_\ts, \tag{5d} \label{eq:density_gradient}
\end{equation*}
where $\vb*{D}_\ts$ denotes the diffusion tensor and $\tau_\ts$ represents the relaxation time of the particle velocity; detailed in \S\ref{sec:kinetic}.


\section{Algorithm}
\label{sec:alg}


Before presenting our method in detail, we first provide an overview of the steps involved in GIC, which is summarized in Figure~\ref{fig:alg}.
Our method couples DEM granules and the fluid through the grid in PIC/FLIP, and employs time integration similar to the symplectic Euler method that the fliud portion follows after the sand portion.

\paragraph{Sand Step}
We subdivide the total time step into $N = \lceil \Delta t / \Delta t' \rceil$ parts and solve for the granules in each substep (Alg. \ref{algorithm2}). 
During each substep, we compute fluid coupling forces (\S \ref{sec:flow_force}, \S \ref{sec:concen-force}) and resolve granule collisions (\S \ref{sec:cag}), updating granule positions to $\vb*{x}_{\ts}^{t+1}$ and transfering momentum exchange $\vb*{F}_{\ts}$ to the grid.

\paragraph{Water Step}
The process of water portion is developed on standard PIC/FLIP method.
Based on the sand position $\vb*{x}_{\ts}^{t+1}$, the volume fraction of granules $\alpha_{\ts}^{t+1}$ is determined, which is used for pressure projection (\S \ref{sec:target}). The density projection is employed after advection and the velocity projection undergoes after applying external forces and the momentum exchange (\S \ref{sec:flow_force}). Finally, near the sand--water interface, the sand get wet (\S \ref{sec:wet}).

\begin{algorithm}
\caption{Granule-In-Cell Time Integrator}\label{algorithm1}

$(\vb*{x}_{\ts}^{t + \Delta t}, \vb*{v}_{\ts}^{t + \Delta t}, \vb*{F}_{\ts})\gets$ SandSubstep($\vb*{x}_{\ts}^t, \vb*{v}_{\ts}^t$) \;
$\alpha_\tf^{t + \Delta t} \gets$ CalTargetFraction($\vb*{x}_{\ts}^{t + \Delta t}$) \Comment*[r]{\S \ref{sec:target}} 
$\vb*{x}_\tf' \gets$ AdvectParticle($\vb*{x}_\tf^t, \vb*{v}_{\mathrm{h}}^t$)\;
$\vb*{x}_\tf^{t + \Delta t} \gets$ DensityProjection($\alpha_\tf^{t + \Delta t}, \vb*{x}_\tf'$) \Comment*[r]{\S \ref{sec:target}}
$\vb*{v}_{\mathrm{h}}' \gets$ P2G($\vb*{x}_\tf^{t + \Delta t}, \vb*{v}_\tf^t$)\;
$\vb*{v}_{\mathrm{h}}'' \gets$ ApplyCoupling($\vb*{F}_{\ts}$) \Comment*[r]{\S \ref{sec:flow_force}}
$\vb*{v}_{\mathrm{h}}^{t + \Delta t} \gets$ VelocityProjection($\alpha_\tf^{t + \Delta t}, \vb*{v}_{\mathrm{h}}''$) \Comment*[r]{\S \ref{sec:target}} 
$\vb*{v}_\tf^{t + \Delta t} \gets$ G2P($\vb*{v}_{\mathrm{h}}^{t + \Delta t}$) \;
WetDEMParticle($\vb*{x}_{\ts}^{t + \Delta t}, \vb*{x}_\tf^{t + \Delta t}, \vb*{v}_{\ts}^{t + \Delta t}, \vb*{v}_\tf^{t + \Delta t}$) \Comment*[r]{\S \ref{sec:wet}} 
\end{algorithm}

\begin{algorithm}
\caption{SandSubstep}\label{algorithm2}
\KwIn{Positions $\vb*{x}_{\ts}^t$, Velocities $\vb*{v}_{\ts}^t$}

$(\vb*{x}_{\ts}^0, \vb*{v}_{\ts}^0) \gets (\vb*{x}_{\ts}^t, \vb*{v}_{\ts}^t)$\;
\For{$i \gets 0$ \KwTo $N$}{
    $\vb*{f}_{\ts}^i \gets$ CalCoupling($\vb*{x}_{\ts}^i, \vb*{v}_{\ts}^i, \Delta t'$) \Comment*[r]{\S \ref{sec:flow_force}, \S\ref{sec:concen-force}}
    $\vb*{F}_{\ts}^i \gets$ TransferCoupling(\,$\vb*{f}_{\ts}^i$)\;
    CollisionHandling(\,$\vb*{f}_{\ts}^i$) \Comment*[r]{\S \ref{sec:cag}}
    $(\vb*{x}_{\ts}^{i+1}, \vb*{v}_{\ts}^{i+1}) \gets$ MoveParticle($\vb*{x}_{\ts}^i, \vb*{v}_{\ts}^i, \vb*{f}_{\ts}^i$)\;
}

\Return ($\vb*{x}_{\ts}^N, \vb*{v}_{\ts}^N$, Sum($\vb*{F}_{\ts}^i$))
\end{algorithm}

\section{Methods}

This section introduces the discretization methods and the computational procedures for calculating interactions between granule and fluid in GIC. \S \ref{sec:target} introduces the projection methods to compute the volume-consistent fluid field in the presence of sand. 
\S \ref{sec:flow_force} and \S\ref{sec:concen-force} introduces the calculation of forces on granules. In \S \ref{sec:wet}, we propose a method to keep fluid trapped in sand through the absorption of PIC particles by granules.

\subsection{Sand as Projection Target}
\label{sec:target}
\begin{figure}
    \centering
    \setlength{\imagelength}{.495\linewidth}
    \includegraphics[width=\imagelength]{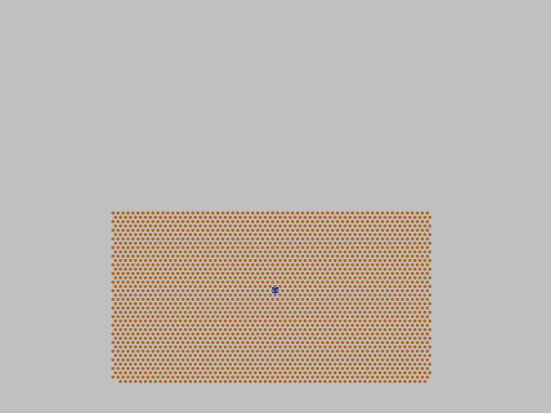}
    \hfill
    \includegraphics[width=\imagelength]{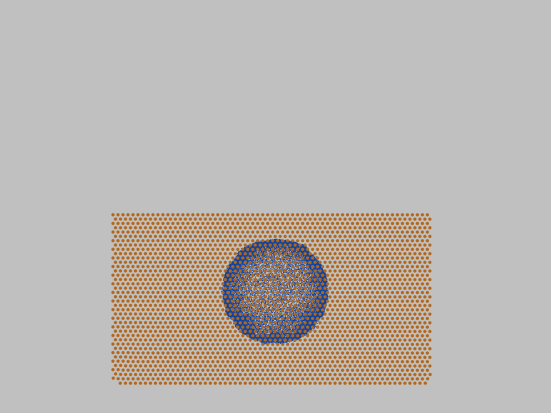}\\
    \vspace{0.01\linewidth}
    \includegraphics[width=\imagelength]{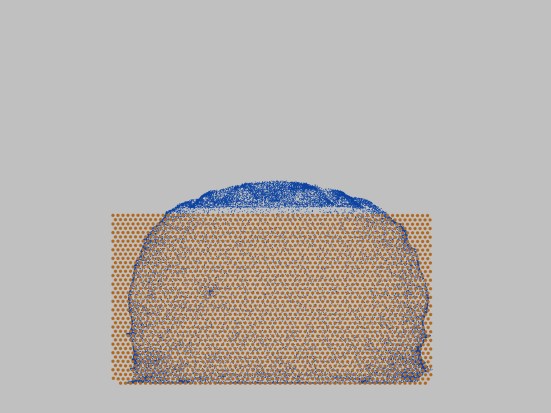}
    \hfill
    \includegraphics[width=\imagelength]{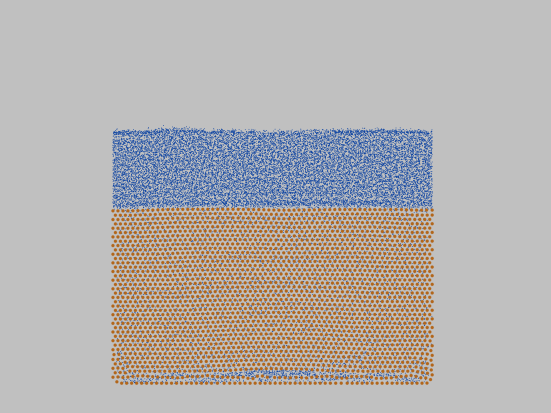}
\vspace{-2em}
\caption{Stability test for our method. We compress 60,000 fluid particles within one cell of $64^2$ grid, accompanied by fixed sand particles. With our projection method, the fluid can recover and move to the outside of the sand in fewer than 20 steps.}
\label{fig:density_2d}    
\end{figure}
In typical incompressible fluid simulation, the mass conservation law is often simplified to the divergence-free velocity field condition, neglecting non-uniform density distributions. However, in sand--water mixtures, the presence of sand creates inhomogeneous space, preventing the application of the divergence-free velocity field condition. As a result, we refer to IDP \cite{kugelstadt2019implicit} and develop our own fraction projection method.

From the Eulerian perspective, each cell is a mixture of sand and water. As written in Eq. \eqref{eq:constraint_0}, within the fluid, we enforce the constraint
\begin{equation}
    \label{eq:constraint}
    \alpha_\ts + \alpha_\tf = 1
\end{equation}
in every cell, where $\alpha_\tf$ denotes the volume fraction of fluid. They are computed at the grid cell centers by mapping the volumes as 
\begin{subequations}
    \begin{align}
    \alpha_\ts\qty(\vb*{x}) &= \frac{1}{V} \sum_i V_{\mathrm{s}_i}N\qty( \vb*{x}_{\mathrm{s}_i}-\vb*{x}),\\
    \alpha_\tf\qty(\vb*{x}) &= \frac{1}{n} \sum_i N\qty( \vb*{x}_{\tf_i}-\vb*{x}),
    \label{eq:alpha_fluid}
    \end{align}
\end{subequations}
where $V$ denotes the volume of a single grid cell, $i$ indexes the particles in the neighborhood of position $\vb*{x}$, $V_{\mathrm{s}_i}$ represents the volume of the $i$-th sand particle, \rv{$N(\cdot)$ is the interpolation kernel} and $n$ denotes the initial number of fluid particles per cell.

\begin{figure*}
    \centering
    \setlength{\abovecaptionskip}{0.cm}

    \subcaptionbox*{}[\linewidth]{
    \includegraphics[width=\linewidth]{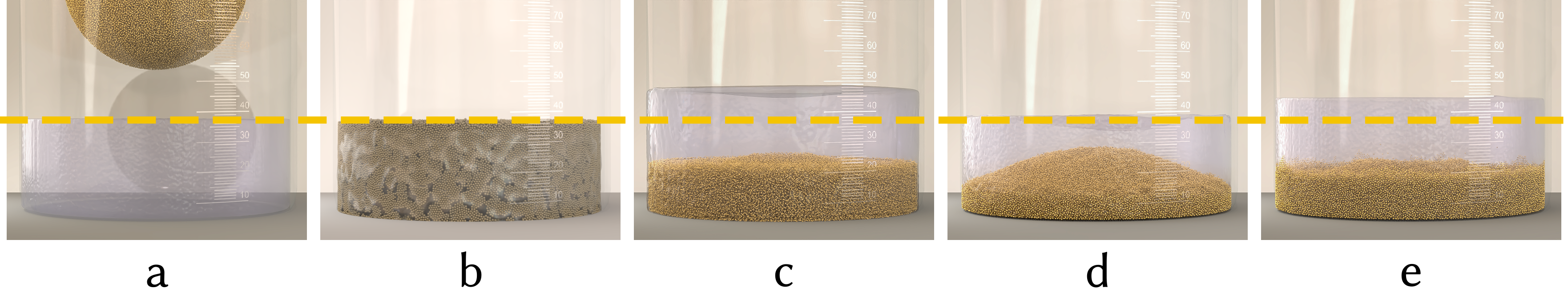}}\\
\vspace{-1.5em}
\caption{Comparison of the volume conservation ability.
In this experiment, a large ball falls into water. From left to right, the first images show the initial state and the followed 4 images are the steady state of SPH-DEM \cite{wang2021visual}, MPM \cite{gao2018animating}, our GIC method without IDP and the full GIC method.
The yellow dash line indicating the initial water level.}
\label{fig:density_3d_1}    
\end{figure*}

\begin{figure}[t]
    \centering
    \setlength{\imagelength}{\linewidth}
    \includegraphics[width=.9\imagelength]{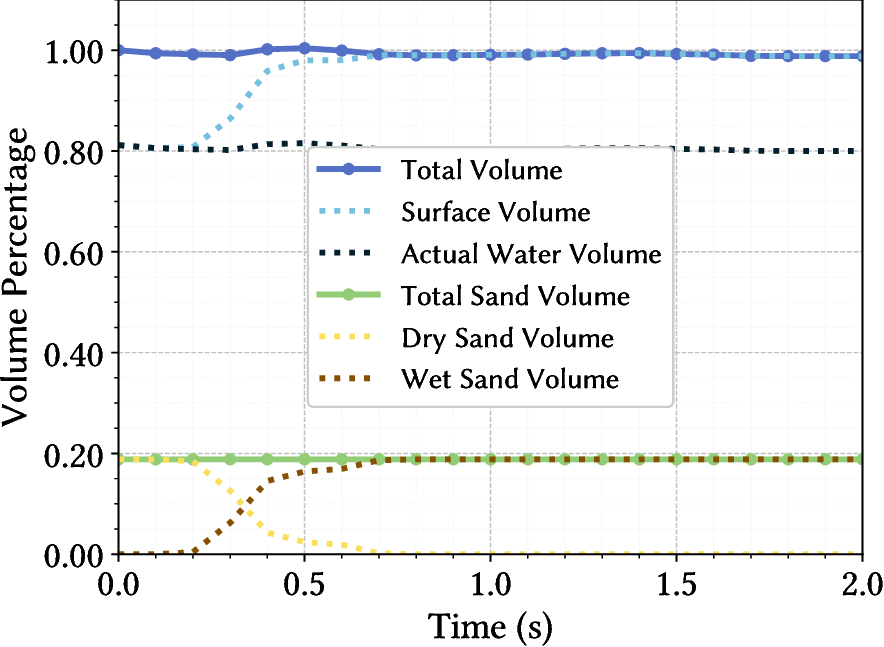}
    \vspace{-.7em}
\caption{Volume for different components of the simulation illustrated in Figure~\ref{fig:density_3d_1} using full GIC method. The initial total volumes is set as $1$. The Surface Volume is obtained by reconstructing the water surface and includes both the Actual Water Volume and Wet Sand Volume. The Total Sand Volume is the sum of Dry Sand Volume and Wet Sand Volume. As shown, our algorithm successfully preserves the volume of both substances.}
\label{fig:density_3d_2}    
\end{figure}

\begin{figure}[t]
    \centering
    \setlength{\imagelength}{\linewidth}
    \includegraphics[width=0.9\imagelength]{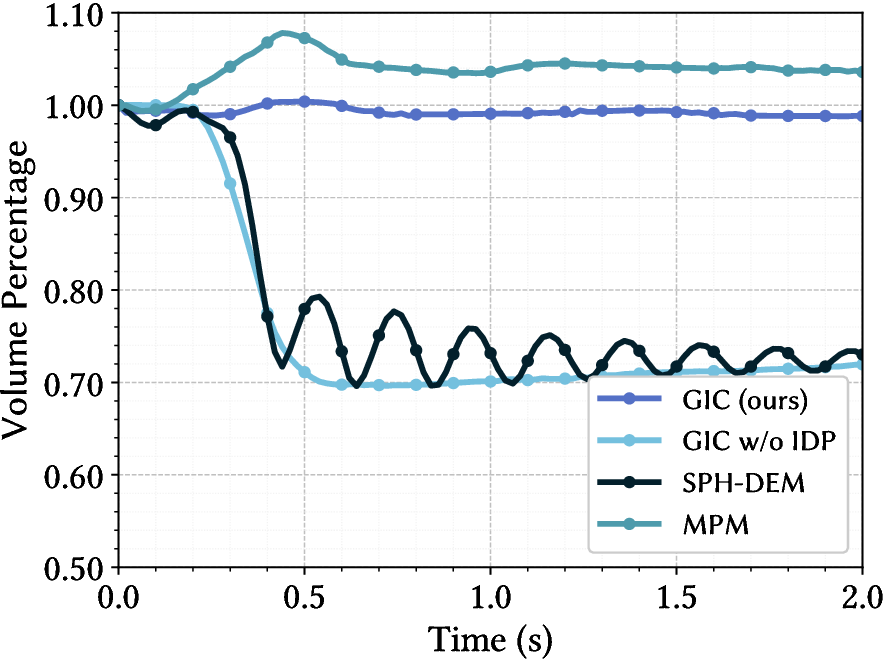}
    \vspace{-1em}
\caption{Comparison of volume percentage for SPH-DEM \cite{wang2021visual}, MPM \cite{gao2018animating} and our method as in Figure~\ref{fig:density_3d_1}. The total fluid volume divided by the initial volume is plotted over time. Our method keeps the relative volume deviation below 1\%.}
\label{fig:density_3d_3}    
\end{figure}

In the case of single-phase fluid, Eq. \eqref{eq:mass} implies a divergence-free velocity field.
However, the movement of sand introduces time variations in the sand volume fraction $\alpha_\ts$, resulting in a non-zero temporal changes in the fluid volume fraction $\pdv*{\alpha_\tf}{t}$.
As in most PIC/FLIP methods, the standard operator splitting method is employed to discretize the Navier-Stokes equation $\eqref{eq:momentum}$,
\begin{equation}
    \label{eq:discret momentum}
   \vb*{v}_\tf' = \vb*{v}_\tf^* - \frac{\Delta t}{\rho_0} \grad p,
\end{equation}
where $\vb*{v}_\tf^*$ denotes the intermediate velocity after the advection and applying other forces except pressure, $\vb*{v}_\tf'$ denotes the velocity at the next time step. To compute the pressure, the time derivative in Eq.~\eqref{eq:mass} is discretized by backward Euler method as
\begin{equation}
    \label{eq:backward Euler}
    \frac{\alpha_\tf' - \alpha_\tf^*}{\Delta t} + \div\qty( \alpha_\tf'\vb*{v}_\tf' ) = 0,
\end{equation}
where $\alpha_\tf'$ denotes the target fluid volume fraction at the next moment and is required to satisfy the constraint \eqref{eq:constraint}. Bringing the velocity of the next time step from Eq. \eqref{eq:discret momentum} to Eq. \eqref{eq:backward Euler} yields
\begin{equation}
    \label{eq:bring}
    \frac{\alpha_\tf' - \alpha_\tf^*}{\Delta t} + \div\qty( \alpha_\tf'\vb*{v}_\tf^* -  \frac{\alpha_\tf' \Delta t}{\rho_0} \grad p) = 0,
\end{equation}
which can be rearranged as
\begin{equation}
    \label{eq:bring2}
      \frac{\Delta t}{\rho_0}\div\qty( \alpha_\tf' \grad p) = \frac{\alpha_\tf' - \alpha_\tf^*}{\Delta t} + \div\qty( \alpha_\tf'\vb*{v}_\tf^*).
\end{equation}

In contrast to solving the Pressure Poisson Equation (PPE) for single phase, the presence of sand introduces spatial variations in the fluid volume fraction, necessitating the inclusion of the gradient of $\alpha_\tf'$, which makes Eq. \eqref{eq:bring2} a variable coefficient Poisson equation.
Despite this complexity, discretizing the equation on the marker-and-cell (MAC) grid results in a diagonally dominant matrix.
The right-hand side is split into two components: one representing the difference between the current and target volume fractions, and the other representing the velocity divergence weighted by the volume fraction. This formulation enables solving the fluid dynamics under a spatially and temporally varying sand volume fraction field. To solve the Poisson equation, both sides are divided by $\alpha_\tf'$, and the equation is split into two terms
\begin{subequations}
    \label{eq:ppe1}
    \begin{align}
      \frac{\Delta t}{\alpha_\tf'\rho_0}\div\qty( \alpha_\tf' \grad p_1) &= \frac{1}{\alpha_\tf'}\div\qty( \alpha_\tf'\vb*{v}_\tf^*),\\
      \frac{\Delta t^2}{\alpha_\tf'\rho_0}\div\qty( \alpha_\tf' \grad p_2) &= 1 - \frac{\alpha_\tf^*}{\alpha_\tf'}, \label{eq:ppel-2}
    \end{align}
\end{subequations}
where $p_1$ is used to update the velocity on the grid with Eq. \eqref{eq:discret momentum} in the standard way, while $p_2$  can be plugged into Eq. \eqref{eq:discret momentum} to get the position changes on the grid as
\begin{equation}
    \label{eq:position change}
    \delta \vb*{x} = (\vb*{v}' - \vb*{v}^*)\Delta t = -\frac{\Delta t^2}{\rho_0}\grad p_2.
\end{equation}
For every fluid particle, its position change is interpolated from $\delta \vb*{x}$ without updating velocity. To avoid 
excessive correction displacements of particles in a single time step, the ratio \rv{$\alpha_\mathrm{f}^*/\alpha_\mathrm{f}'$ is restricted to the range $[0.5, 1.5]$, and the magnitude of the correction $\delta \vb*{x}$ is confined to within one cell width}. 
Additionally, the value of $\alpha_\ts$ must be regulated to prevent repelling fluid by producing negative fraction targets and leading to cavity formation within the fluid. To mitigate this, $\alpha_\ts$ is limited to a maximum of 0.740 in 3D and 0.907 in 2D, corresponding to the proportion of space occupied by densest sphere packing.

At the fluid--air interface, the Dirichlet Boundary Condition (BC) $p = 0$ is employed, allowing the fluid to move into the air freely. At the fluid--solid interfaces, the velocity projection part enforces the Neumann BC $\vb*{v} = \vb*{v}_{\mathrm{b}}$. For the density projection part, the Push-Out Neumann BC is applied, as detailed in \S\ref{Push-Out BC}.

Since the target volume fraction $\alpha_\tf'$ for the next time step is required for both density and velocity projections, the sand phase is computed first during this time step, and its spatial information is then used to solve the fluid projection. The fluid part follows the same procedure as IDP: after advection, the fluid particle positions are corrected by density projection, and then velocity projection is applied to satisfy volume consistency. The specific algorithm is described in \S \ref{sec:alg}.

Actually, the density projection serves to project the current fraction $\alpha_\tf^*$ toward the target fraction $\alpha_\tf'$. Specifically, after the projection, the distribution of fluid particles is adjusted such that the fluid volume fraction complements that of the sand.
Figures~\ref{fig:density_2d} and~\ref{fig:density_3d_1} demonstrate the role of the density projection in adjusting the fluid volume fraction. This technique ensures that the mixture maintains consistent volume, achieving stable and robust volume preservation.

\subsection{Granules under Flow}
\label{sec:flow_force}

In \S \ref{sec:couple}, we introduced three external forces acting on granules in fluid: the pressure gradient force $\vb*{F}_{\mathrm{p}}$; the drag force $\vb*{F}_{\mathrm{d}}$; and the virtual mass force $\vb*{F}_{\mathrm{v}}$. In this section, we will explain how these forces are calculated in GIC.
Since the sand portion is calculated prior to the fluid in each step, the fluid information used in each sand substep is derived from the previous fluid time step.
\paragraph{Pressure Gradient Force}
Calculating the pressure gradient force requires the pressure gradient from water, as detailed in \S \ref{sec:target}. By interpolating from the grid, the gradient at the position $\vb*{x}_\ts$ is obtained and used to compute the force acting on the particle through Eq.~\eqref{eq:gradient-force}.
\paragraph{Virtual Mass Force}
To compute the virtual mass effect, the term related to the acceleration of the granule is integrated into the motion term, allowing the dynamic equation to be expressed as
\begin{equation}
    \label{eq:sand kine}
    \qty(m_\ts + \frac{2}{3}\pi r^3\rho_0) \dv {\vb*{v}_\ts}{t} = \frac{2}{3}\pi r^3\rho_0 \dv {\vb*{v}_\tf}{t} + \vb*{F}_{\mathrm{o}},
\end{equation}
where $m_\ts$ denotes the mass of the granule, $\vb*{F}_{\mathrm{o}}$ denotes all the other forces applied on it. From Eq. \eqref{eq:sand kine}, it can be observed that the virtual mass force acts as a form of inertia for granules in the fluid. To obtain the time derivative of the fluid velocity $\dv*{\vb*{v}_\tf}{t}$, the Navier-Stokes equation \eqref{eq:momentum} is rewritten as
\begin{equation}
    \label{eq:momentum2}
    \dv{\qty(\alpha_\tf\vb*{v}_\tf)}{t} = \alpha_\tf \vb*{b}_\tf - \frac{\alpha_\tf}{\rho_0}\grad p + \frac{1}{\rho_0}\vb*{F}_\ts.
\end{equation}
When further expanding the left-hand term, the material derivative, it yields 
\begin{equation}
    \label{eq:momentum3}
    \frac{\vb*{v}_\tf}{\alpha_\tf}\dv{\alpha_\tf}{t} + \dv{\vb*{v}_\tf}{t} = \vb*{b}_\tf - \frac{1}{\rho_0}\grad p + \frac{1}{\rho_0 \alpha_\tf}\vb*{F}_\ts.
\end{equation}
Eq. \eqref{eq:mass} shows that $\dv*{\alpha_\tf}{t} = 0$, enabling $\dv*{\vb*{v}_\tf}{t}$ to be expressed as
\begin{equation}
    \label{eq:momentum4}
    \dv{\vb*{v}_\tf}{t} = \vb*{b}_\tf - \frac{1}{\rho_0}\grad p + \frac{1}{\rho_0 \alpha_\tf}\vb*{F}_\ts,
\end{equation}
whose right-hand side is precisely the velocity increment obtained from the grid calculations in the FLIP framework.
\paragraph{Drag Force}
The drag force can be directly calculated by interpolating the fluid velocity at the position $\vb*{x}_\ts$ from the grid, and then applying it to compute the drag force.
\paragraph{Exchange Momentum}
Since the sand phase is calculated before the fluid phase at each time step, the fluid information from the previous time step at time $t$ is used during the particle sub-time step $t + i \cdot \Delta t' \to t + (i+1) \cdot \Delta t'$. Throughout the sand time step, all the forces exerted on a particle by the fluid are summed and projected onto the grid as
\begin{equation}
    \vb*{F}_\ts\qty(\vb*{x}) =-\frac{\Delta t'}{\Delta t} \sum_{i}\sum_{j}\vb*{f}_{\mathrm{s}_j}^i N\qty(\vb*{x}_{\mathrm{s}_j}^i-\vb*{x}),
\end{equation}
where $\vb*{F}_\ts$ is the total force transferred to the grid by the granules during the time step, $i$ represents all sub-steps within a single time step, $j$ indexes the granules in the neighborhood of position $\vb*{x}$ on the grid, $\vb*{f}_{\mathrm{s}_j}^i$ denotes the total force exerted on granules $j$ during the $i$-th sub-step, and $N$ is the interpolation kernel. When calculating the fluid state at time $t + \Delta t$, these forces are added as external forces to the fluid velocity, along with the body force on grid.
\subsection{Wetting Granule in Cell}
\label{sec:wet}

\begin{figure}
    \centering
    \setlength{\imagelength}{.495\linewidth}
    \includegraphics[width=\imagelength]{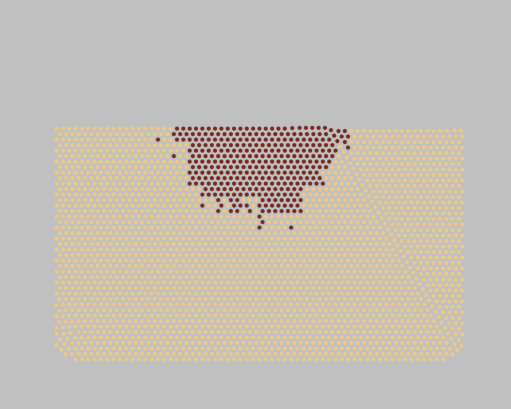}
    \hfill
    \includegraphics[width=\imagelength]{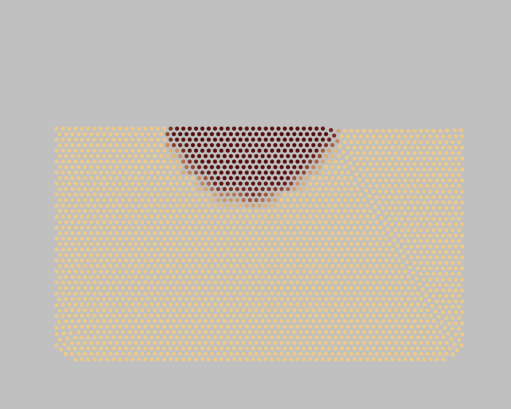}
    \vspace{-2em}
\caption{Absorption of Water. A drop of water falls on a pile of sand, simulated with a grid of resolution $32^2$.
The left figure demonstrates the trivial traversal of unsaturated granules to absorb PIC particles, while the right figure utilizes grid-based interpolation for water deficit calculation.}
\label{fig:absorb}    
\end{figure}
Sand can absorb and retain water within its porous structure due to capillary action. For example, when a pile of wet but unsaturated sand is placed on the ground without external forces, the water will remain trapped rather than flowing away.
Our model simulates the absorption process as a phase change, converting fluid particles in the PIC framework into water content within DEM granules.
This section describes the approach to handling this phase change, along with the corresponding momentum and mass transfer mechanism.


The volume of water absorbed is proportional to the granule's volume, defined by the ratio $r_i$, with a maximum absorption limit $r_{\max}$. The absorbed water is treated as additional mass for the granule as
 \begin{equation}
    m_{\mathrm{s}_i}= V_{\mathrm{s}_i}(\rho_\ts + r_i\rho_0),
\end{equation}
where $m_{\ts_i}$ is the mass of the $i$-th granule.
To convert the phase of water from the PIC fluid particles to the ratio on granules, a straightforward approach involves traversing through each granule and checking for nearby fluid particles. When nearby particles are detected, the granule absorbs and removes them from the fluid, converting these particles into the ratio $r_i$  and transferring momentum.
However, this method leads to uneven absorption, as the processing order of granules determines which granules absorb particles first, causing imbalances in water distribution and momentum transfer, as shown in Figure~\ref{fig:absorb}.
\begin{figure}
    \centering
    \setlength{\imagelength}{.495\linewidth}

    \includegraphics[width=\imagelength]{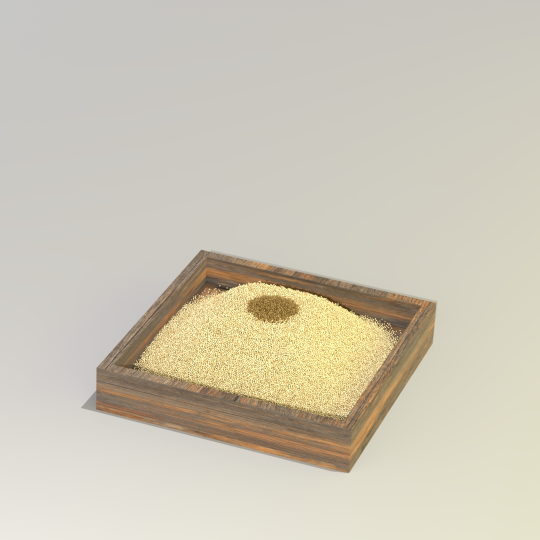}
    \hfill
    \includegraphics[width=\imagelength]{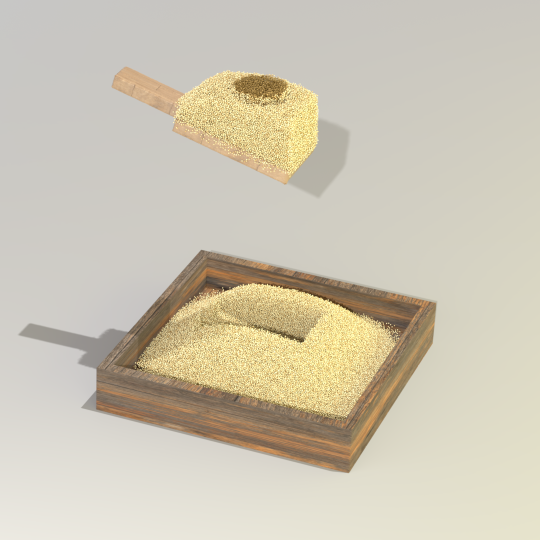} \\
    \vspace{0.01\linewidth}
    \includegraphics[width=\imagelength]{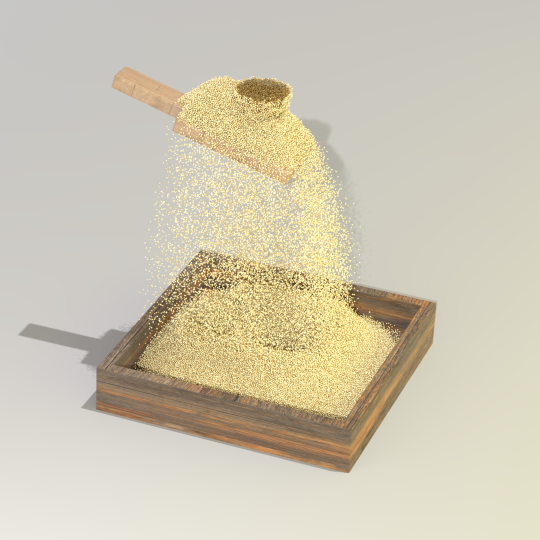}
    \hfill
    \includegraphics[width=\imagelength]{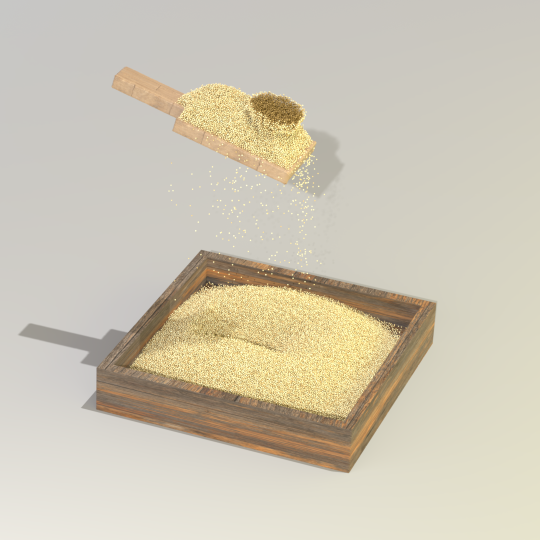}
\vspace{-2em}
\caption{Cat Litter box. This experiment simulates a cat litter box with a small central area wetted. After scooping, the dry sand falls while the wet clump remains on the shovel. The four figures depict the litter state at $t=0, t = 0.4\ \mathrm{s}, t = 4\ \mathrm{s}$, respectively.}
\label{fig:cat}    
\end{figure}

To address this, a strategy is adopted where the moisture deficit $r_{\max{}} - r_i$ of each granule is projected onto the grid. This deficit determines the number of fluid particles to be removed in each cell, with particles randomly selected and their momentum and volume projected onto the grid. The absorbed ratio is interpolated from the grid and added to the granule. Water absorption by sand is modeled as a perfectly inelastic collision, and in the next sand time step, the momentum from this process is gradually added into the granule's momentum as
\begin{equation}
    m_{\mathrm{s}_i}\vb*{v}_{\ts_i} = m_{\ts_i}\vb*{v}_{\ts_i} + \frac{\Delta t'}{\Delta t}\vb*{p}_{\mathrm{absorb}}.
\end{equation}

This approach ensures a uniform distribution of absorbed water among the granules, as illustrated in Figure~\ref{fig:absorb}. When calculating the forces acting on the particles in the fluid, the absorbed water is also treated as an additional volume. Furthermore, when performing the density projection, we include this additional volume in $\alpha_\ts$. However, since the actual radius of the wetted granules does not increase, their dense packing still follows the original radius of the granules. The absorbed water is considered to occupy a portion of the voids. Therefore, when clamping $\alpha_\ts$, its upper limit should be increased to $0.740 \cdot (1 + r_{\max{}})$ in 3D and $0.907 \cdot (1 + r_{\max{}})$ in 2D.

\subsection{Sediment Concentration Gradient Force}
\label{sec:concen-force}

\begin{figure*}
    \centering
    \setlength{\imagelength}{.198\linewidth}
    \includegraphics[width=\imagelength]{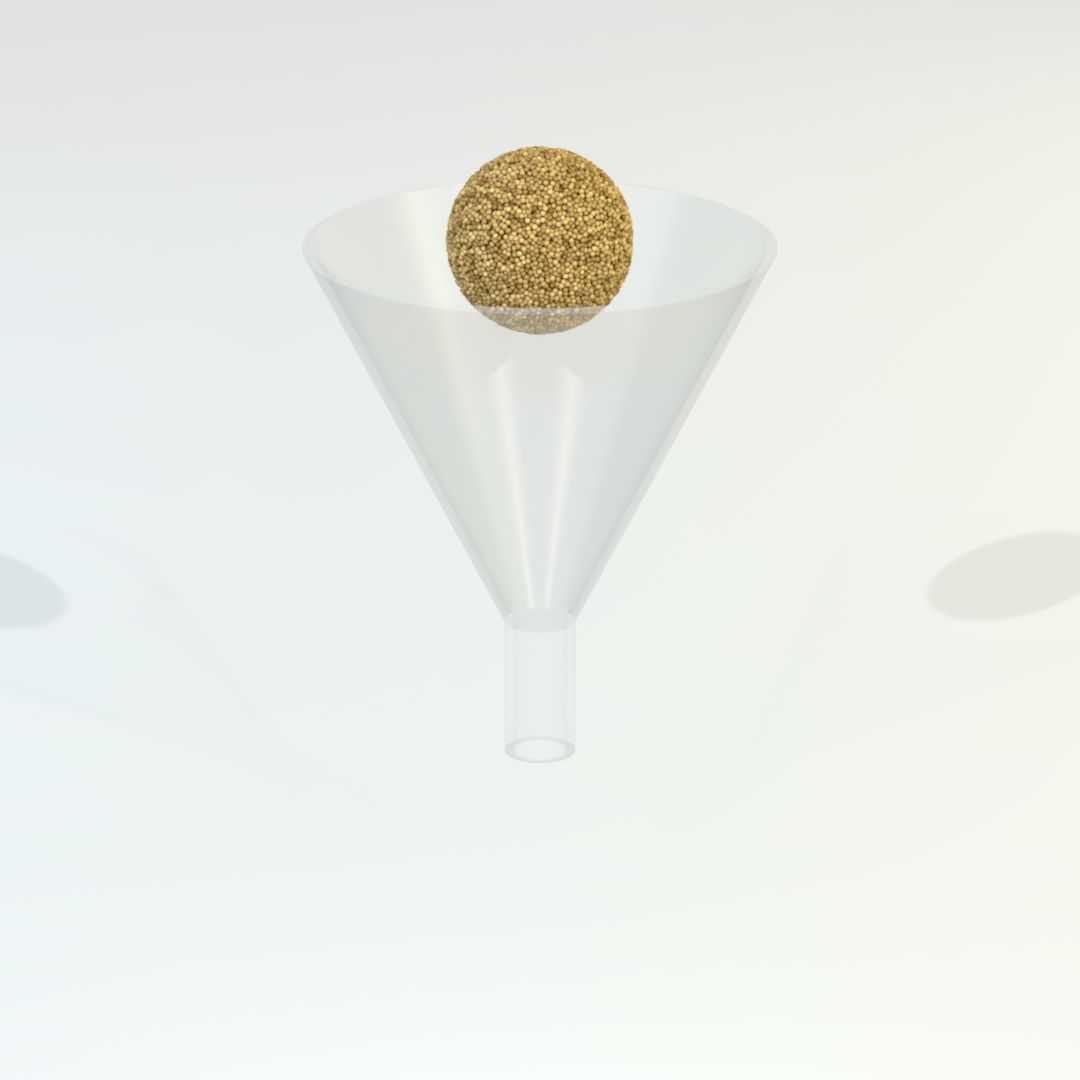}\hfill
\includegraphics[width=\imagelength]{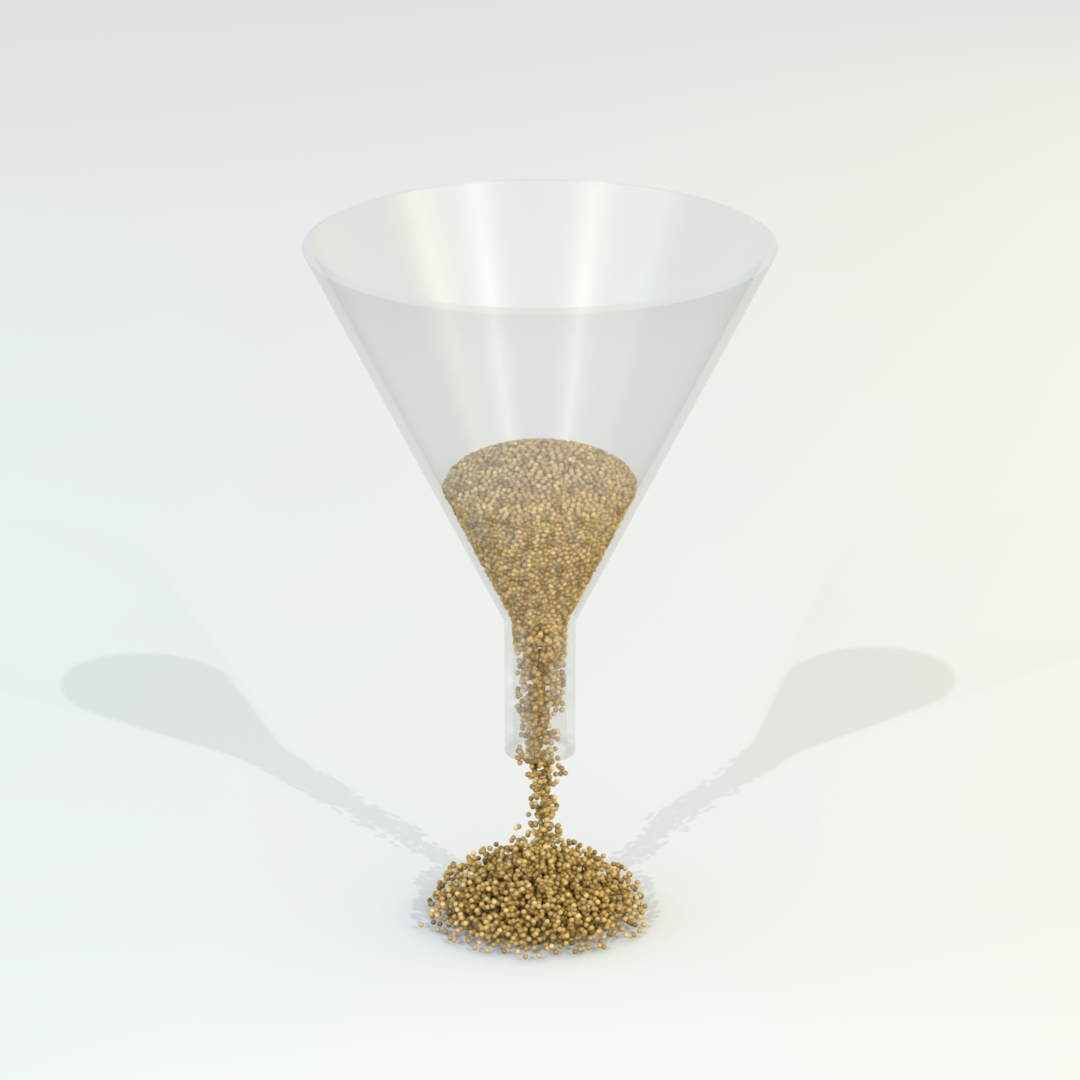}\hfill
\includegraphics[width=\imagelength]{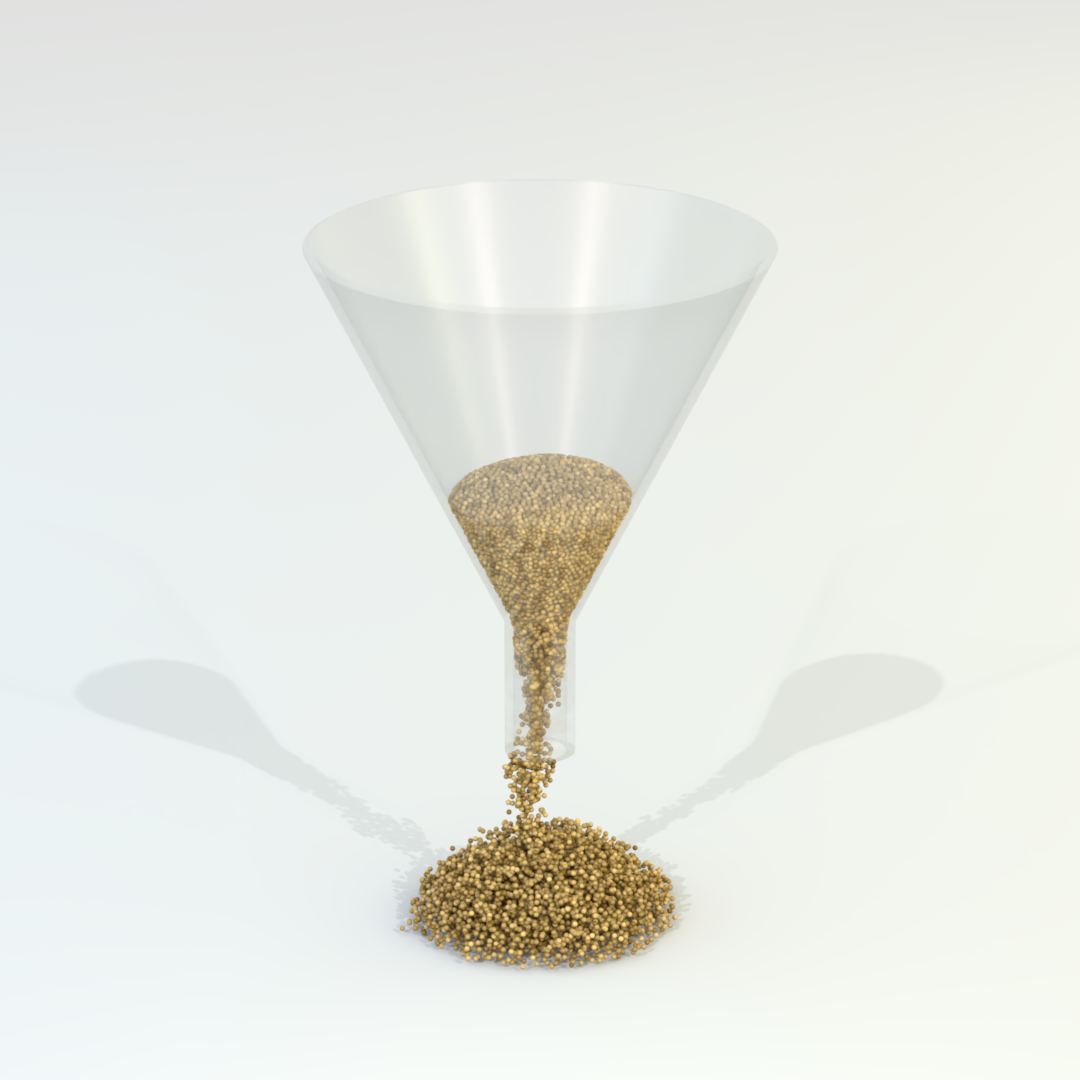}\hfill
\includegraphics[width=\imagelength]{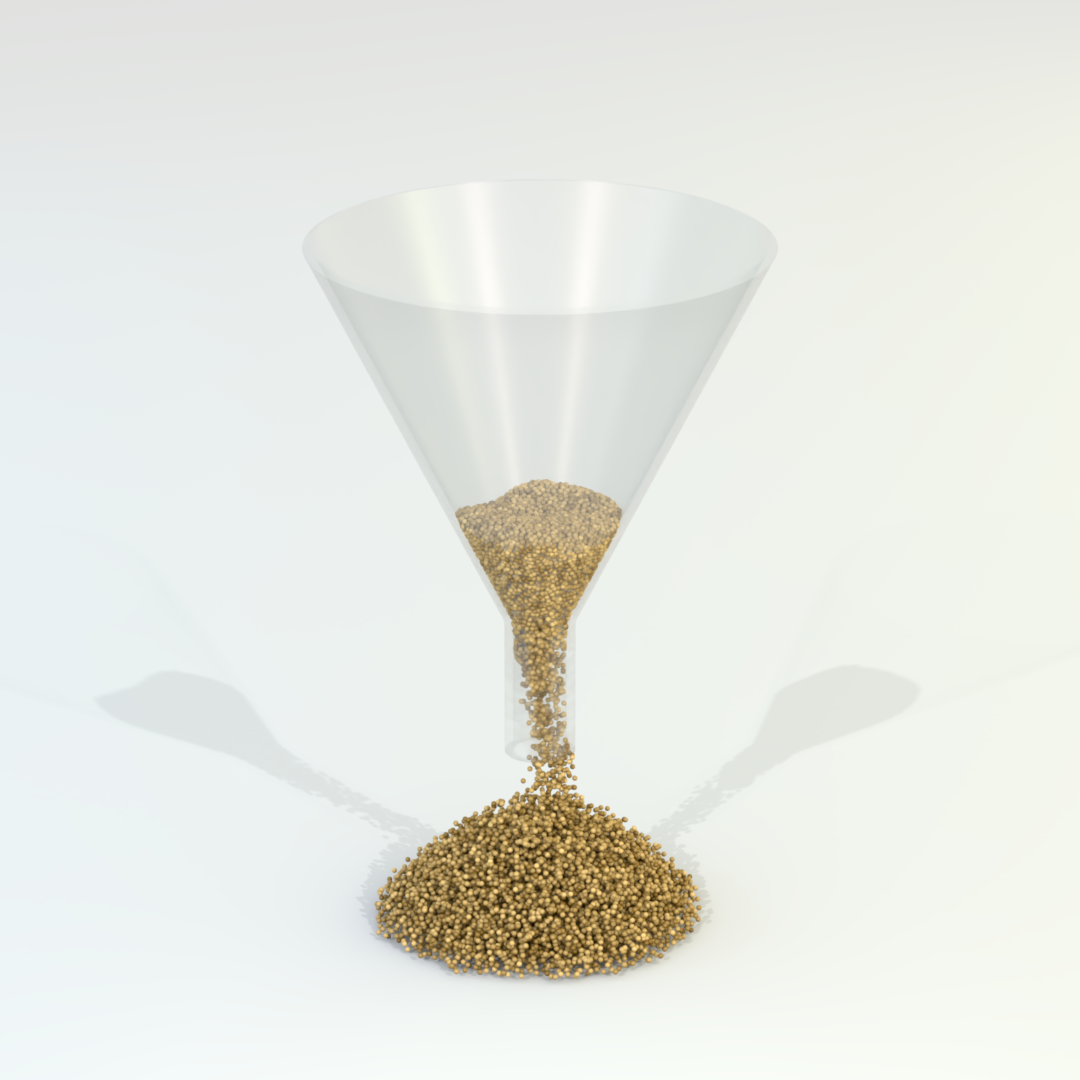}\hfill
\includegraphics[width=\imagelength]{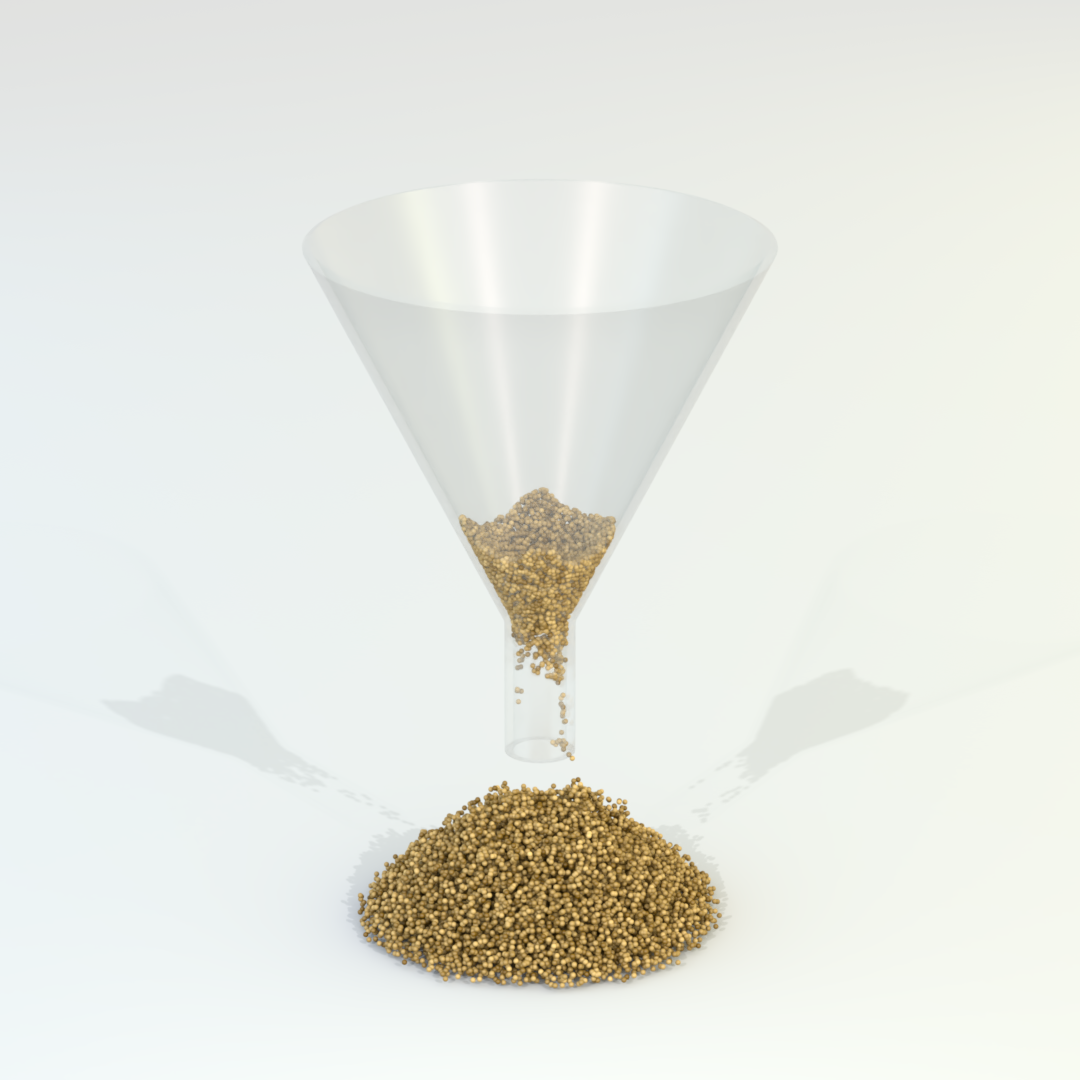}\\
\vspace{0.0025\linewidth}
\includegraphics[width=\imagelength]{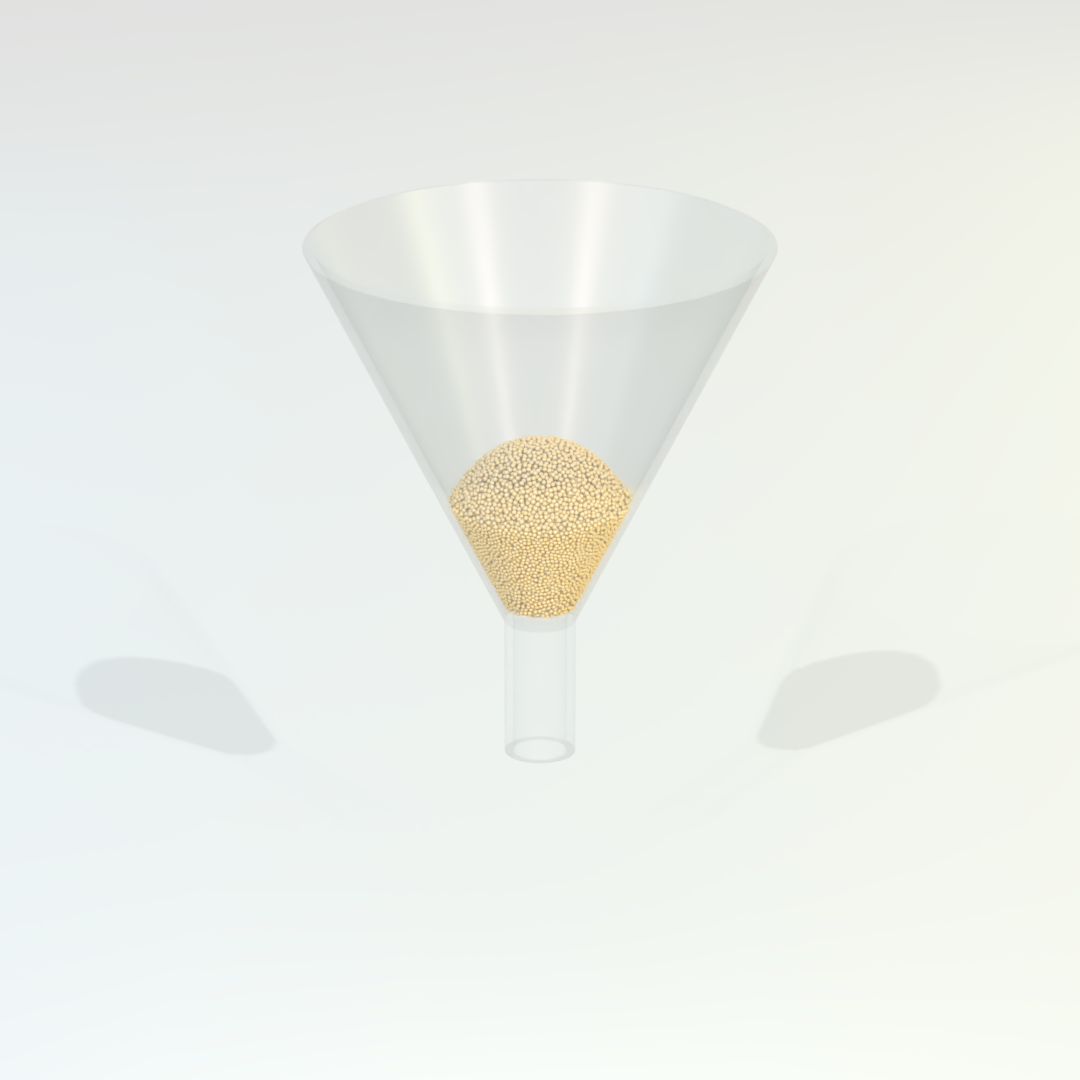}\hfill
\includegraphics[width=\imagelength]{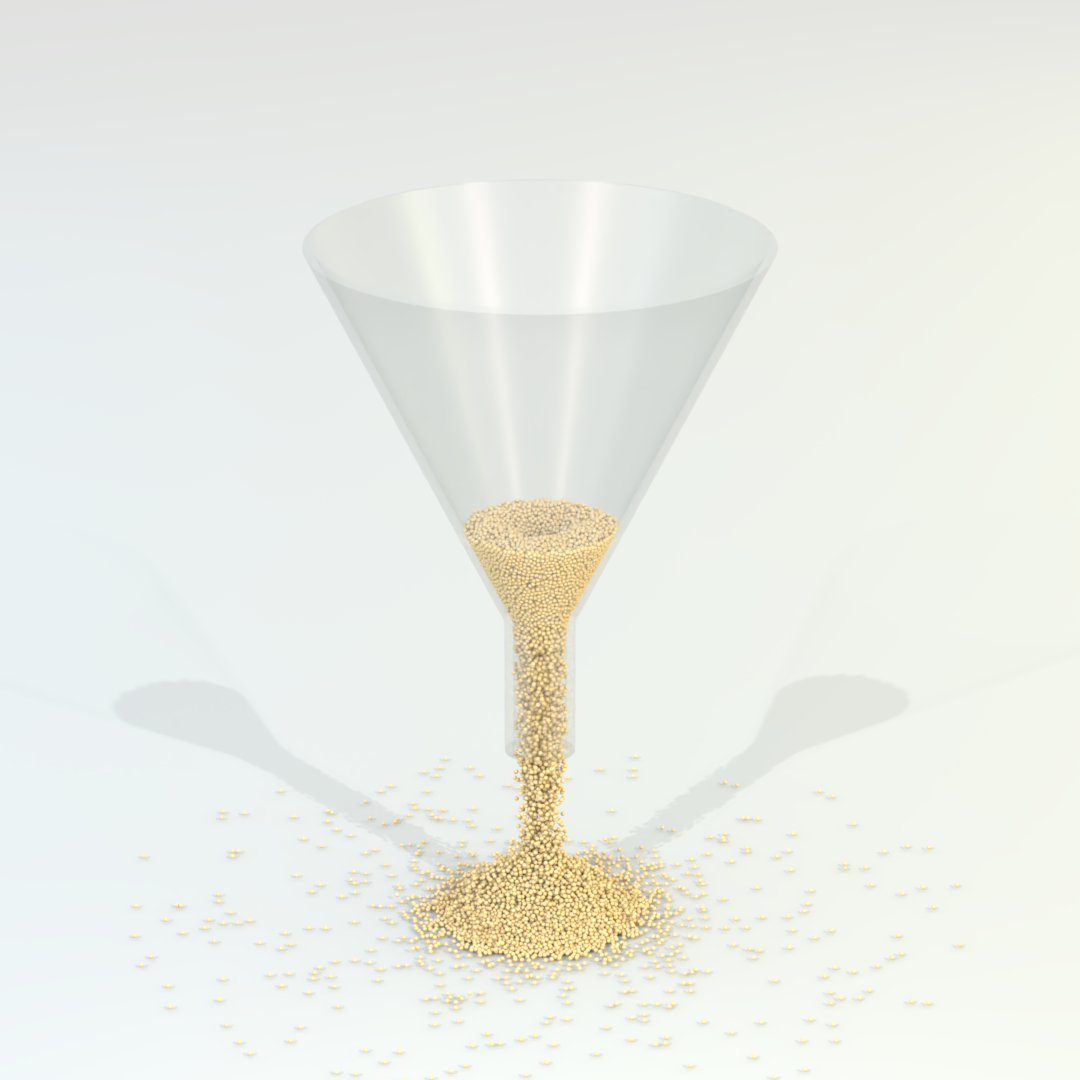}\hfill
\includegraphics[width=\imagelength]{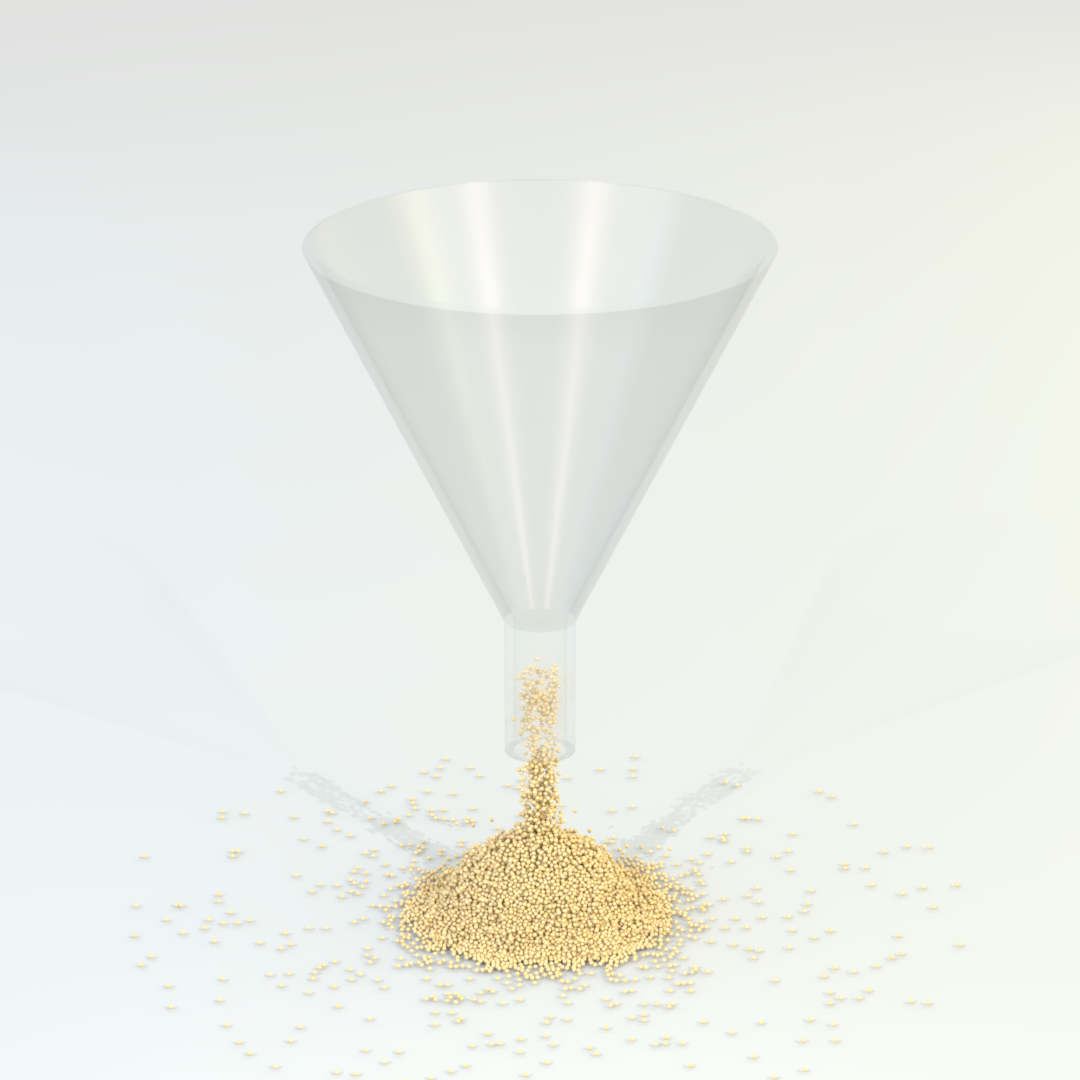}\hfill
\includegraphics[width=\imagelength]{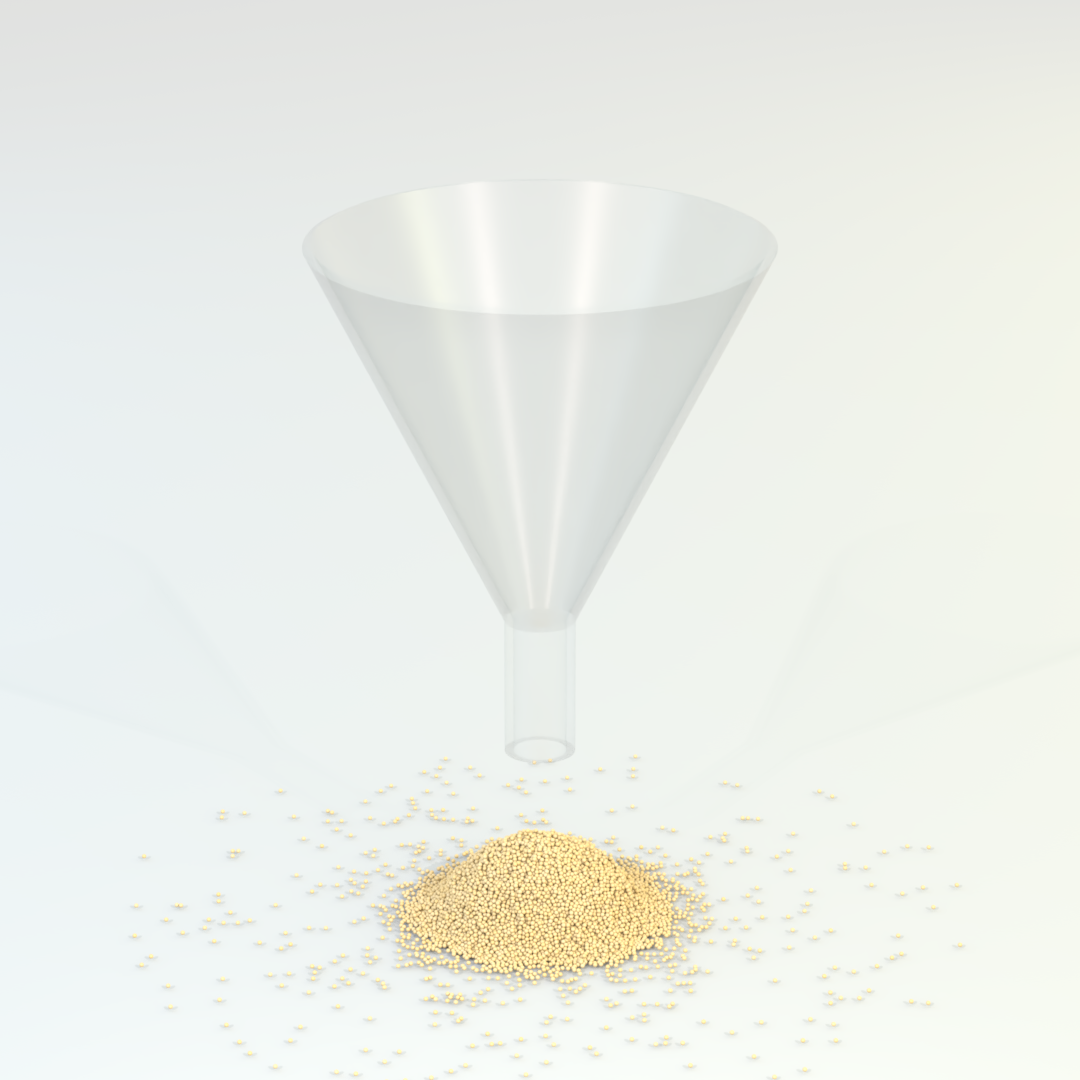}\hfill
\includegraphics[width=\imagelength]{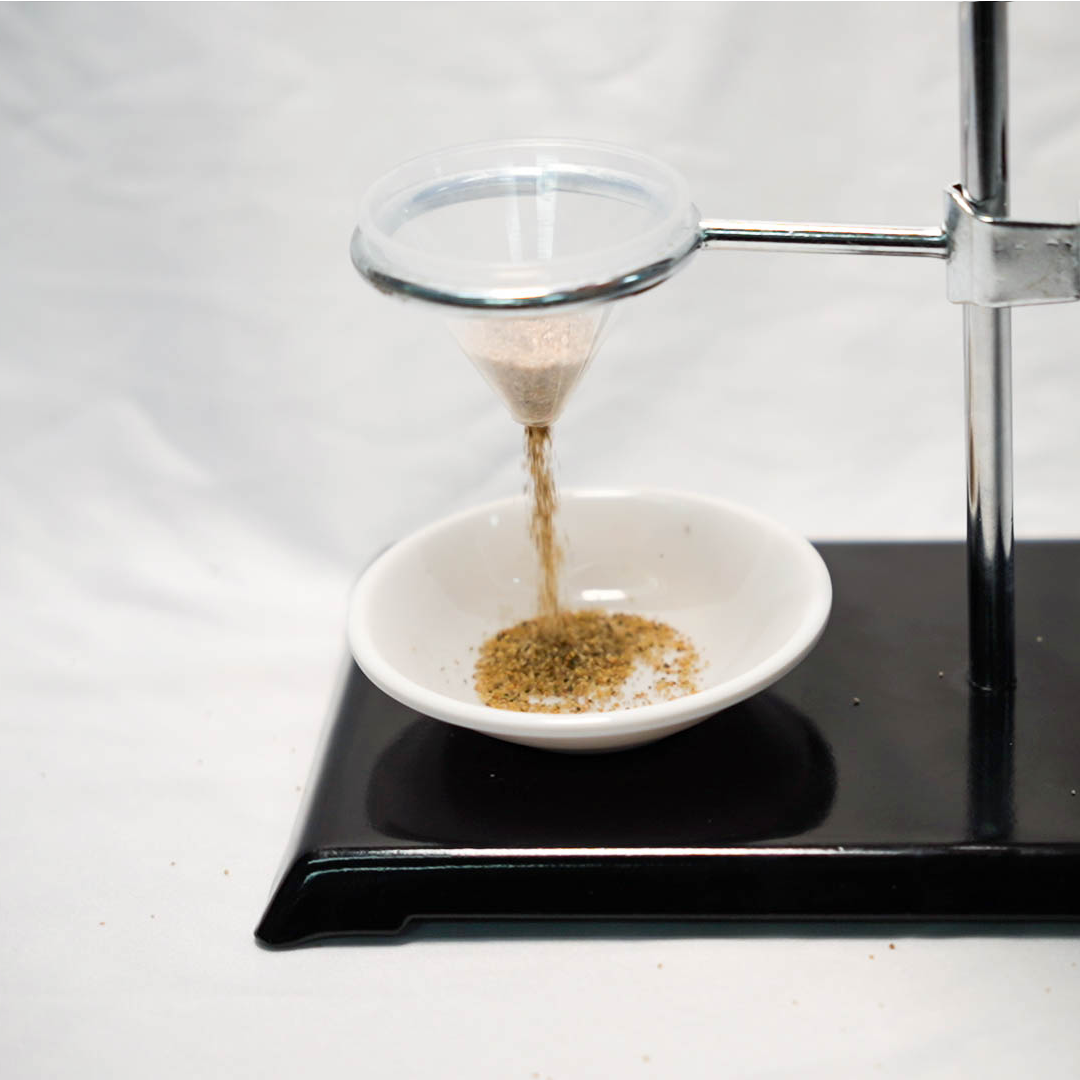} \\ 
\vspace{0.0025\linewidth}
\includegraphics[width=\imagelength]{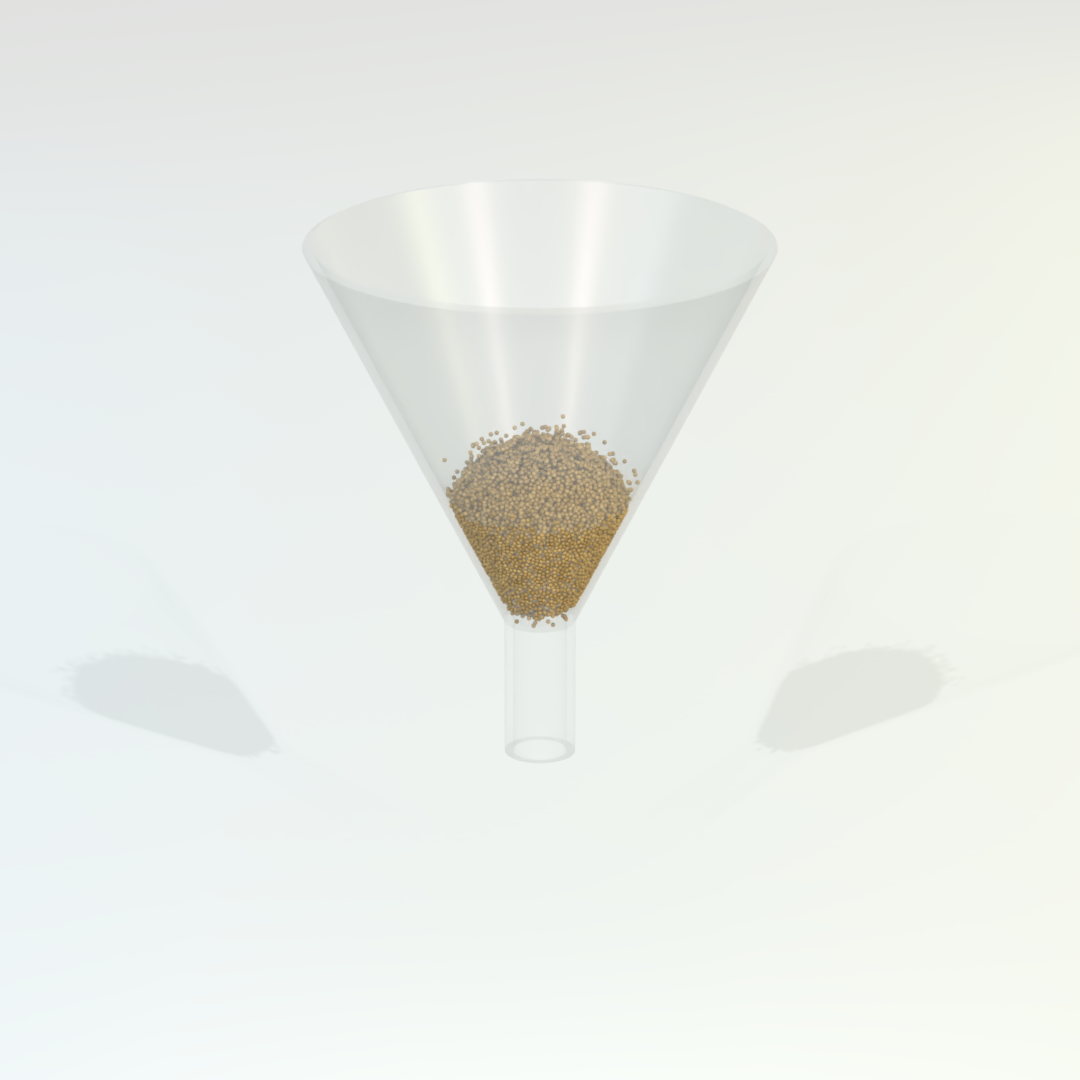}\hfill
\includegraphics[width=\imagelength]{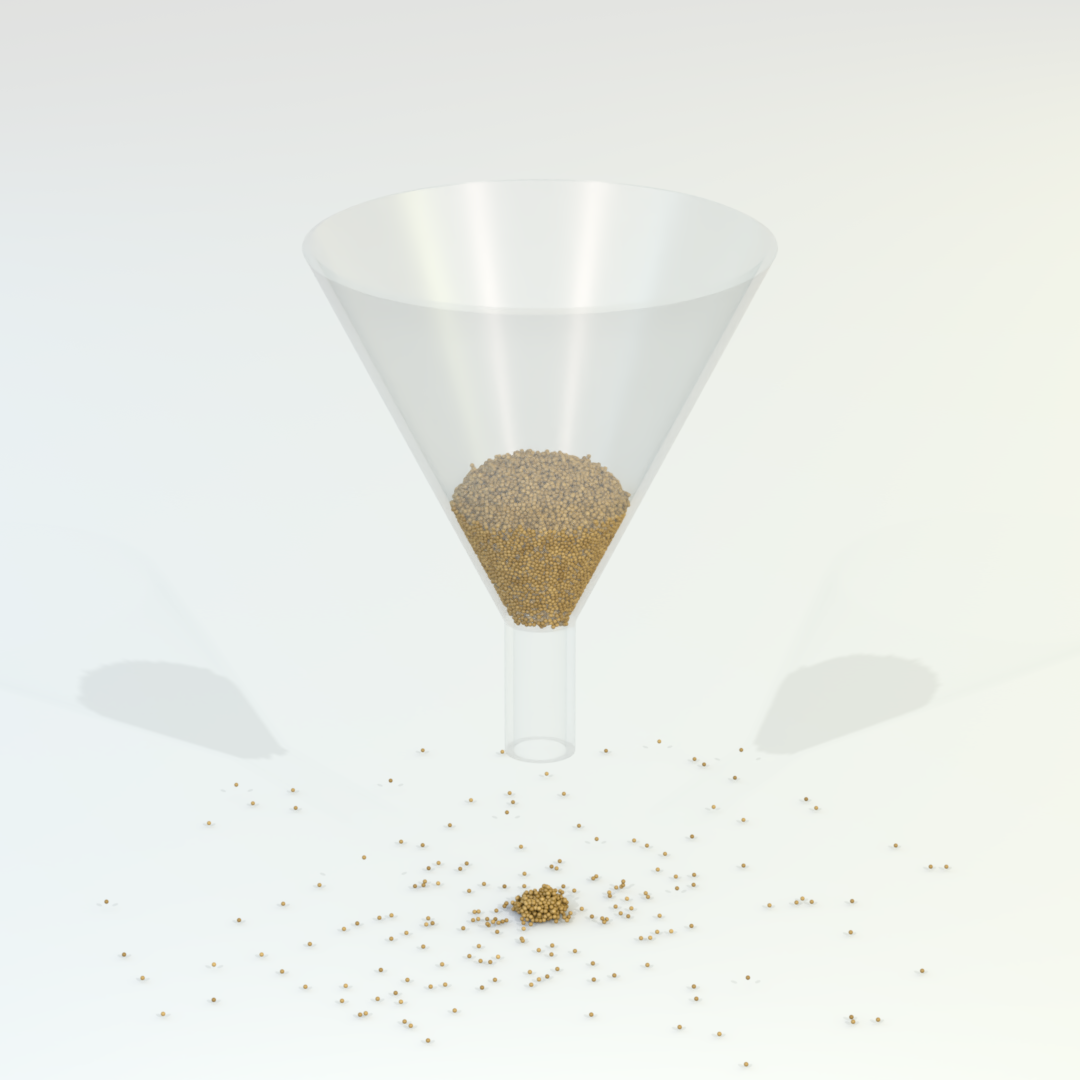}\hfill
\includegraphics[width=\imagelength]{figs/wet-1.png}\hfill
\includegraphics[width=\imagelength]{figs/wet-1.png}\hfill
\includegraphics[width=\imagelength]{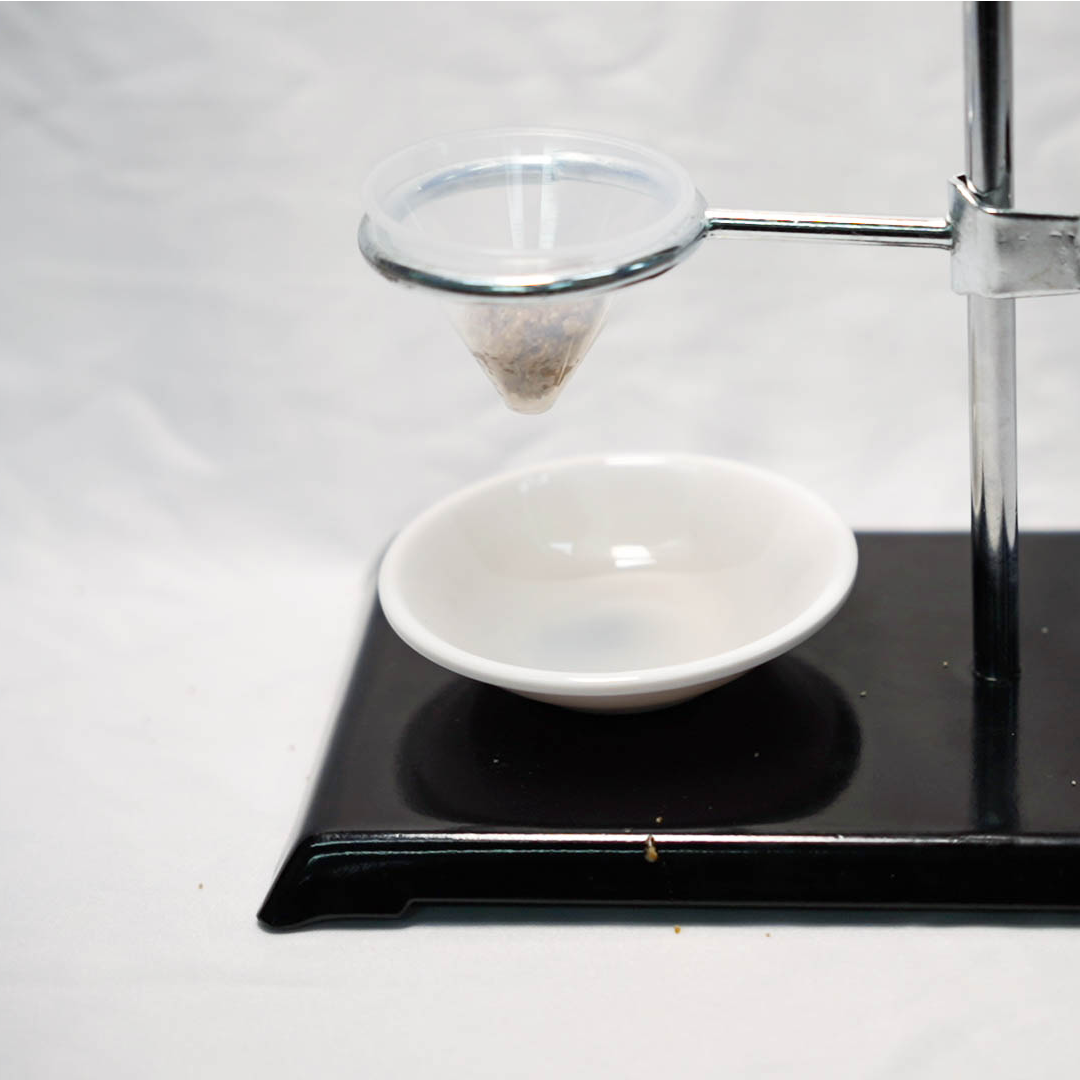} \\
\vspace{0.0025\linewidth}
\includegraphics[width=\imagelength]{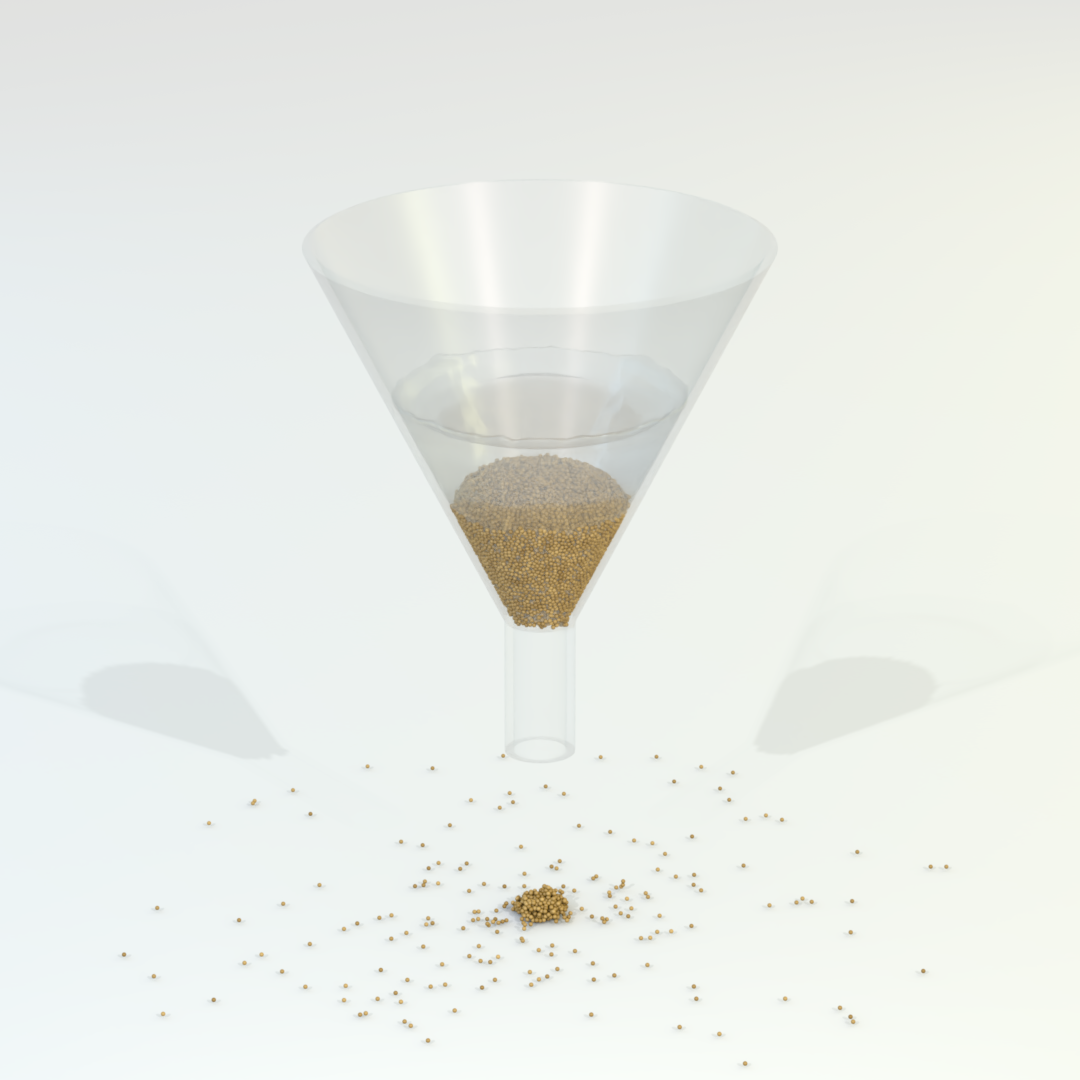}\hfill
\includegraphics[width=\imagelength]{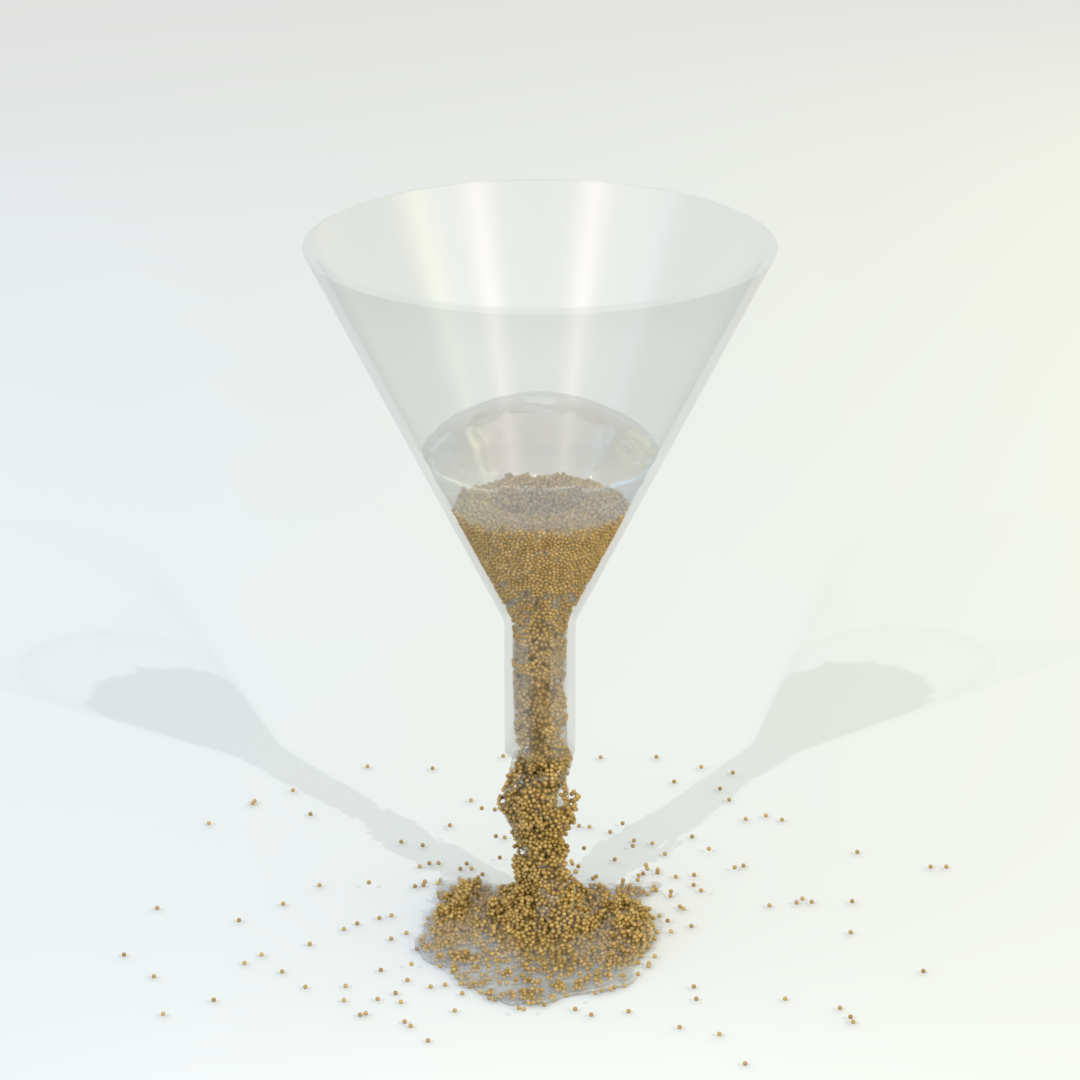}\hfill
\includegraphics[width=\imagelength]{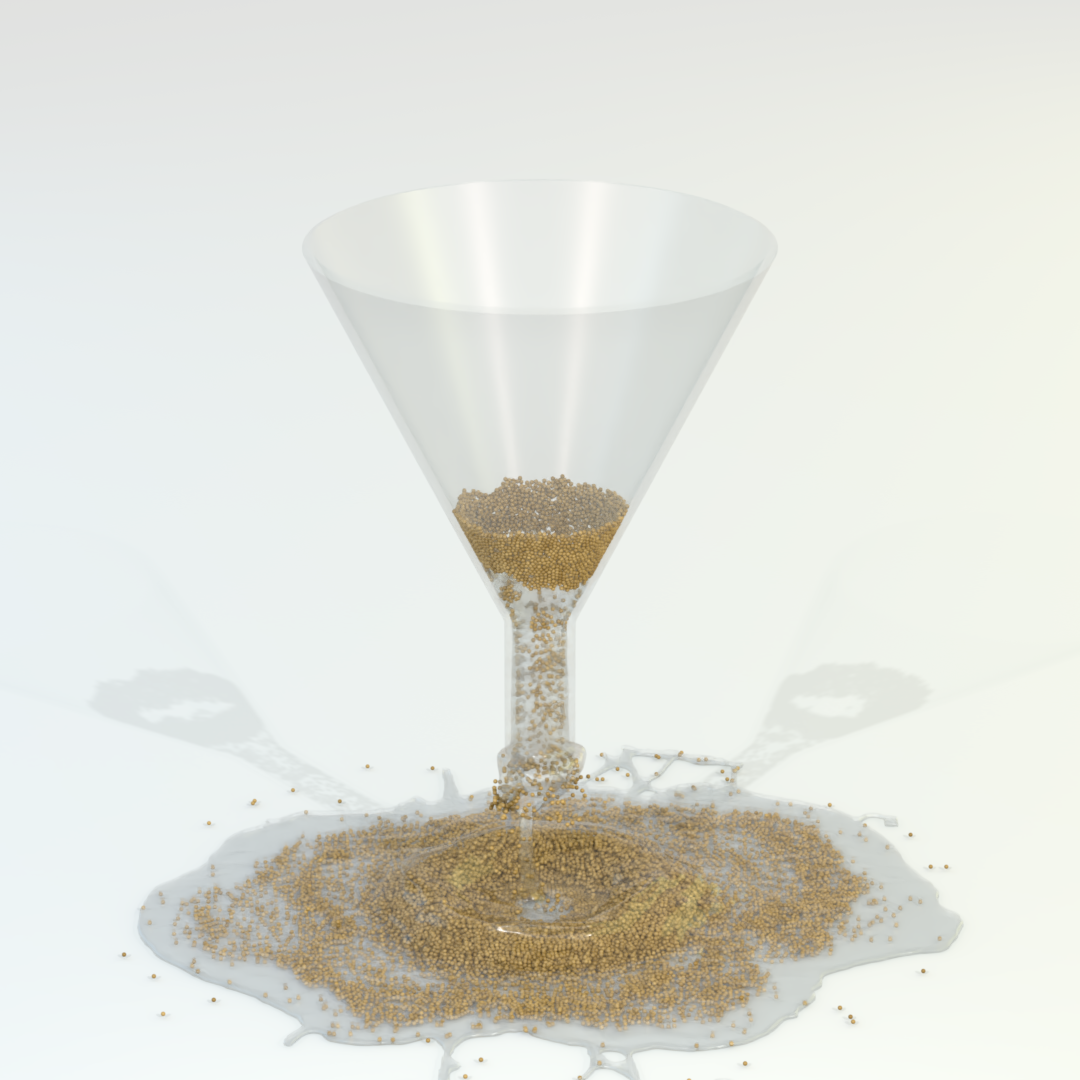}\hfill
\includegraphics[width=\imagelength]{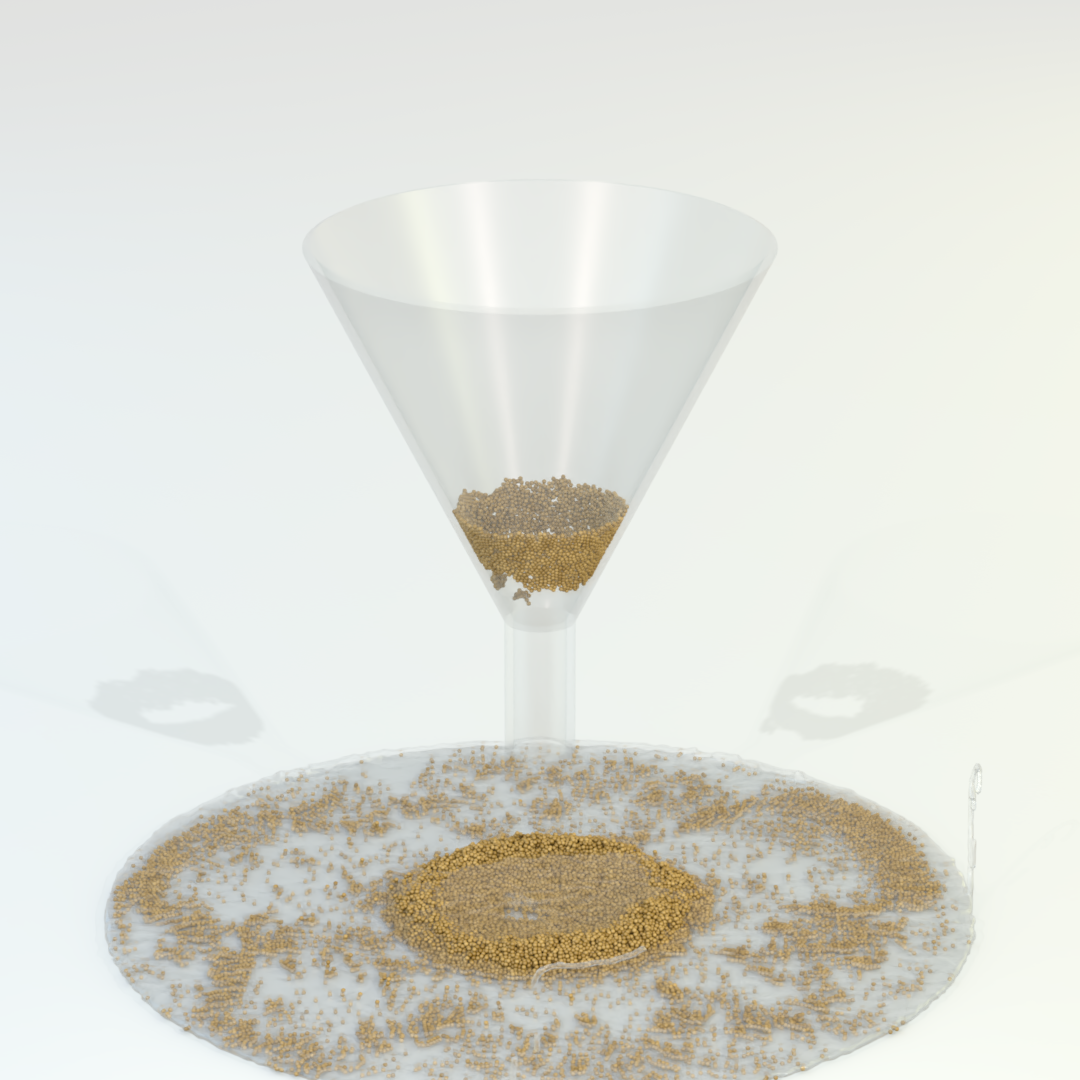}\hfill
\includegraphics[width=\imagelength]{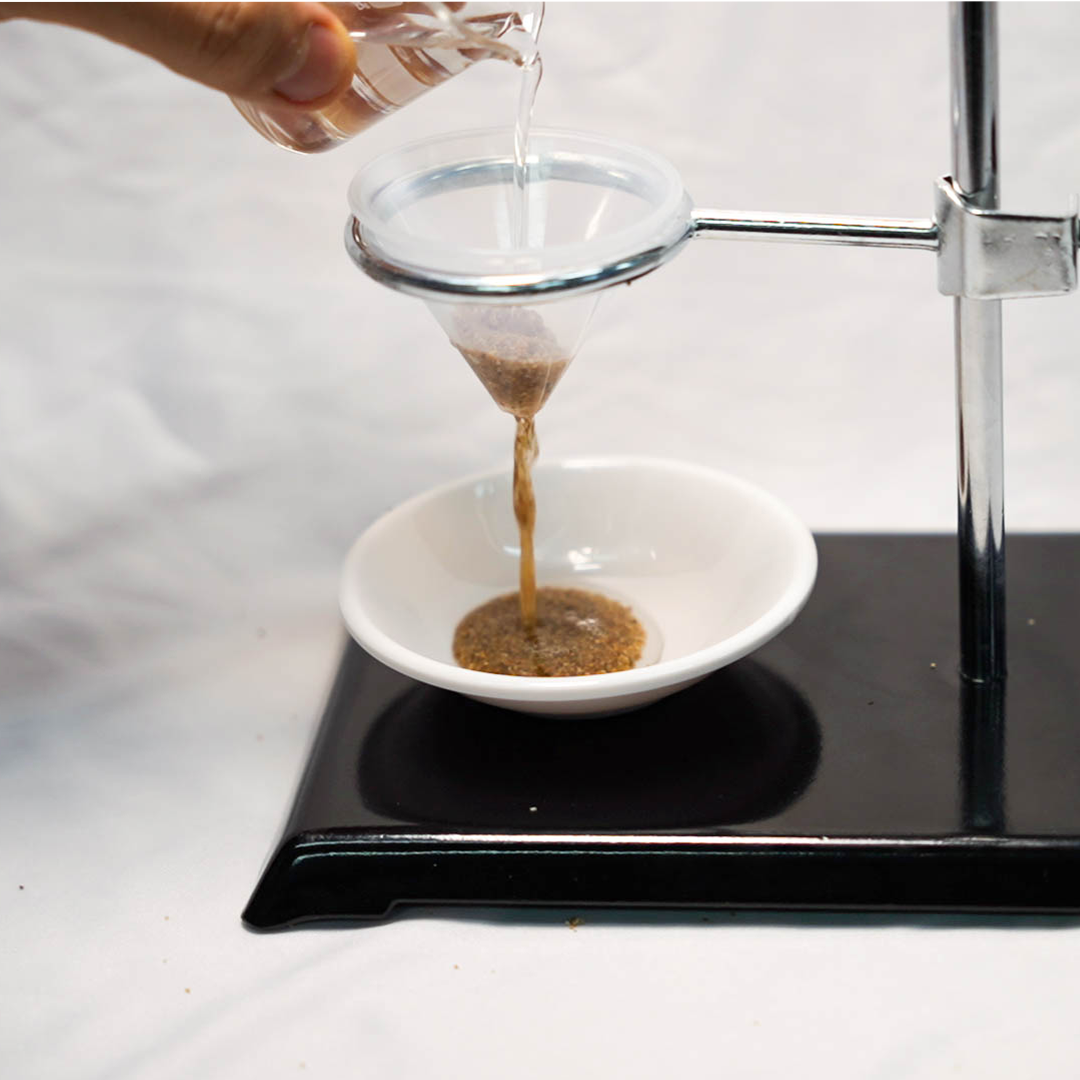}\\
\vspace{-.5em}
\caption{\rv{
Sand through funnel. In this figure we compare the simulation of sand with different moisture ratios going through a funnel using our method and the MPM method \cite{tampubolon2017multi}. The first row shows the result of MPM with wet sand at $t = 0, 1, 2, 4, 6~\mathrm{s}$. The following three rows shows the result with our method for dry sand, the wet sand and adding water to the stationary wet sand respectively. The first four columns in these rows show the simulated sand states at $t = 0.5, 2.5, 4.5, 6~\mathrm{s}$, and the fifth column is the key state captured in the real experiment. A notable difference is that in MPM method, wet sand does not get stuck in the funnel.
 }}
\label{fig:funnel}    
\end{figure*}
\paragraph{Force of Concentration Gradient}
In \S \ref{sec:couple}, we introduced that ensemble averaging in mesoscopic systems results in a concentration gradient force, as expressed in Eq. \eqref{eq:density_gradient}. This equation emphasizes the concentration gradient of sand, necessitating consideration of the asymmetry caused by the distribution of granules.
To capture this asymmetry, the force is discretized on the granules. However, due to their rigid spherical boundaries at the mesoscopic level, directly calculating density by counting granules introduces significant error. Since this force arises only when sand and water coexist, and under such conditions $\alpha_\ts = 1 - \alpha_\tf$ holds, Eq. \eqref{eq:density_gradient} is rewritten as
\begin{equation}
    \vb*{F}_{\alpha} = \frac{\rho_\ts \vb*{D}_\ts}{\alpha_{\tf} \tau_\ts} \vdot \grad \alpha_\tf.
\end{equation}
Since the original equation calculates the concentration gradient near the granule, the total water fraction must be considered, encompassing both free water in the surrounding space and the portion bound within the granule. A normalization factor reflecting the proportion of sand in the space is also incorporated. This correction is represented by $\tilde{\alpha}_\tf$, as
\begin{equation}
    \tilde{\alpha}_{\tf_i} = r_i + \frac{\alpha_{\tf_i}}{1 - \alpha_{\tf_i}}.
\end{equation}
Reconstructing the spatial density field in SPH style, the value of density field at $\vb*{x}$ is interpolated by granule with the kernel $W$ as 
\begin{equation}
    \tilde{\alpha}_\tf(\vb*{x}) = \sum_i \qty(r_i + \frac{\alpha_{\tf_i}}{1 - \alpha_{\tf_i}}) W(\vb*{x} - \vb*{x}_{s_i}).
\end{equation}
Therefore, the density gradient can be written as 
\begin{equation}
    \grad \tilde{\alpha}_\tf(\vb*{x}) = \sum_i \qty(r_i + \frac{\alpha_{\tf_i}}{1 - \alpha_{\tf_i}}) W'(\vb*{x} - \vb*{x}_{\ts_i}) \frac{\vb*{x} - \vb*{x}_{\ts_i}}{\norm{\vb*{x} - \vb*{x}_{\ts_i}}}.
\end{equation}
Using the linear kernel function, the force is expressed as 
\begin{equation}
    \vb*{F}_{\alpha}(\vb*{x}) = \frac{\rho_\ts \vb*{D}_\ts}{\alpha_{\tf} \tau_\ts}\vdot \qty[\sum_i \qty(r_i + \frac{\alpha_{\tf_i}}{1 - \alpha_{\tf_i}})\frac{\vb*{x}_{\ts_i} - \vb*{x}}{\norm{\vb*{x}_{\ts_i} - \vb*{x}}}].
\end{equation}
The coefficient $(\rho_\ts \vb*{D}_\ts)/(\alpha_{\tf} \tau_\ts)$ in front of this force is related to the materials of the two media and also to the distribution of the granules in space. Thus, the force from the $i$-th granule to the $j$-th granule is rewritten as
\begin{equation}
    \vb*{F}_{ij} = -A\qty(\norm{\vb*{x}_{\ts_i} - \vb*{x}_{\ts_j}}, \mathrm{sr}_i)\frac{\vb*{x}_{\ts_i} - \vb*{x}_{\ts_j}}{\norm{\vb*{x}_{\ts_i} - \vb*{x}_{\ts_j}}},
\end{equation}
where $A$ is a function related to the relative distance between granules and the moisture ratio $\mathrm{sr}_i = r_i + \alpha_{\tf_i} / (1 - \alpha_{\tf_i})$. To ensure that the forces between two granules satisfy Newton's third law, the interaction is symmetrized as
\begin{equation}
    \vb*{F}_{ij} = -A\qty(\norm{\vb*{x}_{\ts_i} - \vb*{x}_{\ts_j}}, \frac{\mathrm{sr}_i+\mathrm{sr}_j}{2})\frac{\vb*{x}_{\ts_i} - \vb*{x}_{\ts_j}}{\norm{\vb*{x}_{\ts_i} - \vb*{x}_{\ts_j}}}. \label{eq:symmetry-capillary}
\end{equation}
As a result, a function $A$ is required to model the forces between granules resulting from fluid-mediated interactions and capillary effects. A lot of works about models for liquid--solid two-phase flow has been developed \cite{wu2000mathematical, hibiki2004one}. To be compatible with the situation where the sand is wet but not immersed into the fluid, the liquid bridge model, representing capillary action, is selected.

\paragraph{Capillary Force}
Capillary action is one of the key factors explaining the differences in cohesion and flow properties between wet and dry sand. In existing research, the capillary action in sand is often modeled as the attractive force between granules caused by liquid bridges \cite{rabinovich2005capillary, yang2021capillary}. This interaction aligns with our microscopic definition of the concentration gradient force, allowing its computation within the capillary force framework. The capillary force is decomposed into two components: one dependent on the moisture ratio of the granules and the other on the distance between them. 

The concentration gradient force is expected to satisfy two properties. First, it should be a short-range force, meaning that the force should diminish to zero when the distance between particles exceeds a certain threshold. Additionally, the magnitude of the force should decrease as the distance between particles increases. Second, as the moisture ratio of the sand increases, the property of sand undergoes a transition from a plastic solid to a fluid-like state. To achieve that, the capillary force should be zero when the moisture ratio is zero, reach a maximum at a specific moisture level, and then gradually decrease as the moisture ratio continues to rise.  

To fulfill these conditions, the moisture and distance terms in the concentration gradient force are adjusted. For the distance term, the energy of the solid surface under the liquid bridge, as proposed by \citet{israelachvili2011intermolecular}, is adopted to calculate the force. The energy of the liquid bridge between granule $i$ and $j$ is given by
\begin{equation}
    E = -2\pi\sigma r\qty(-s_{ij}+\sqrt{s_{ij}^2+\frac{2V^*}{\pi r}})\cos\qty(\theta),
\end{equation}
where $\sigma$ is the surface tension coefficient, $r$ is the particle radius, $\theta$ is the contact angle, $s_{ij} = \norm*{\vb*{x}_j - \vb*{x}_i} - 2r$ is the separation distance between two granules, $V^*$ is the liquid bridge volume, which is set to be 0.01\% of the granule volume.
Experimental studies \cite{rabinovich2005capillary} indicate that the contact angle $\theta$ is relatively small. With approximation $\cos \theta = 1$,
the surface tension force
is given by
\begin{equation}
    F^{\mathrm{st}}_{ij}= -2\pi\sigma r \qty(\qty(1+\frac{2V^*}{\pi rs_{ij}^2})^{-\frac{1}{2}}-1),
\end{equation}
where the effective range of capillary forces is defined by a rupture distance $d_{\mathrm{r}}$, calculated in \cite{willett2000capillary} as
\begin{equation}
    d_{\mathrm{r}} = \qty(V^{*})^{\frac{1}{3}} + 0.1 \qty(V^{*})^{\frac{2}{3}}.
\end{equation}


Next, the moisture-dependent component of the capillary force is addressed by modeling a capillary force curve that varies with the moisture ratio, assuming the force is zero in two extreme cases: when the moisture content is $0$, indicating a completely dry state, and when the moisture ratio is $1$, indicating a completely saturated state. This reflects the absence of capillary effects in a space entirely devoid of water or fully filled with water. The capillary force reaches its maximum value when the moisture ratio equal to the maximum absorption ratio of particles, which means there is no free fluid particles in the space.
We fit this relationship through a Bézier curve, following the approach used in SPH-DEM, denoted by  $\Gamma((\mathrm{sr}_i + \mathrm{sr}_j)/2)$. Consequently, the concentration gradient force between two particles is expressed as
\begin{equation}
    \vb*{F}_{\alpha,ij} =
    \begin{cases}
    -\Gamma\qty(\frac{1}{2}(\mathrm{sr}_i + \mathrm{sr}_j))F^{\mathrm{st}}_{ij}\frac{x_i- x_j}{\norm*{x_i - x_j}} , & \text{if } 0 < s_{ij} < d_{\mathrm{r}}, \\
    0, & \text{else}.
    \end{cases}
\end{equation}

\begin{table*}[t]
    \centering
    \caption {
    Granule-In-Cell performance and parameters on all samples tested.
    ``Avg. Time'' represents the average runtime per fluid time step. For the sand-only test, the substep time is set to $1000$ single-step sand solving times.
    }
    \vspace{-1em}
    \tabcolsep=.25em
    \begin{threeparttable}
        \begin{tabular}{c|c|c|c|c|c|c|c|c|c|c}
        \hline
        \multirow{2}*{Case} & \multicolumn{4}{c|}{General Settings} & \multicolumn{2}{c|}{Particles} & \multicolumn{3}{c|}{Avg. Time (s)} & \multirow{2}*{Total Time (s)} \\
        \cline{2-10}
        & Figure & Resolution & Frames & Time Steps & Sand & Water & Sand & IDP & Overall & \\
        \hline
        Big Ball & Fig.~\ref{fig:density_3d_1} & $128 \times 256 \times 128$ & $100$ & $4071$ & $2.5 \times 10^5$ & $1.32 \times 10^7$ & $4.06$ & $1.541$ & $15.04$ & $6.1 \times 10^4$ \\
        Big Ball (w/o IDP) & Fig.~\ref{fig:density_3d_1} & $128 \times 256\times 128$ & $100$ & $2179$ & $2.5\times 10^5$ & $1.32 \times 10^7$ & $5.79$ & - & $16.85$ &  $3.7 \times 10^4$ \\
        Small Ball (Virtual Mass) & Fig.~\ref{fig:add} & $128 \times 256\times 128$ & $200$ & $1999$ & $2.5\times 10^5$ & $1.32 \times 10^7$ & $2.88$ & $2.31$ & $18.03$ &  $3.6 \times 10^4$ \\
        Stir (Ours)& Fig.~\ref{fig:stir} & $128 \times 192\times 128$ & $400$ & $7394$ & $1.18 \times 10^5$ & $1.83 \times 10^7$ & $4.14$ & $8.69$ & $18.34$ & $1.36 \times 10^5$\\
        \rv{Stir (MPM)}& Fig.~\ref{fig:stir} & $128 \times 192\times 128$ & $300$ & $2150$ & $1.18 \times 10^5$ & $1.83 \times 10^7$ & $5.44$ & $10.73$ & $16.17$ & $3.5 \times 10^5$\\
        Small Ball (3.5) & Fig.~\ref{fig:small} & $128 \times 256\times 128$ & $100$ & $2573$ & $2.0 \times 10^4$  & $1.93 \times 10^7$ & $1.377$ & $9.49$ & $17.65$  & $4.5 \times 10^4$  \\
        Small Ball (1.5) & Fig.~\ref{fig:small} & $128 \times 256\times 128$ & $100$ & $3452$ & $2.0 \times 10^4$  & $1.93 \times 10^7$ & $1.560$ & $9.49$ & $18.07$  & $6.2 \times 10^4$  \\
        Small Ball (0.2) & Fig.~\ref{fig:small} & $128 \times 256\times 128$ & $100$ & $803$ & $2.0 \times 10^4$  & $1.93 \times 10^7$ & $10.02$ & $7.84$  & $23.6$  & $1.89 \times 10^4$  \\
        Cat Litter & Fig.~\ref{fig:cat} & $128 \times 192\times 128$  & $300$ & $3412$ & $2.5 \times 10^5$ & $1.67 \times 10^4$ & $13.13$ & $0.01510$ & $13.73$  & $4.7\times 10^4$  \\
        Funnel (coarse) (Ours)& Fig.~\ref{fig:funnel2} & $128 \times 192\times 128$ & $100$ & $100$ & $1.19\times 10^3$ & - & $2.94$ & - & $2.94$ & $2.9 \times 10^2$\\
        \rv{Funnel (coarse) (MPM)}& Fig.~\ref{fig:funnel2} & $128 \times 192\times 128$ & $300$ & $300$ & $1.19\times 10^3$ & - & $52.10$ & - & $52.10$ & $1.56 \times 10^4$\\
        Funnel (fine) & Fig.~\ref{fig:funnel2} & $128 \times 192\times 128$ & $200$ & $4200$ & $4.7\times 10^4$ & - & $2.08$ & - & $2.08$ & $8.7 \times 10^3$\\
        Funnel (dry) & Fig.~\ref{fig:funnel} & $128 \times 192\times 128$ & $300$ & $5400$ & $2.8\times 10^4$ & - & $1.689$ & - & $1.689$ & $9.1 \times 10^3$\\
        Funnel (wet) (Ours)& Fig.~\ref{fig:funnel} & $128 \times 192\times 128$ & $300$ & $8895$ & $2.8\times 10^4$ & $3.1 \times 10^5$ & $1.918$ & $0.1005$ & $2.94$ & $2.6 \times 10^4$ \\
        \rv{Funnel (wet) (MPM)}& Fig.~\ref{fig:funnel} & $128 \times 192\times 128$ & $300$ & $300$ & $2.8\times 10^4$ & - & $53.20$ & - & $53.20$ & $1.60 \times 10^4$ \\
        Dam Breaking & Fig.~\ref{fig:dam} & $192 \times 192 \times 128$ & $500$ & $8385$ & $2.2 \times 10^5$ & $9.7 \times 10^6$ & $3.35$ & $5.54$ & $10.94$ & $9.2 \times 10^4$\\
        Sandfall & Fig.~\ref{fig:sandfall} & $128 \times 128 \times 128$ & $350$ & $2006$ & $1.20 \times 10^6$ & $1.48 \times 10^7$ & $32.2$ & $1.641$ & $43.6$ & $8.7 \times 10^5$\\
        \hline
        \end{tabular}        
    \end{threeparttable}
    \label{tab:statistics}
    \vspace{-0.5em}
\end{table*}

\section{Experiments}

\begin{figure*}
    \centering
    \setlength{\imagelength}{.245\linewidth}
\includegraphics[width=\imagelength]{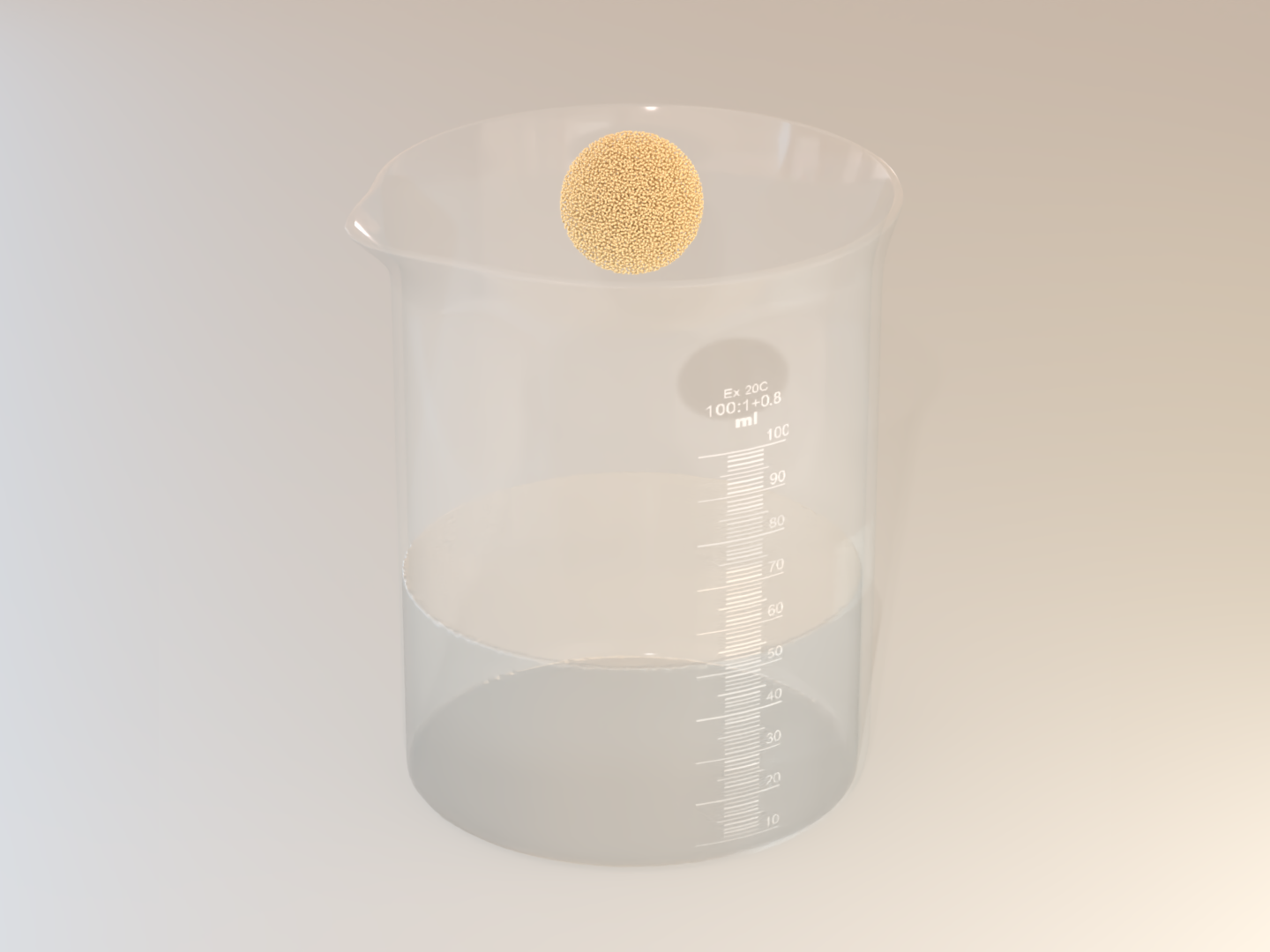}
\hfill
\includegraphics[width=\imagelength]{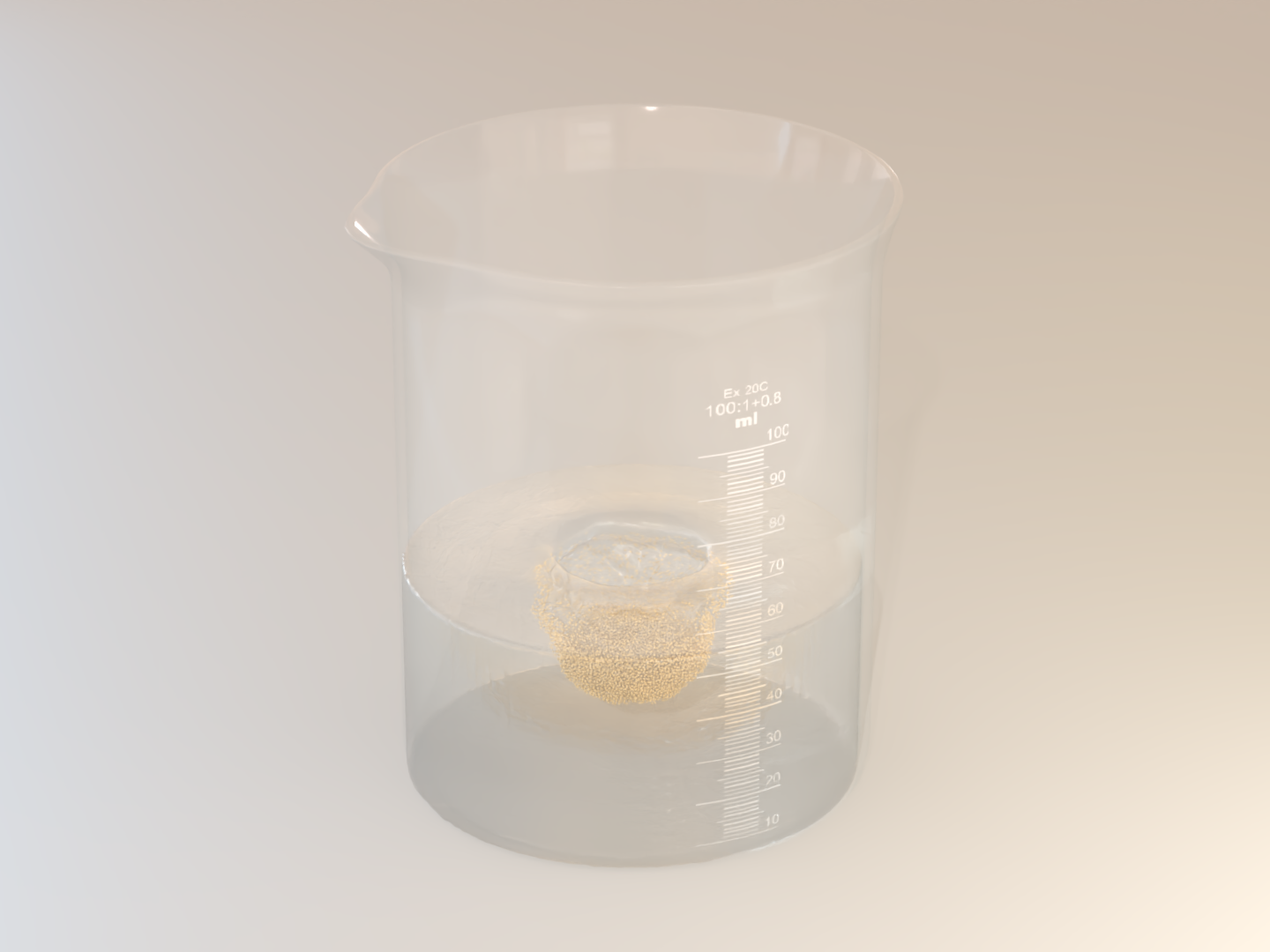}
\hfill
\includegraphics[width=\imagelength]{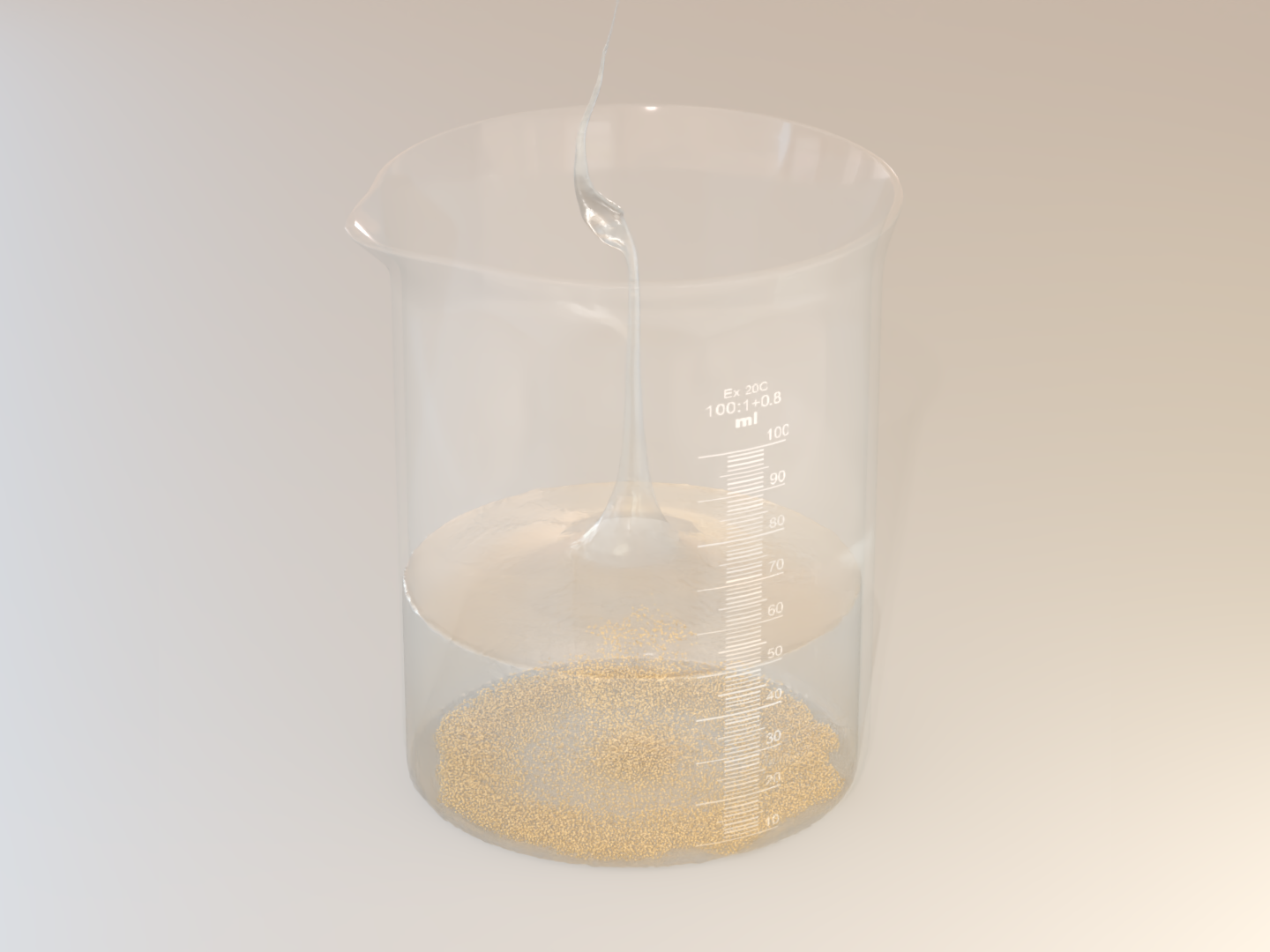}
\hfill
\includegraphics[width=\imagelength]{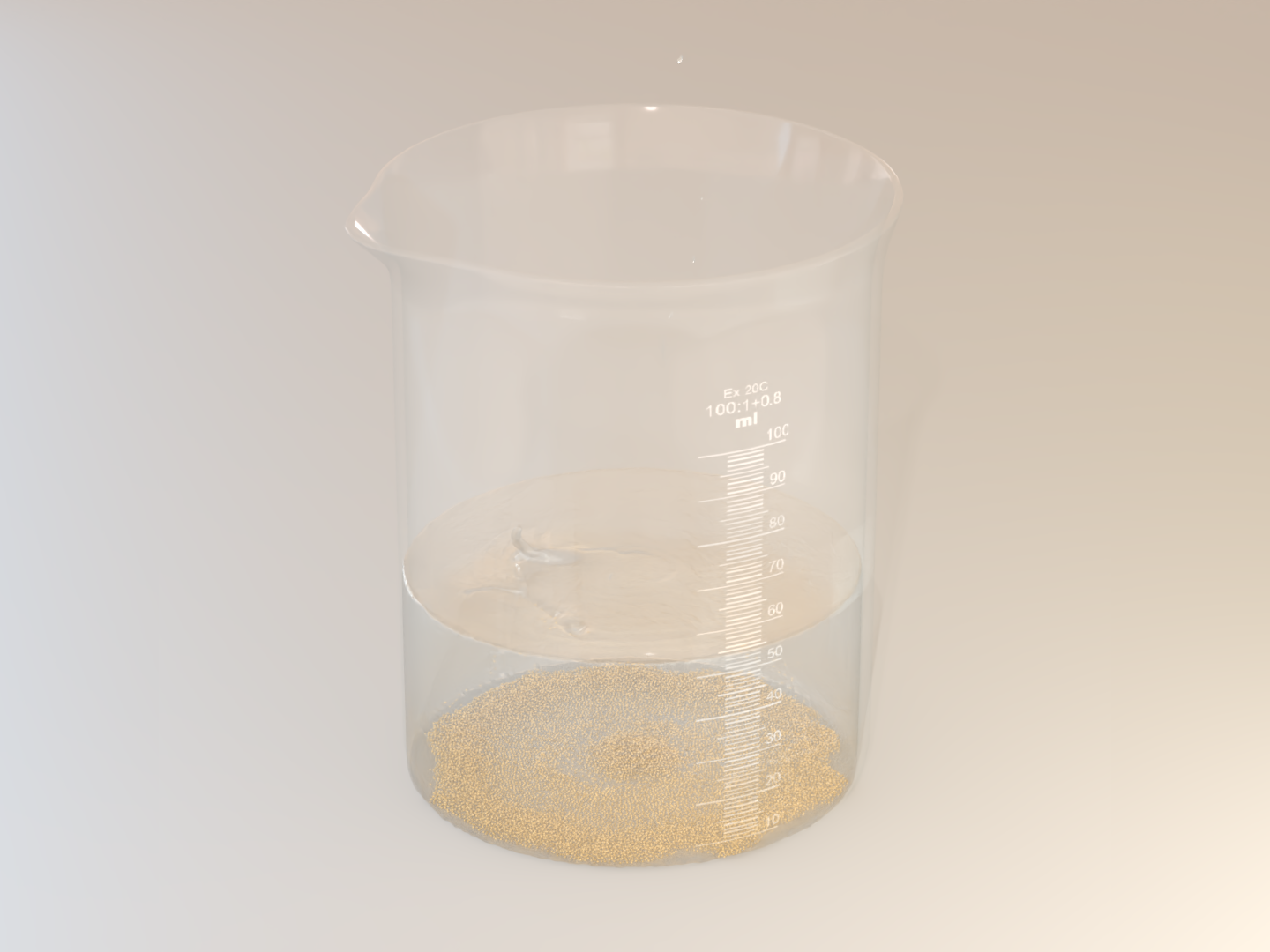}\\ 
\vspace{.00667\linewidth}
\includegraphics[width=\imagelength]{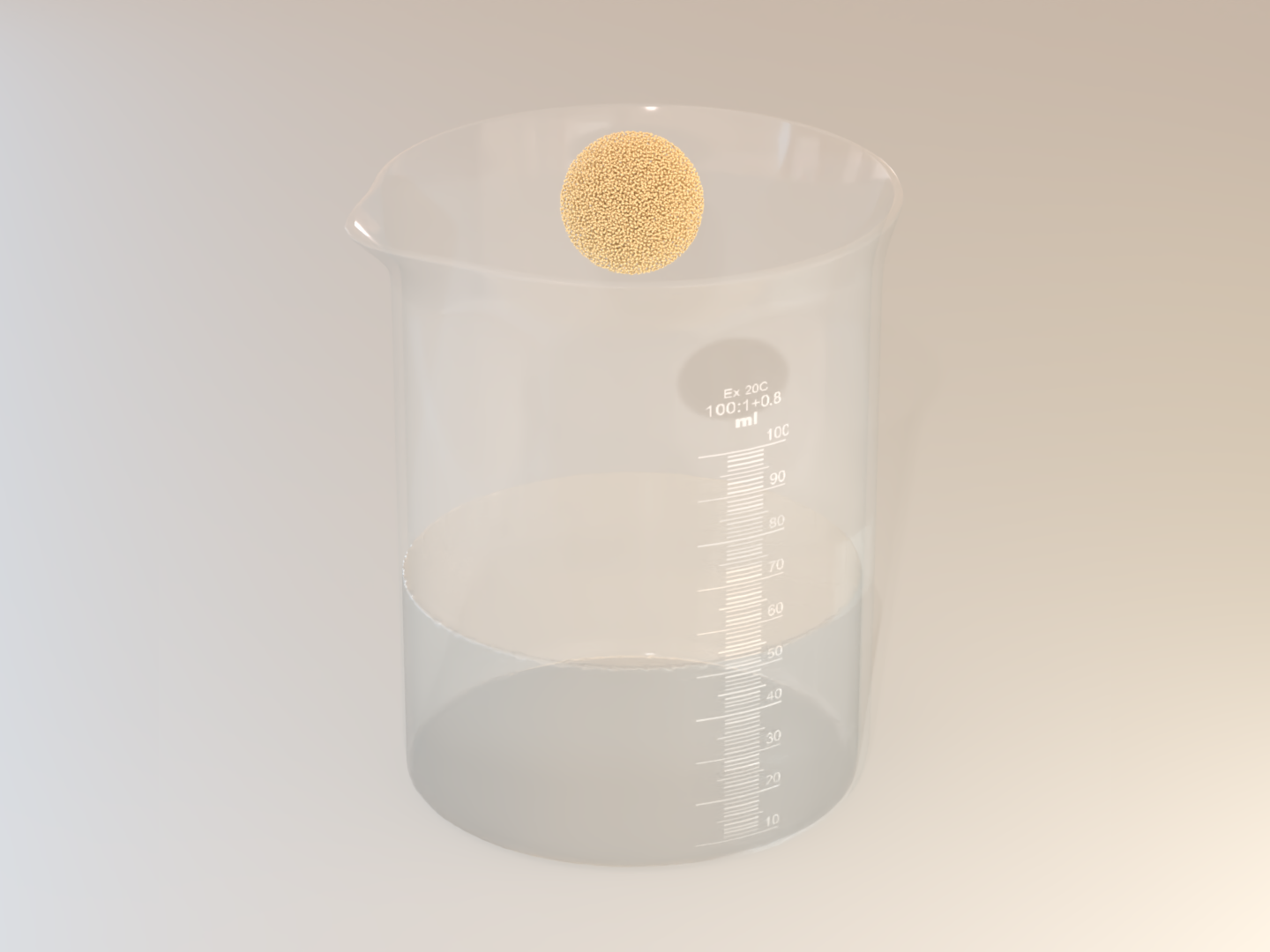}
\hfill
\includegraphics[width=\imagelength]{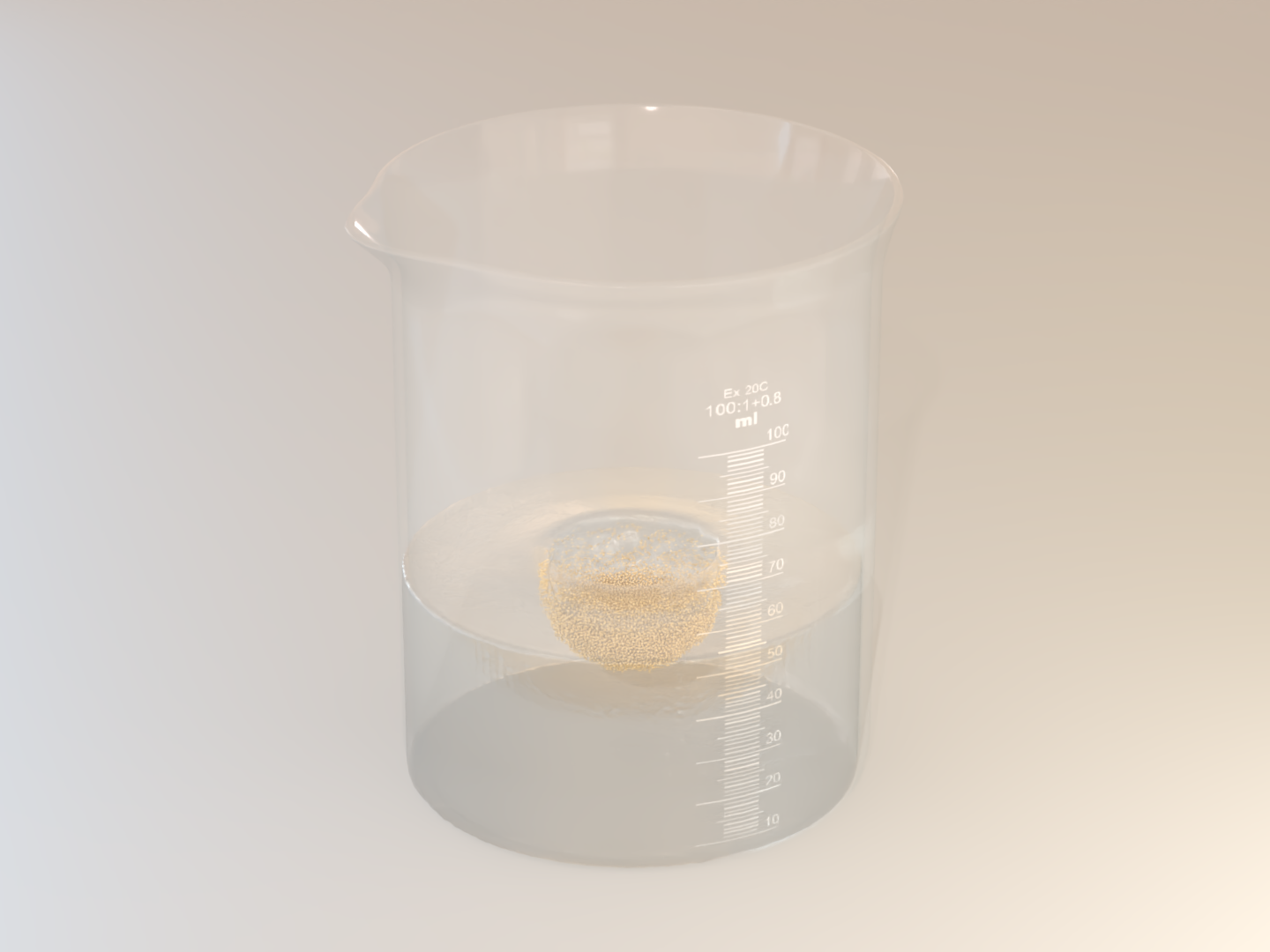}
\hfill
\includegraphics[width=\imagelength]{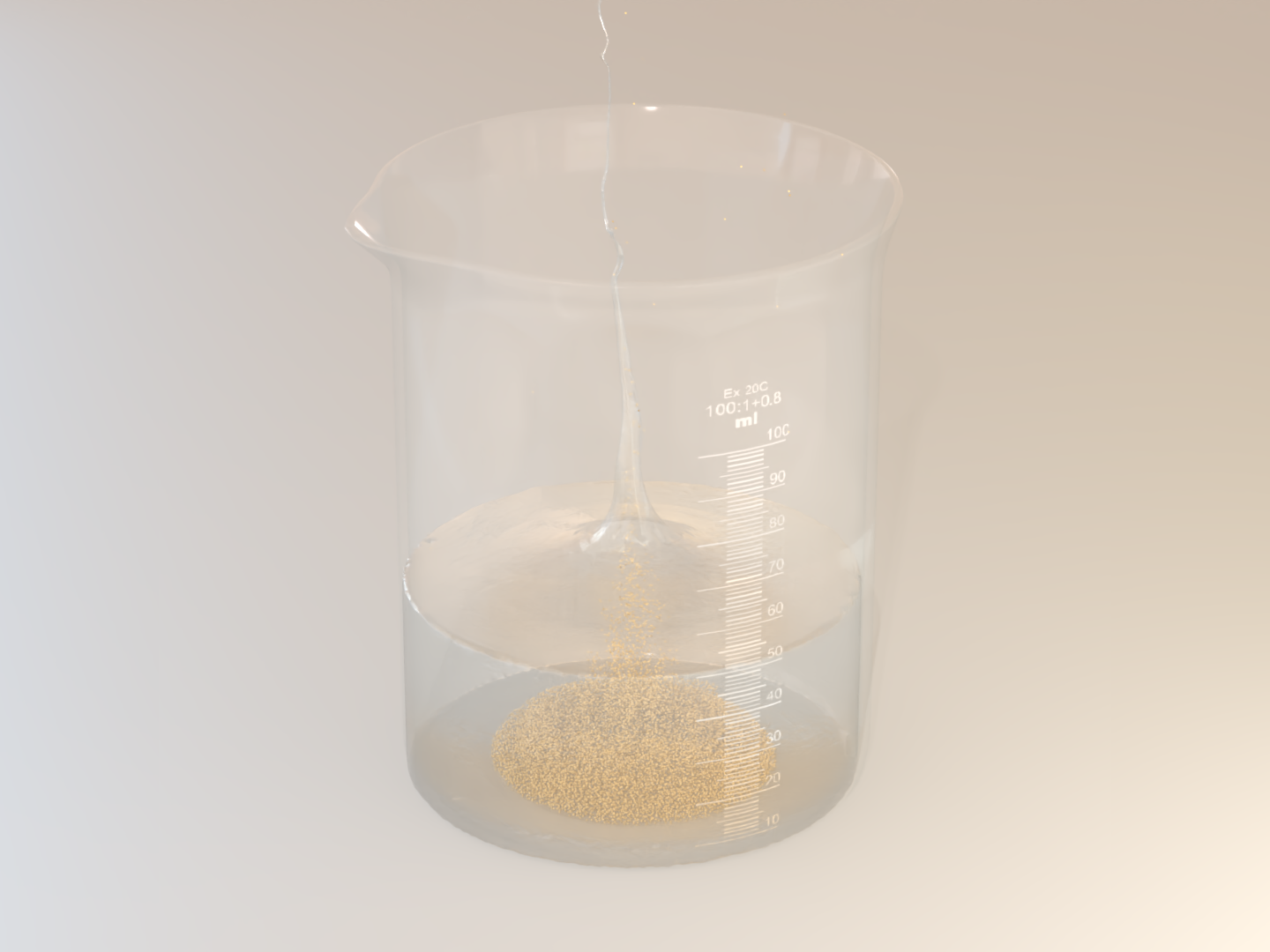}
\hfill
\includegraphics[width=\imagelength]{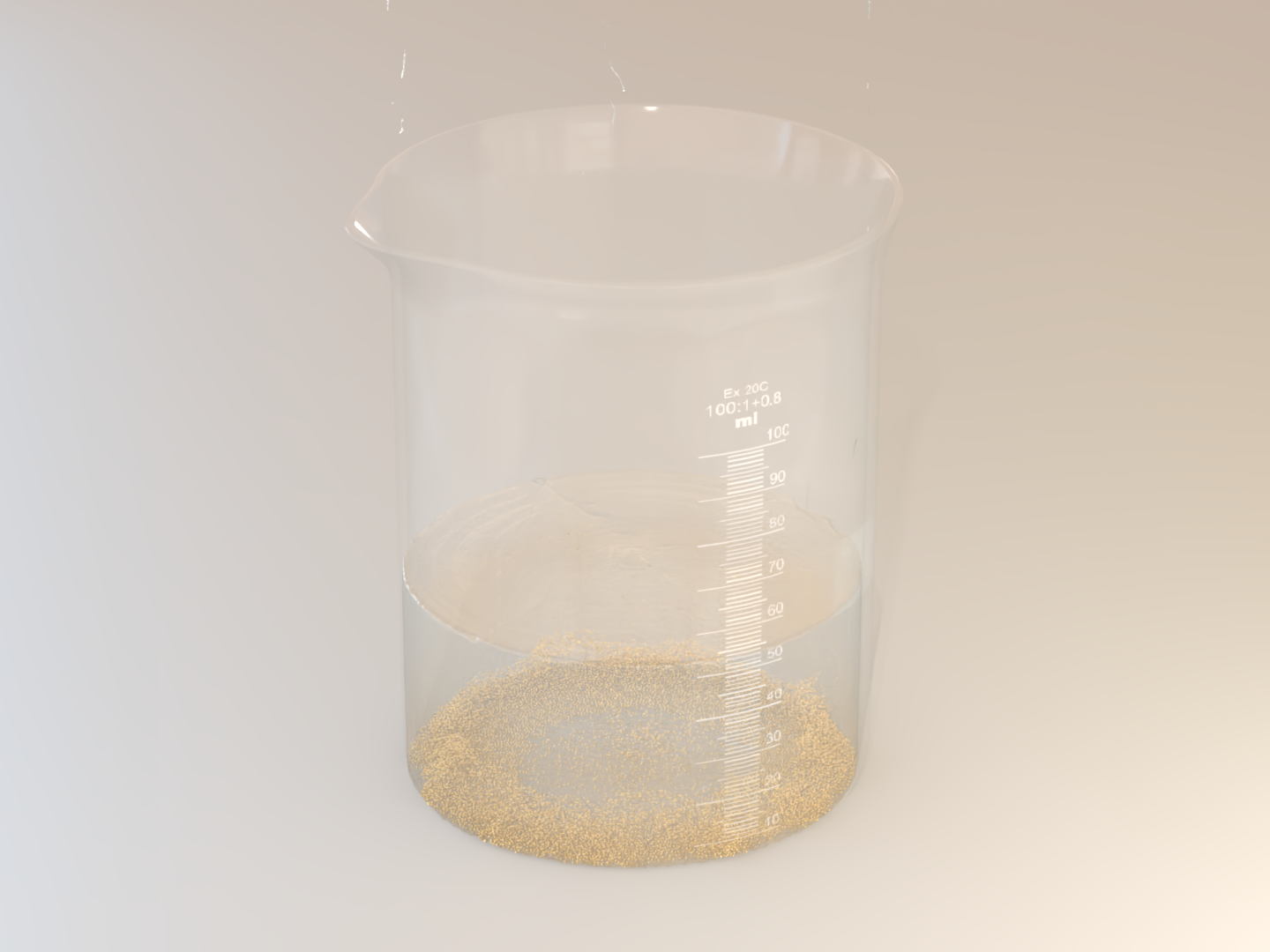}\\
\vspace{.00667\linewidth}
\includegraphics[width=\imagelength]{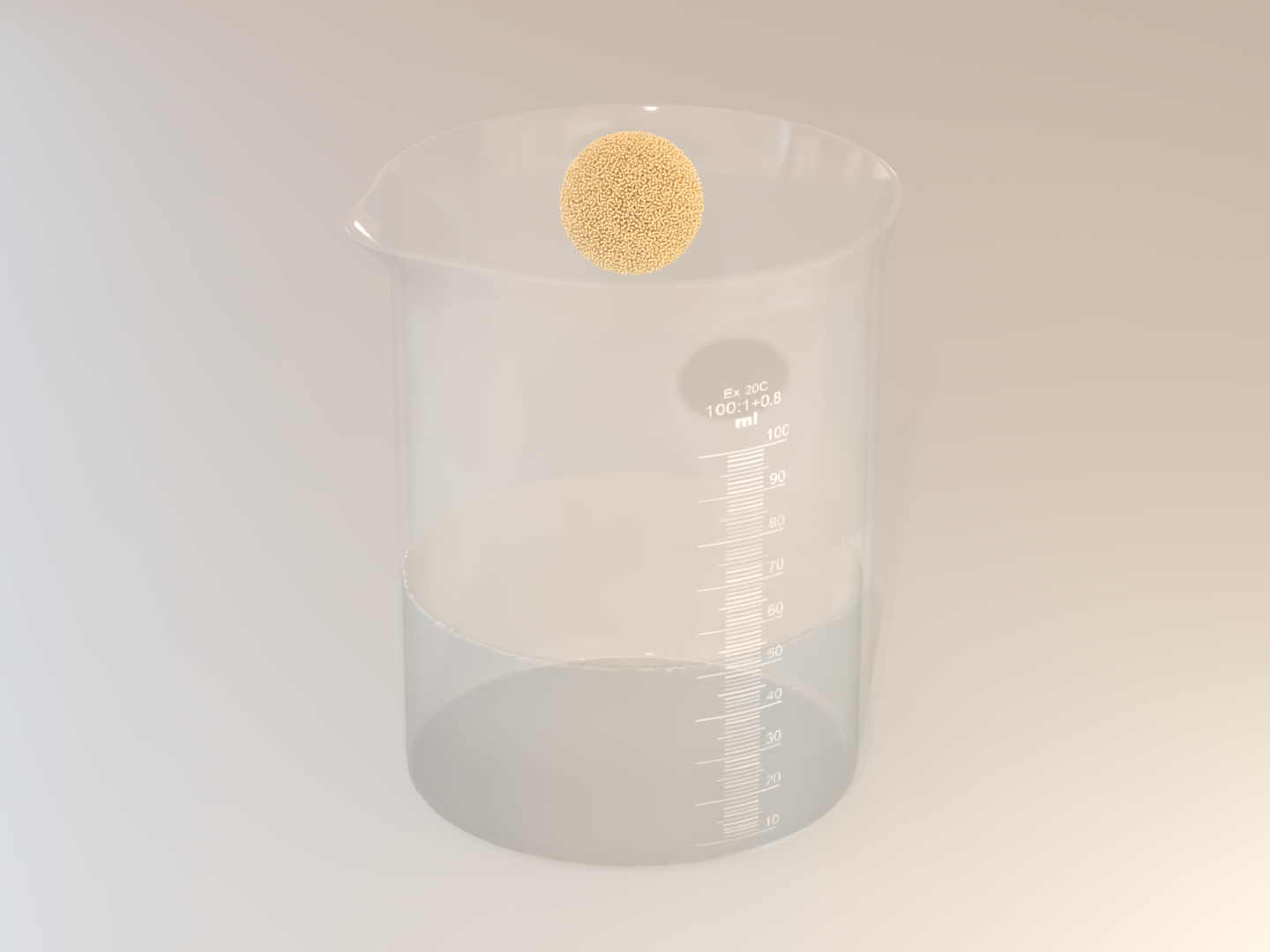}
\hfill
\includegraphics[width=\imagelength]{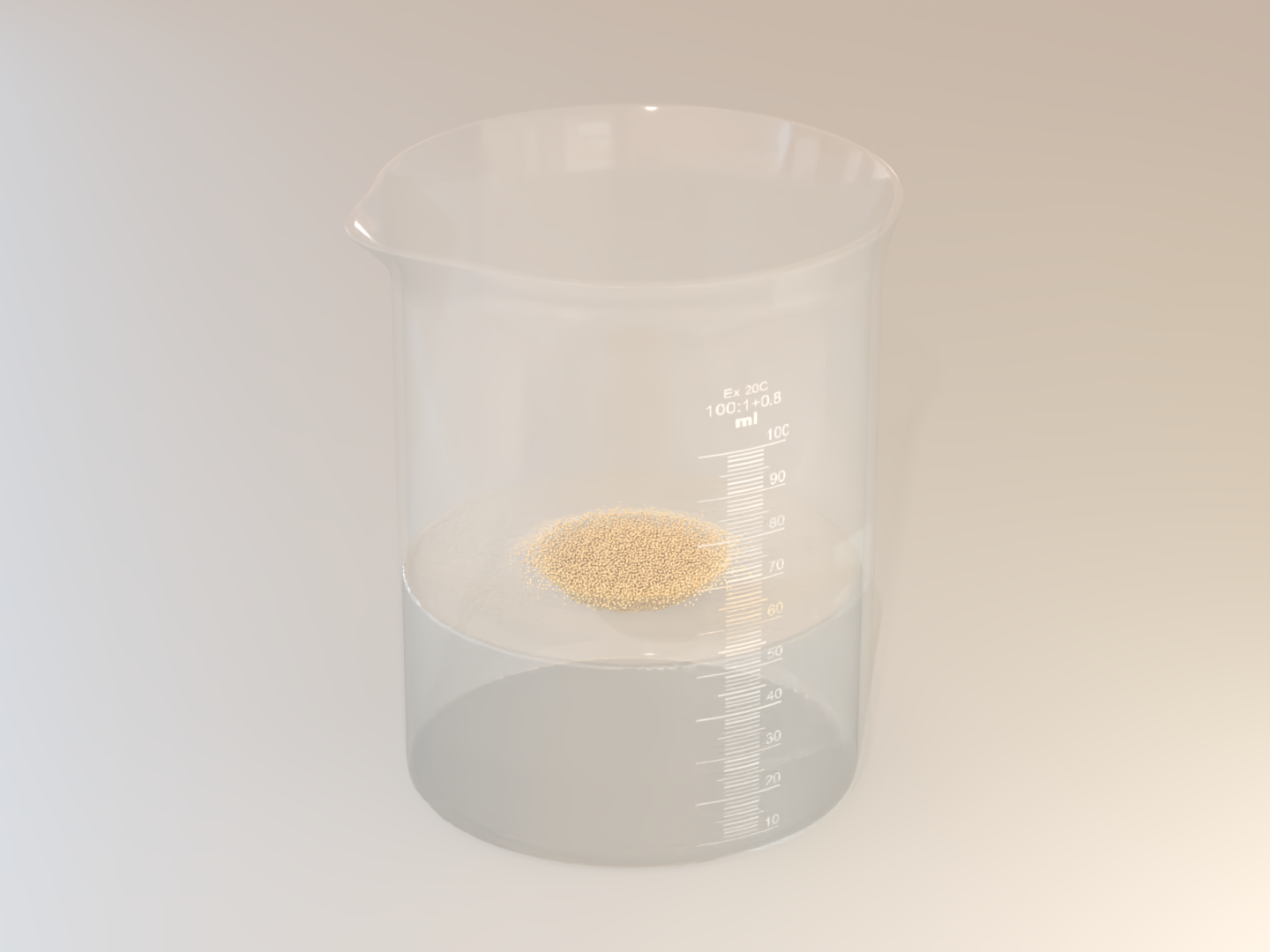}
\hfill
\includegraphics[width=\imagelength]{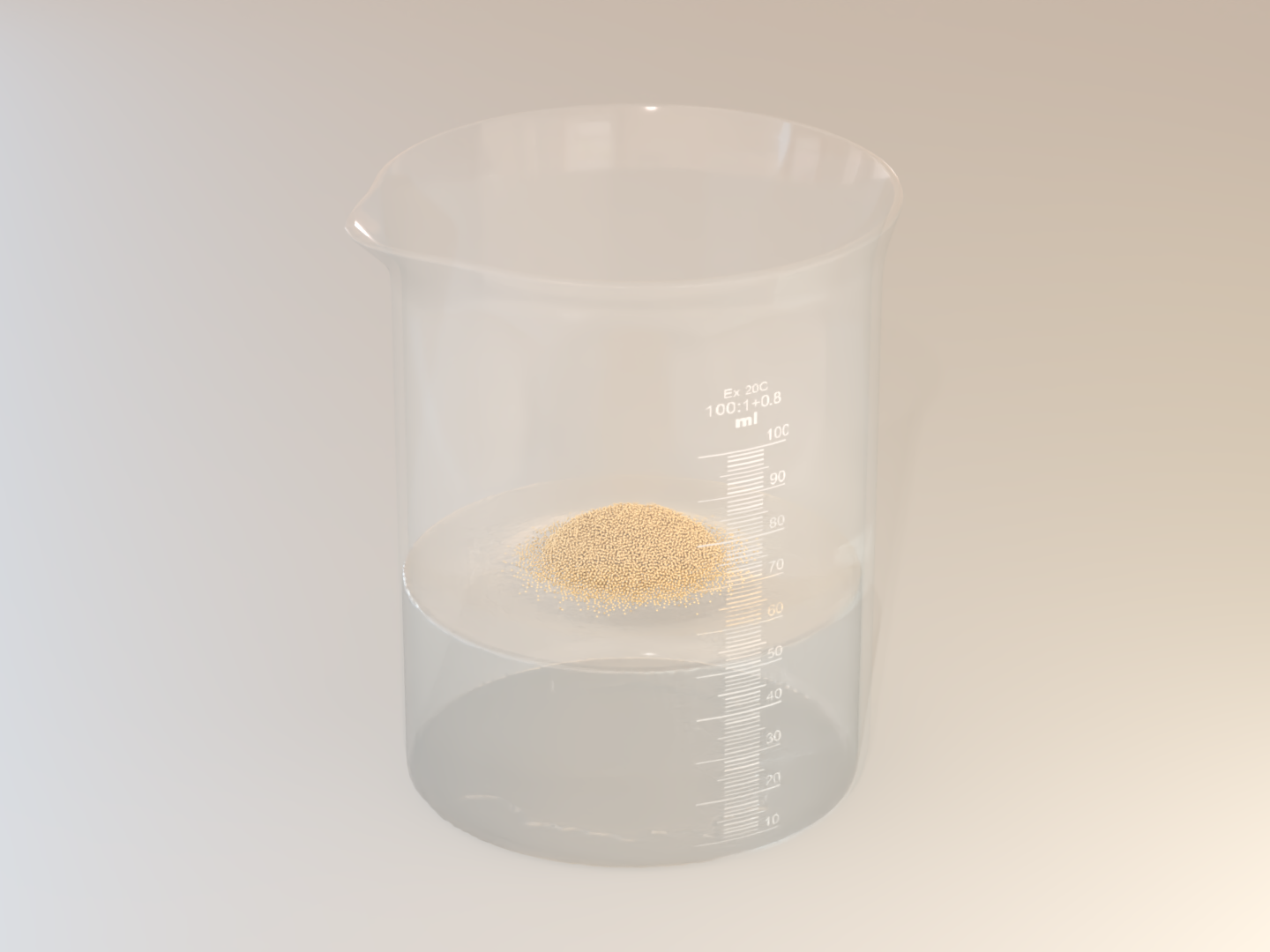}
\hfill
\includegraphics[width=\imagelength]{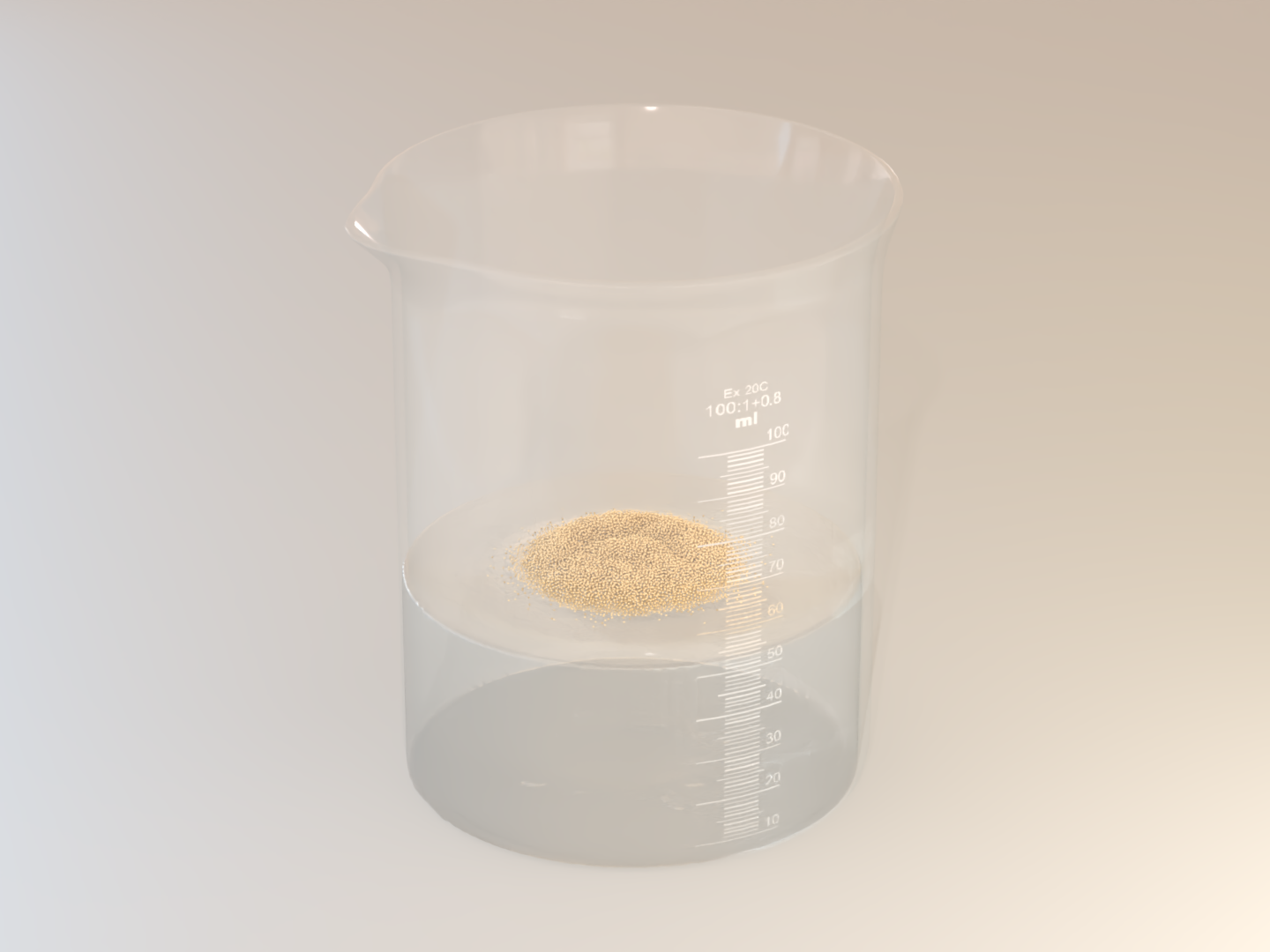}
\vspace{-1em}
\caption{Small Ball. In this figure we show the behaviour of balls with different densities falling into water. The three rows correspond to the densities $\rho = 3.5, 1.5, 0.2 \times 10^3\ \mathrm{kg}/\mathrm{m}^3$ from top to bottom, and the four column correspond to the densities $t = 0, 0.5, 1, 2\ \mathrm{s}$ from left to right repectively.}
\label{fig:small}    
\end{figure*}

\begin{figure}
    \centering
    \setlength{\imagelength}{.495\linewidth}
\includegraphics[width=\imagelength]{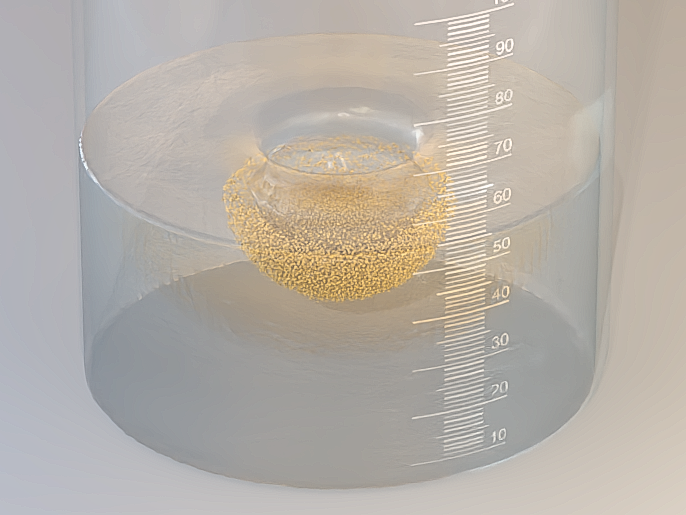}
\includegraphics[width=\imagelength]{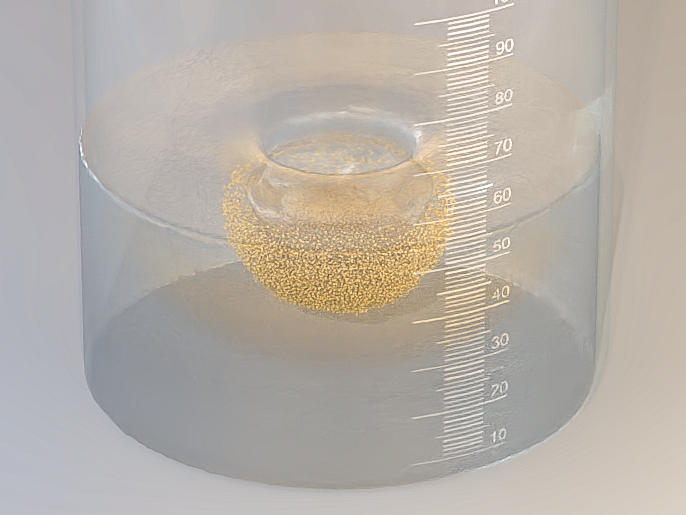} \\
\vspace{0.01\linewidth}
\includegraphics[width=\imagelength]{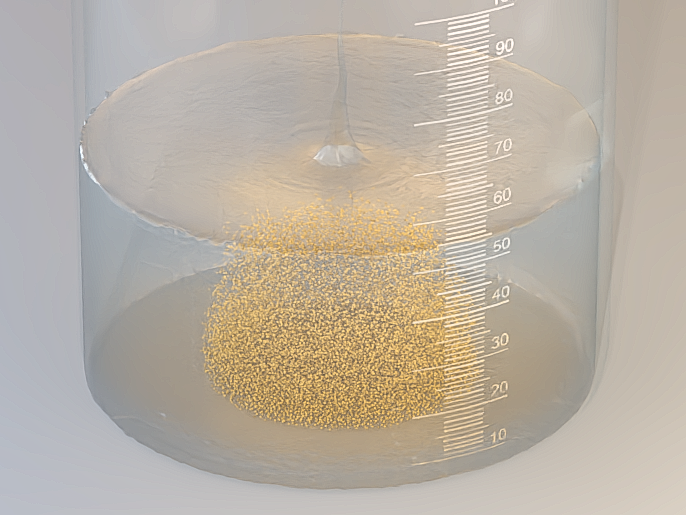}
\includegraphics[width=\imagelength]{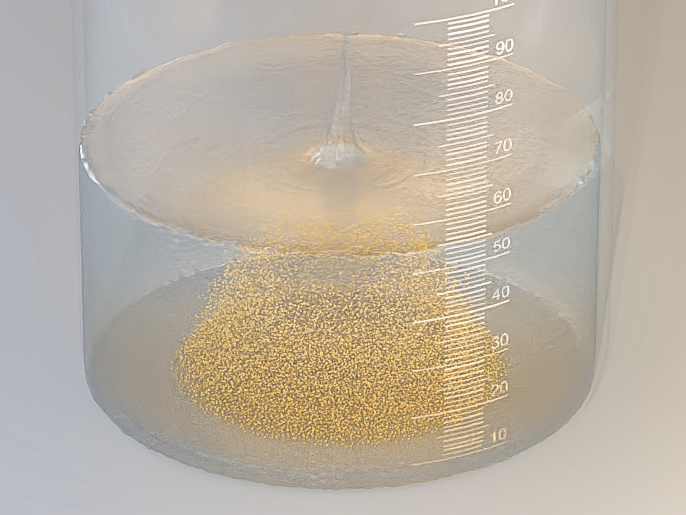}\\
\vspace{0.01\linewidth}
\includegraphics[width=\imagelength]{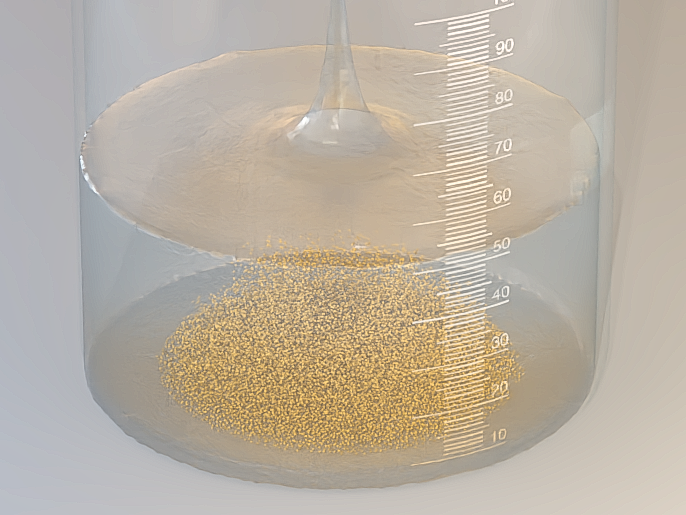}
\includegraphics[width=\imagelength]{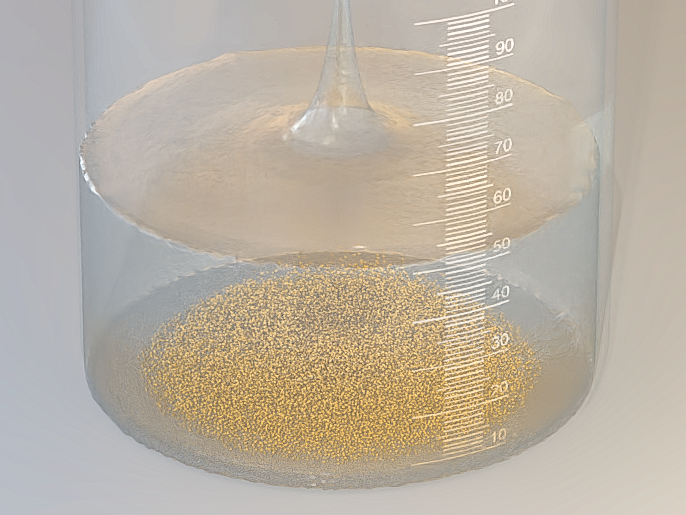}
\vspace{-2em}
\caption{Virtual Mass Force. In this figure, we show the virtual mass effect. The two column have the same initial conditions for the ball, but left one has the Virtual mass force added, while the right one does not. The three rows correspond to $t = 0.5, 0.75, 0.875\  \mathrm{s}$ respectively.}
\label{fig:add}    
\end{figure}

We implemented the GIC method 
based on C++ and used Intel oneAPI Threading Building Blocks (oneTBB) for parallel acceleration on the CPU. All the tests were run on an AMD EPYC 9K84 processor (80 core, 3.7 GHz, 160 GB RAM).

\paragraph{Parameters} In our samples, the density of water $\rho_\tf = 10^3\ \mathrm{kg}/\mathrm{m}^3$, the density of sand $\rho_{\ts} = 2.5\times 10^3\ \mathrm{kg}/\mathrm{m}^3$, Young's modulus of the sand $E = 10^6\ \mathrm{Pa}$, and Poisson's ratio $\nu = 0.3$. Grid spacing $h = 7.8 \times 10^{-4}\ \mathrm{m}$, sand particle radius $r = 3.9 \times 10^{-4}\ \mathrm{m}$, surface tension coefficient $\sigma = 0.07\ \mathrm{N}/\mathrm{m}$ and coefficient of drag between sand and water $\mu = 0.44$ \cite{schiller1933uber}.

\paragraph{Time steps} The time step for fluid $\Delta t$ in our method is determined by the Courant--Friedrichs--Lewy (CFL) condition \cite{Lewy1928}, while for granules $\Delta t'$ is determined by the Rayleigh criterion \cite{tavarez2007discrete}. For stiff granules, typically $\Delta t' \sim \Delta t / 100$. The Rayleigh time step $\Delta t'$ for sand is written as
\begin{equation}
    \Delta t' = \frac{1}{2}\sqrt{\frac{m_\ts}{k_{\mathrm{n}}}},
\end{equation}
where $m_\ts$ denotes the mass of granule, $k_{\mathrm{n}} = Er$ denotes the normal stiffness coefficients in Eq. \eqref{eq:k_n}. And to prevent a large gap between the fluid time step and the sub-time steps for the sand, an upperbound for the fluid time step is adopted as
\begin{equation}
    \Delta t = \min \qty{\frac{\Delta x}{\max(\norm{\vb*{v}_\tf},\norm{\vb*{v}_\ts})}, 1000 \cdot \Delta t'}.
\end{equation}
Dimensional information, performance statistics and other relevant parameters for all our samples are presented in Table~\ref{tab:statistics}.

\paragraph{Comparison experiments} The simulation and rendering of SPH-DEM method are based on the released code from \citet{wang2021visual}. For the lack of open-source MPM-based implementation for sand--water mixtures, we implemented the algorithm as \citet{gao2018animating}, employing APIC for water and MPM for sand.

\subsection{Ablation Study}
\paragraph{Implicit Density Projection}
Figure~\ref{fig:density_3d_1}, \ref{fig:density_3d_2} and  \ref{fig:density_3d_3} shows the ability to maintain the total volume by density projection. In this test, the maximum water absorption ratio of sand is set to $0$. In the experiment, when employing the density projection method, the fluid level rises while the total volume stays unchanged; without IDP, water can occupy the same space as sand, causing the fluid level to remain at its initial height.
SPH-DEM uses the equation of state related to the fluid, neglating the presence of the granule, causing the total volume oscillation and reducing to the initial volume of fluid.
Even though the phase parameter is included in the transport equation in MPM, cavity forms inside the fluid due to the lack of recovery strategy. As undergoing much bigger distortion from oupling with the sand, the total volume increases after the crash from the sand ball.

\paragraph{Virtual Mass Effect}
Figure~\ref{fig:add} illustrates the virtual mass effect. In this test, the drag force coefficient is set to $\mu = 0.044$ to emphasize the virtual mass effect. This force serves as a kind of inertia that the overall motion of the sand ball gets slower, while the trajectory is almost unaffected.

\subsection{Simulation}

\paragraph{Small Ball into Water}
Figure~\ref{fig:small} demonstrates the effect of the pressure gradient force. In this test, a sand ball of different densities falls into the water. Since the gradient of pressure in static water is only related to the density of the fluid, the motion of the denser sand ball is much less susceptible to this force. From the figure, the denser ball tends to pile up in the middle of the beaker, while the less dense ball floats on the surface. Since the less dense ball is only saturated with water at its bottom, isolating the upper layer from the water, the top of floating sand pile stays dry.

\begin{figure}
    \centering
    \setlength{\imagelength}{.495\linewidth}

\includegraphics[width=\imagelength]{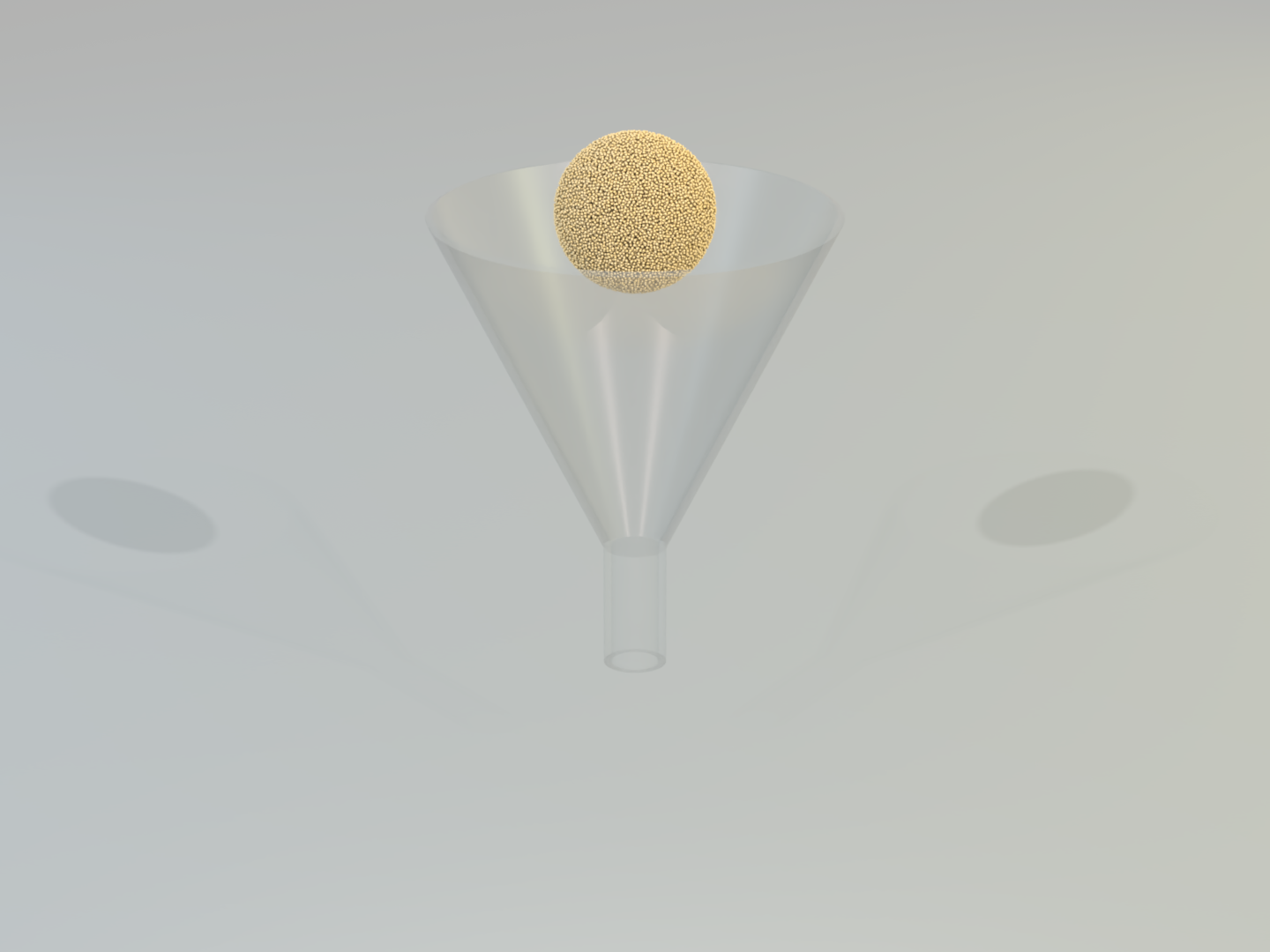}
\includegraphics[width=\imagelength]{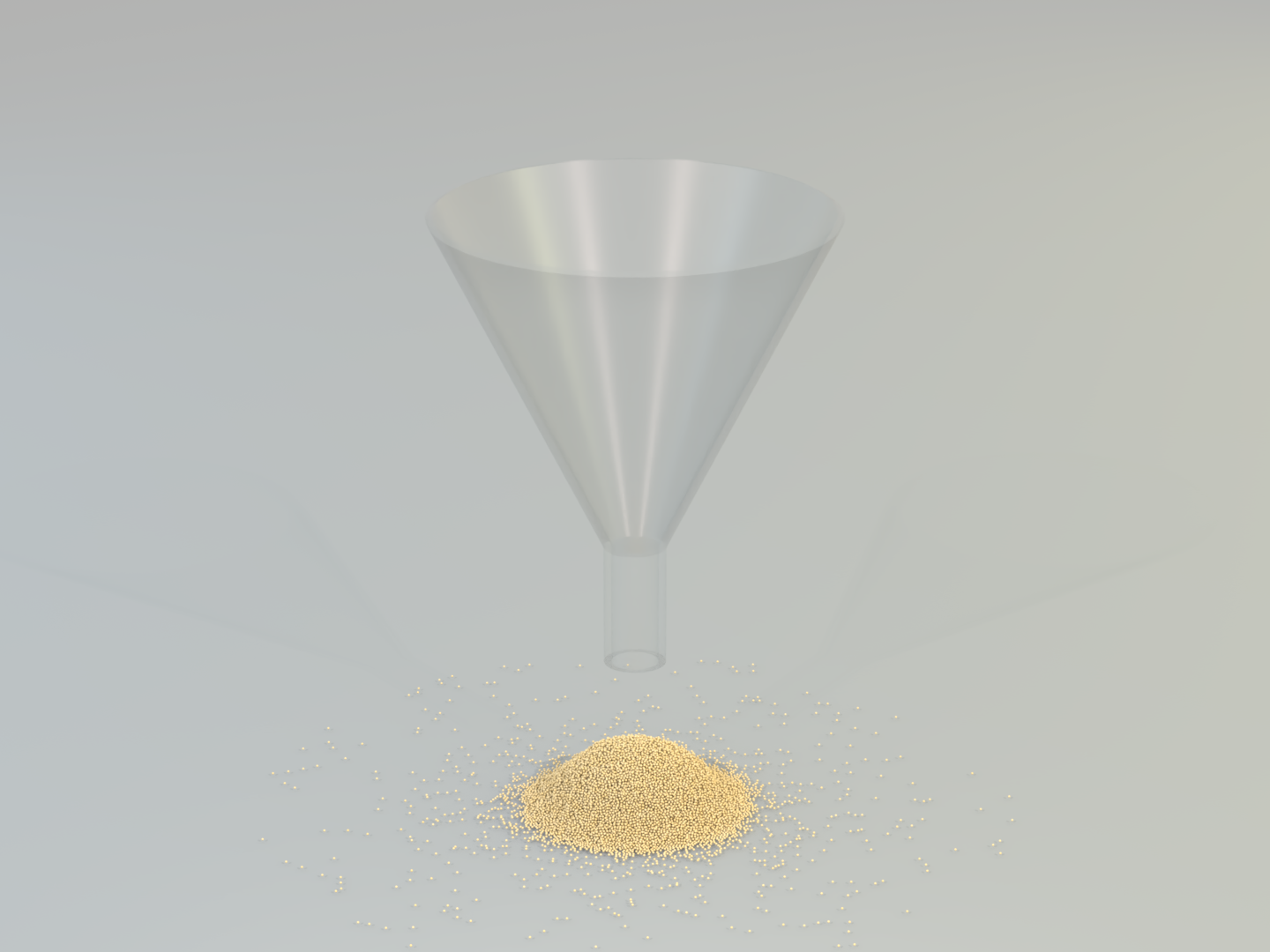}
\\
\vspace{0.005\linewidth}
\includegraphics[width=\imagelength]{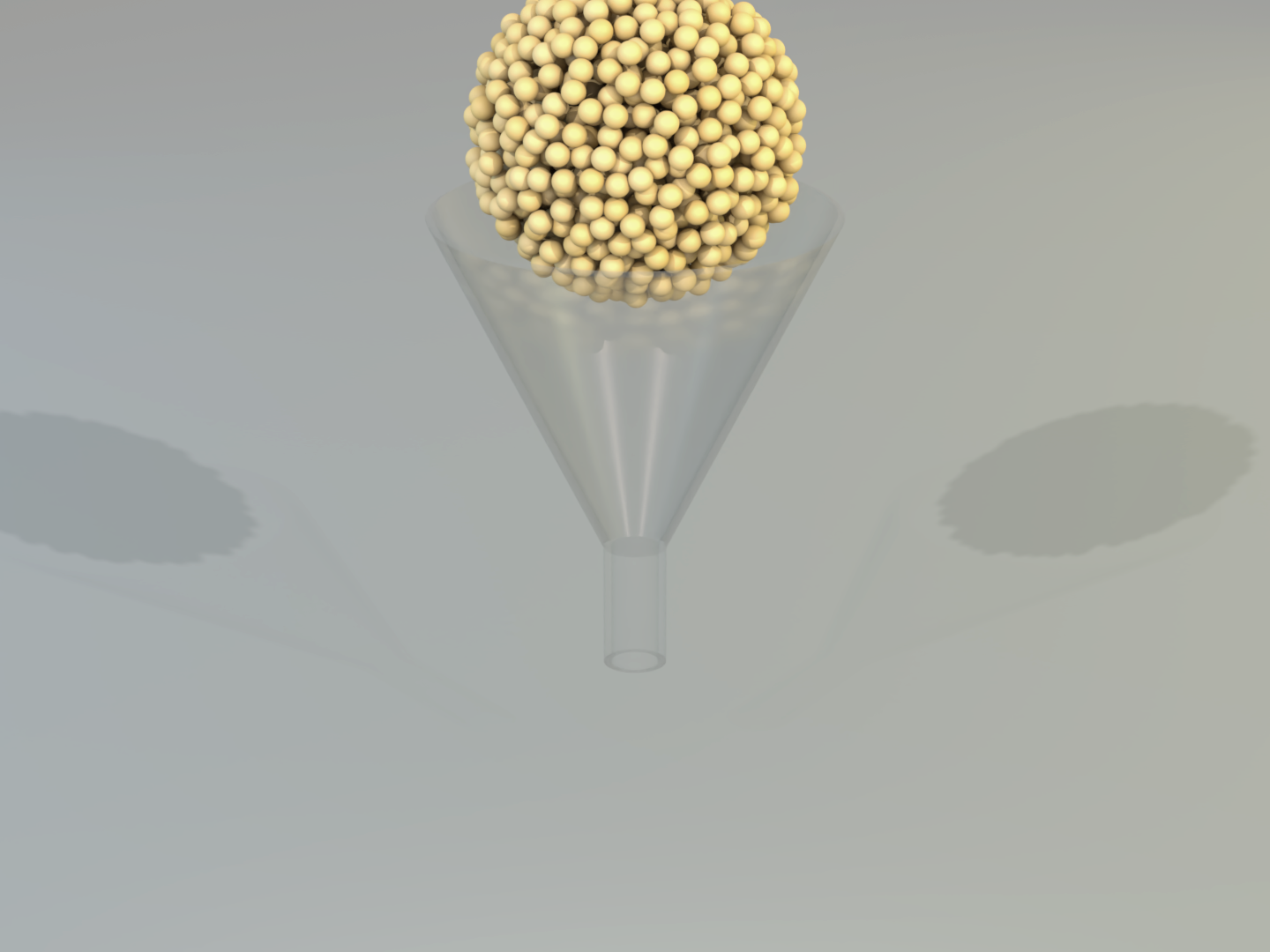}
\includegraphics[width=\imagelength]{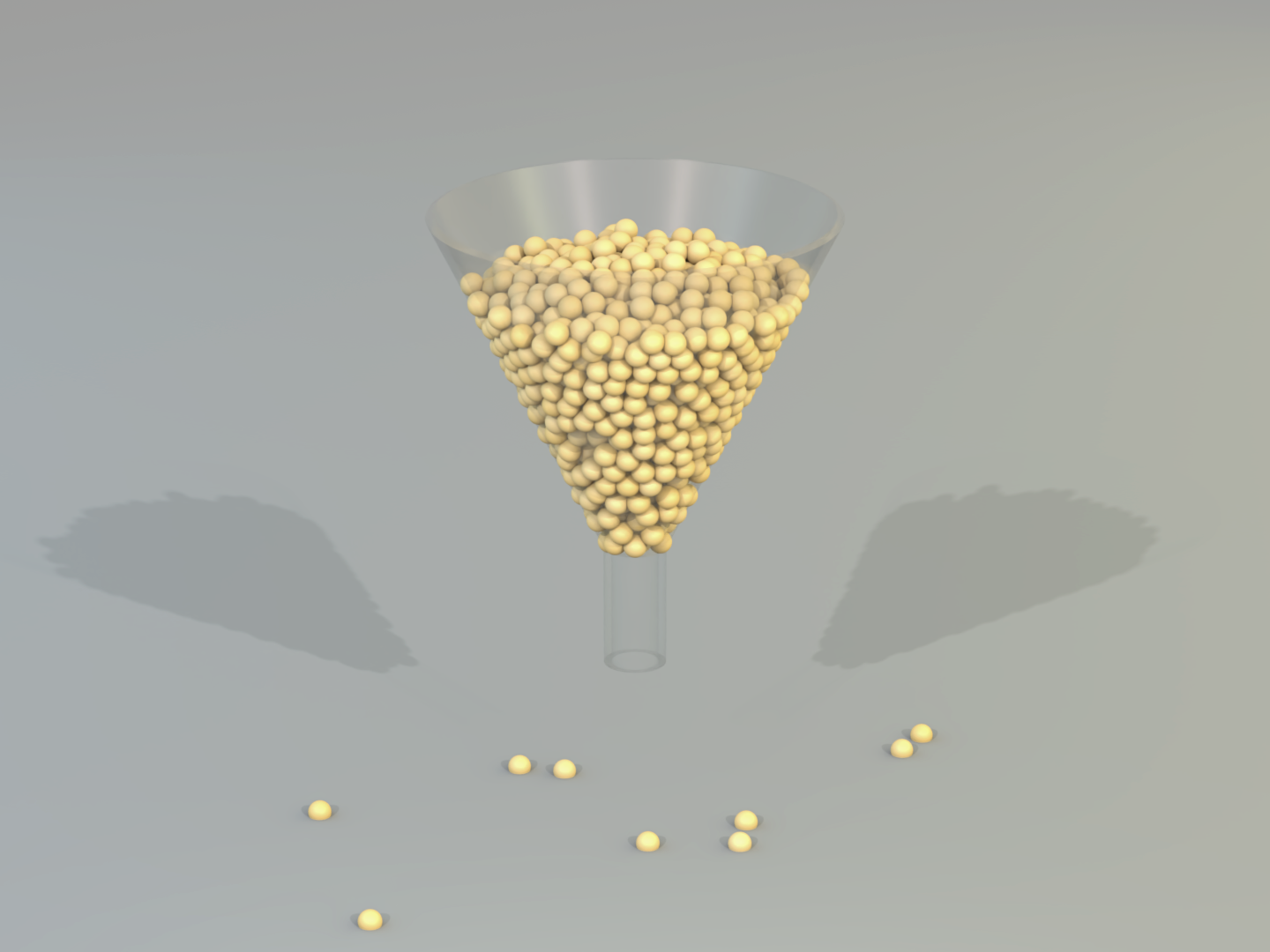} 
\\
\vspace{0.005\linewidth}
\includegraphics[width=\imagelength]{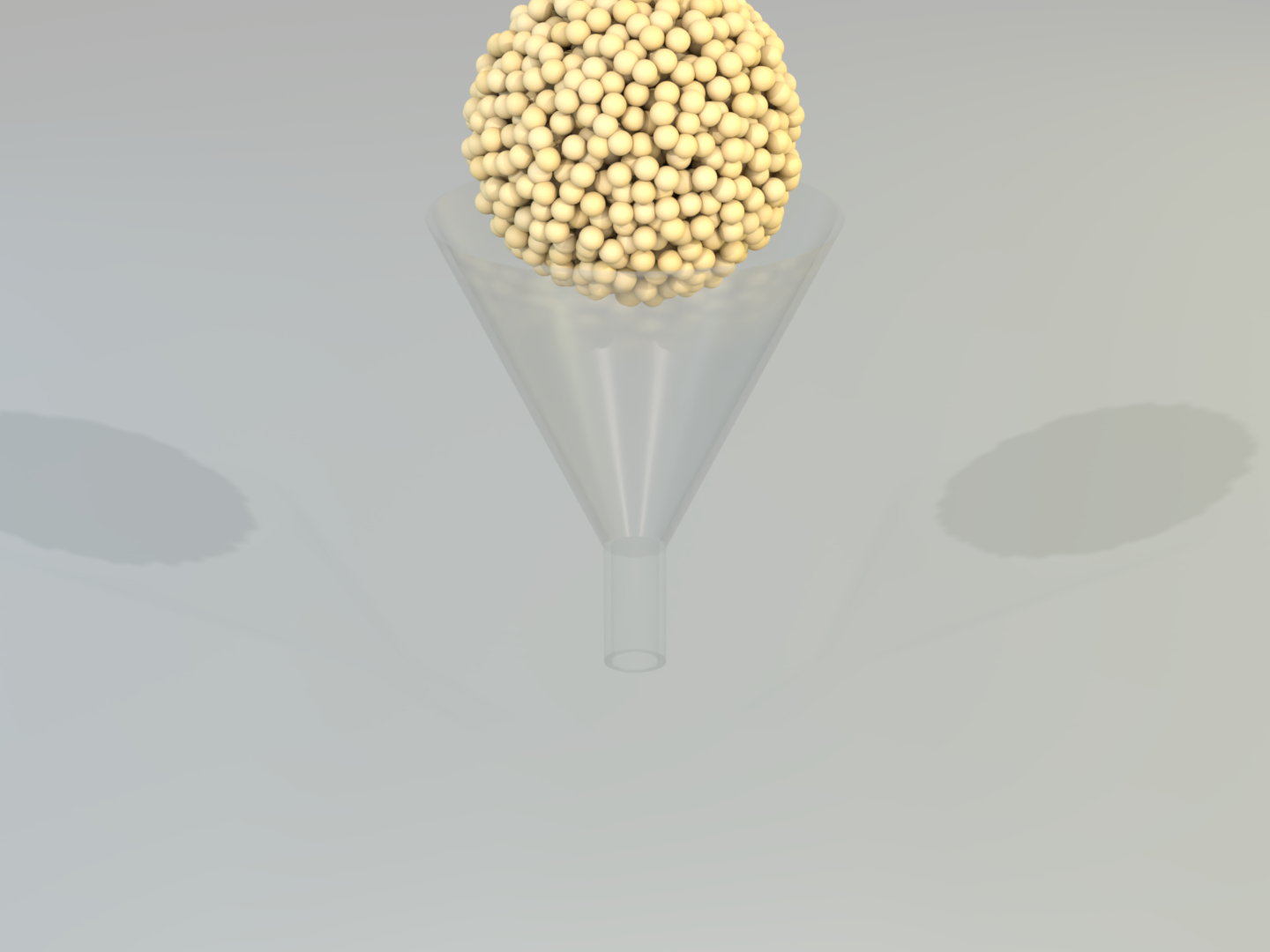}
\includegraphics[width=\imagelength]{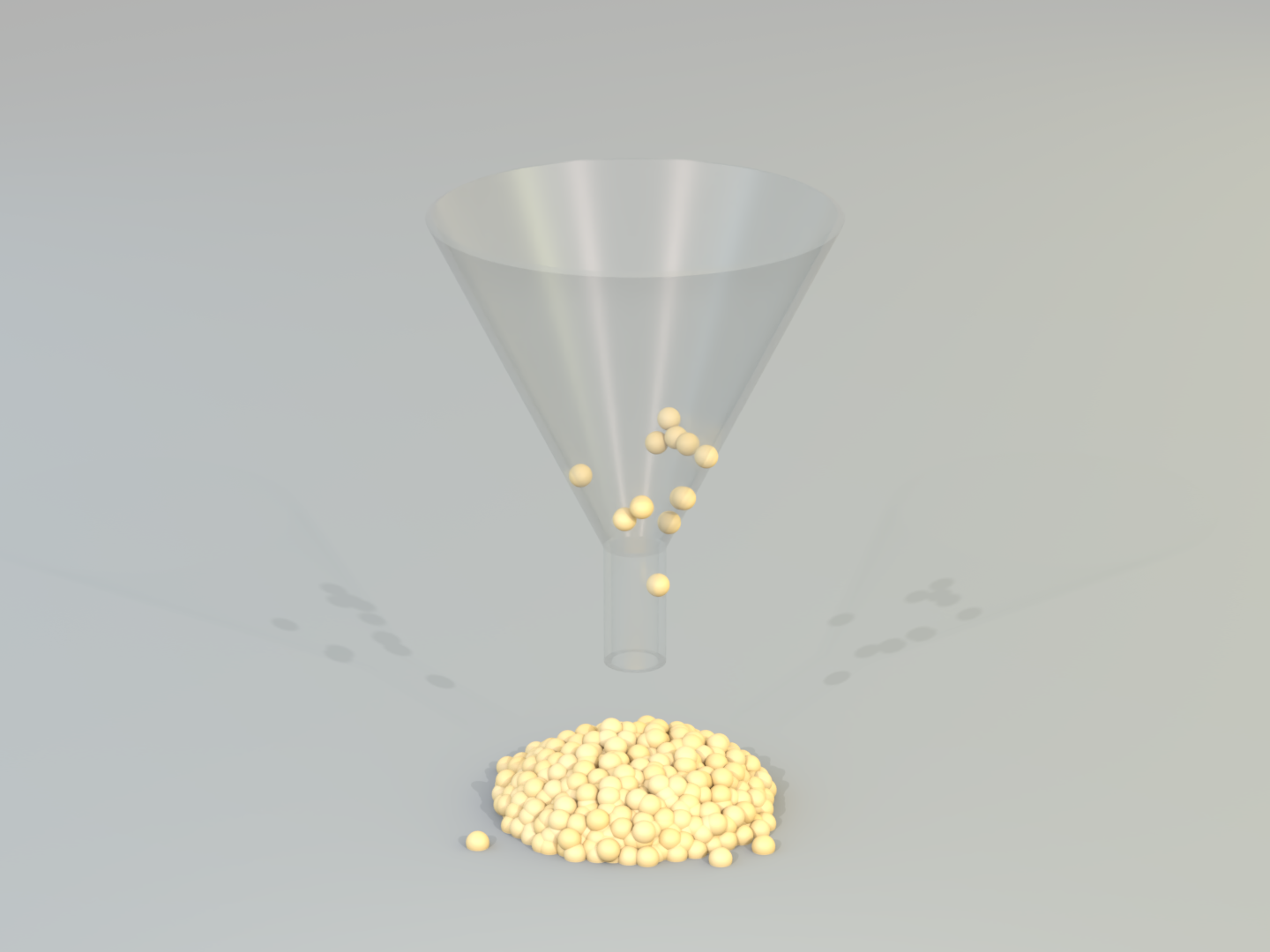}




\caption{\rv{Sand through funnel. In this figure, we show the behavior of sand of different particle size going through the funnel. The three rows shows the sand with radius $r = 1.95 \times 10^{-4}\ \mathrm{m},
1.37 \times 10^{-3}\ \mathrm{m}$ simulated with our method, and $r = 1.37 \times 10^{-3}\ \mathrm{m}$ simulated using the method of \citet{klar2016drucker}, respectively. The two columns show the initial state and the final steady state. As can be seen from the figure, our algorithm successfully demonstrates radius-dependent clogging.}}
\label{fig:funnel2}  
\end{figure}

\paragraph{Funnel}
In Figure~\ref{fig:funnel2}, we simulate a sand ball with different granular radius falling through a funnel with a half-angle of $\arctan(0.5)$ and an opening radius of $0.03\ \mathrm{m}$. Although the diameter of the granules is smaller than the funnel mouth, the sand ball with larger particle size still blocks the funnel due to the friction, while the finer one pass smoothly. 
\rv{
This result is in contrast with the MPM approach used by \citet{klar2016drucker}, where boundary conditions are imposed through the background grid. As a consequence, the outflow through the funnel mouth is constrained by the cell size rather than the physical particle size. 
When using a particle radius identical to our larger granules, their simulation allows the material to flow through away.
This comparison highlights DEM's advantage in capturing behaviors inherently related to particle-scale effects.
Additionally, as shown in Table~\ref{tab:statistics}, the MPM is considerably more computationally expensive for the coarse sand funnel experiment. The primary reason is that the computational cost is tied to the grid, not just the number of particles. Although the number of sand particles is low, the system maintains a high grid resolution to respect the boundary, leading to prolonged computation. This clearly demonstrates the DEM's advantage for simulating sand particles with a radius much larger than the grid cell size, as its computation is particle-based and avoids this grid-dependent cost.
}

In Figure~\ref{fig:funnel}, we test the concentration gradient force by simulating the sand with different moisture ratio levels as it flow through the funnel. To be noted, sand particle radius $r = 2.3 \times 10^{-4}\ \mathrm{m}$ in this experiment.
Dry sand, with no concentration gradient forces, flows freely through the funnel.
Conversely, when saturated with absorbed water, the sand reaches the maximum concentration gradient force. This causes the sand particles to clump together, displaying enough cohesion to support themselves near the funnel opening rather than falling out.
\rv{
In comparison, while the MPM method by \citet{tampubolon2017multi} incorporates a Drucker–Prager model that can control the value of cohesion, it fails to reproduce the solid-like blocking behavior observed in reality. Their method does not allow the sand to form a stable arch, and the sand continues to flow through the funnel.
}
Noting that although all the sand stays inside the funnel in the real experiment, there is a small pile falls out in our simulation, this discrepancy arises because the sand we throw into the funnel in the real sensoria has a flat bottom, whereas the bottom of the small ball in the simulation is convex. After the small pile falls out, the sand remaining in the funnel also develops a flat bottom, which further indicates the realism of our simulation.
\rv{After adding more water, the spatial density distribution of water becomes uniform, leading to $\mathrm{sr} = 1$ for any sand particle in Eq. \eqref{eq:symmetry-capillary} when calculating capillary force. Consequently, the concentration gradient force is eliminated and the granules flow out of the funnel again.}
Our algorithm effectively addresses the challenge of obtaining interactions under relatively constant deformation near the funnel opening, which is a limitation faced by continuous representation models.

\paragraph{Stir}
In Figure~\ref{fig:stir} we simulate the tea leaf paradox, which demonstrates the dynamics of particles in a rotating fluid.
When tea is stirred in a cup, the leaves will migrate to the center bottom of the cup rather than being pushed to the edge by centrifugal force \cite{einstein1926ursache}.
That is because as the fluid rotates, a parabolic surface forms due to centripetal force, creating a pressure gradient from the edge toward the center.
Near the bottom, friction slows the fluid, making the inward pressure gradient stronger than the centrifugal force required and generating an inward flow. This secondary flow pulls particles toward the center, causing them to spiral along the bottom of the cup.
In our tests, there is a flat sand bed on the bottom initially. As we start stirring, the fluid forms a concave surface with the secondary flow pushing the granules toward the center. After stopping the stirring, the fluid returns to its initial level, and the granules are collected at the center.
\rv{
In comparison, the MPM-based approach by \citet{gao2018animating} models granular material as a continuous medium, which prevents sand grains from dispersing naturally under fluid agitation. Furthermore, their approach uses two separate grids to represent the fluid and granular phases. This treatment inhibits the generation of inward flow on the sand bed. As a result, the motion of the sand appears to be lifted by a pressure distribution rather than being carried by the secondary flow.
This experiment demonstrates the superiority of our method in capturing rich dynamic details of sand and achieving greater physical realism compared to previous approaches.
}


\begin{figure}
    \centering
    \setlength{\imagelength}{.495\linewidth}

    \includegraphics[width=\imagelength]{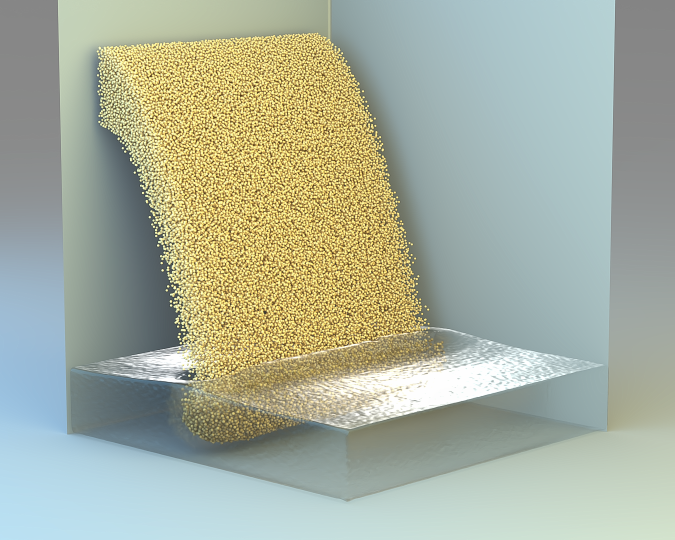}
    \hfill
    \includegraphics[width=\imagelength]{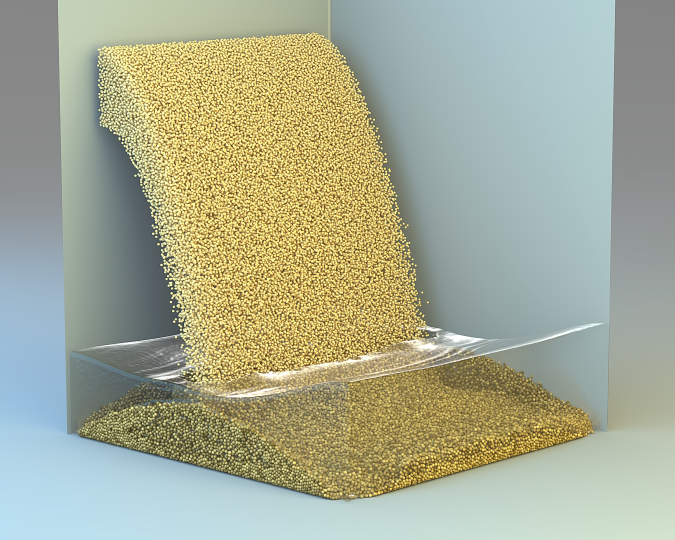} \\
    \vspace{0.01\linewidth}
    \includegraphics[width=\imagelength]{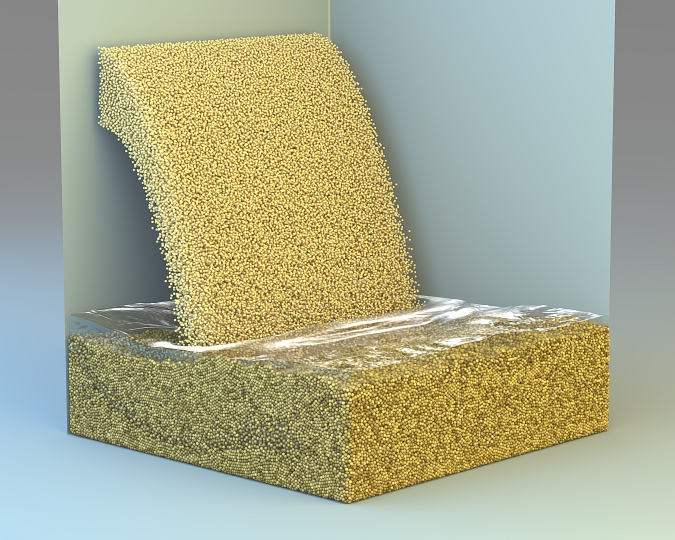}
    \hfill
    \includegraphics[width=\imagelength]{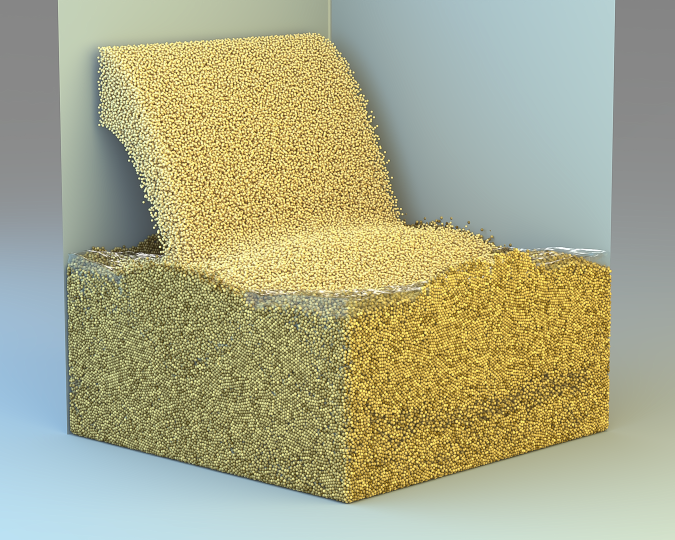}
\vspace{-2em}
\caption{Sandfall. This experiment shows a sandfall plunges into the water and absorbed most of the water. The four figures depict the state at $t=0.5\ \mathrm{s},t=1.0\ \mathrm{s}, t = 2.5\ \mathrm{s}, t = 5\ \mathrm{s}$, respectively.}
\label{fig:sandfall}    
\end{figure}

\begin{figure*}
    \centering
    \setlength{\imagelength}{.33\linewidth}
\includegraphics[width=\imagelength]{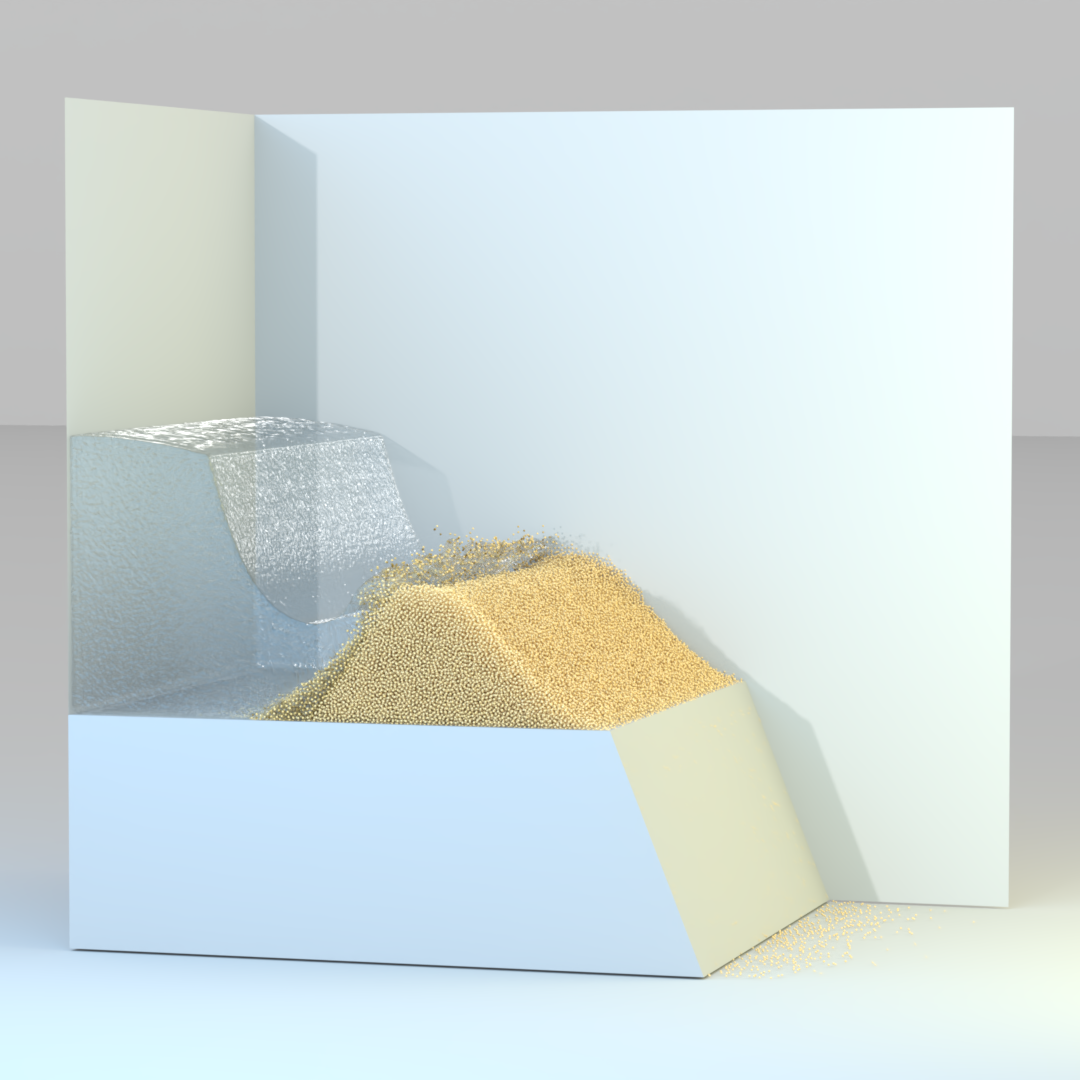}
\hfill
\includegraphics[width=\imagelength]{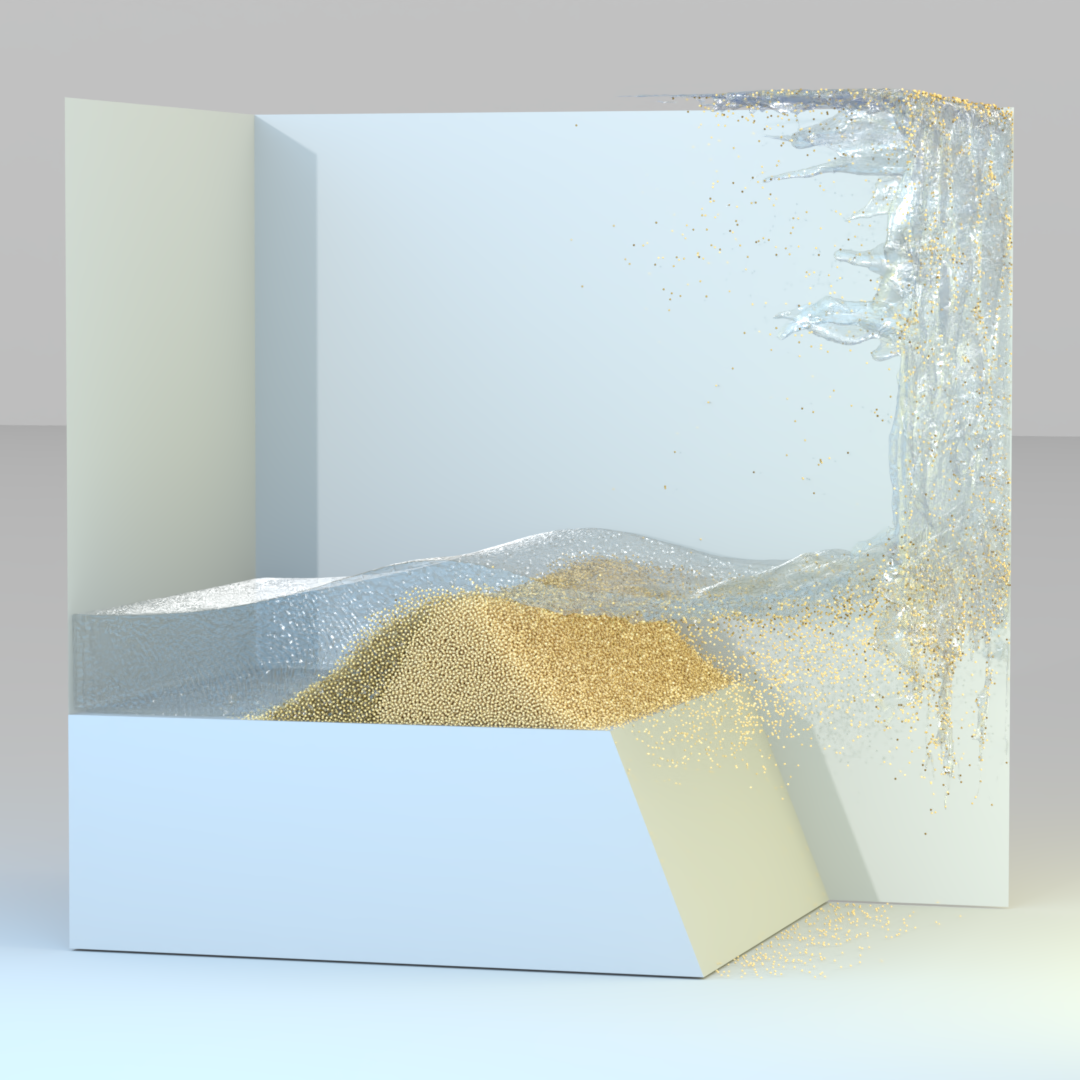}
\hfill
\includegraphics[width=\imagelength]{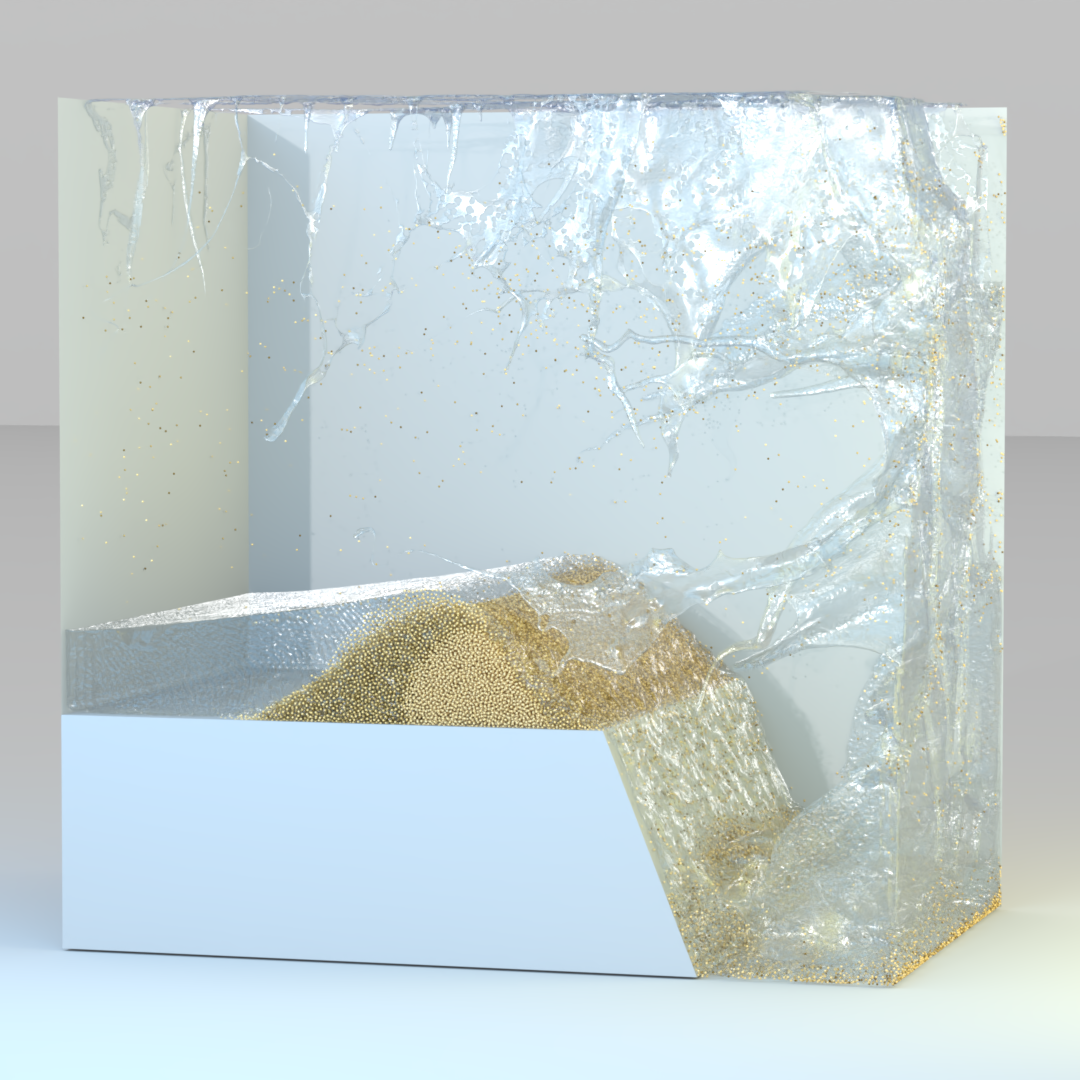}
\vspace{0.01\linewidth}
\includegraphics[width=\imagelength]{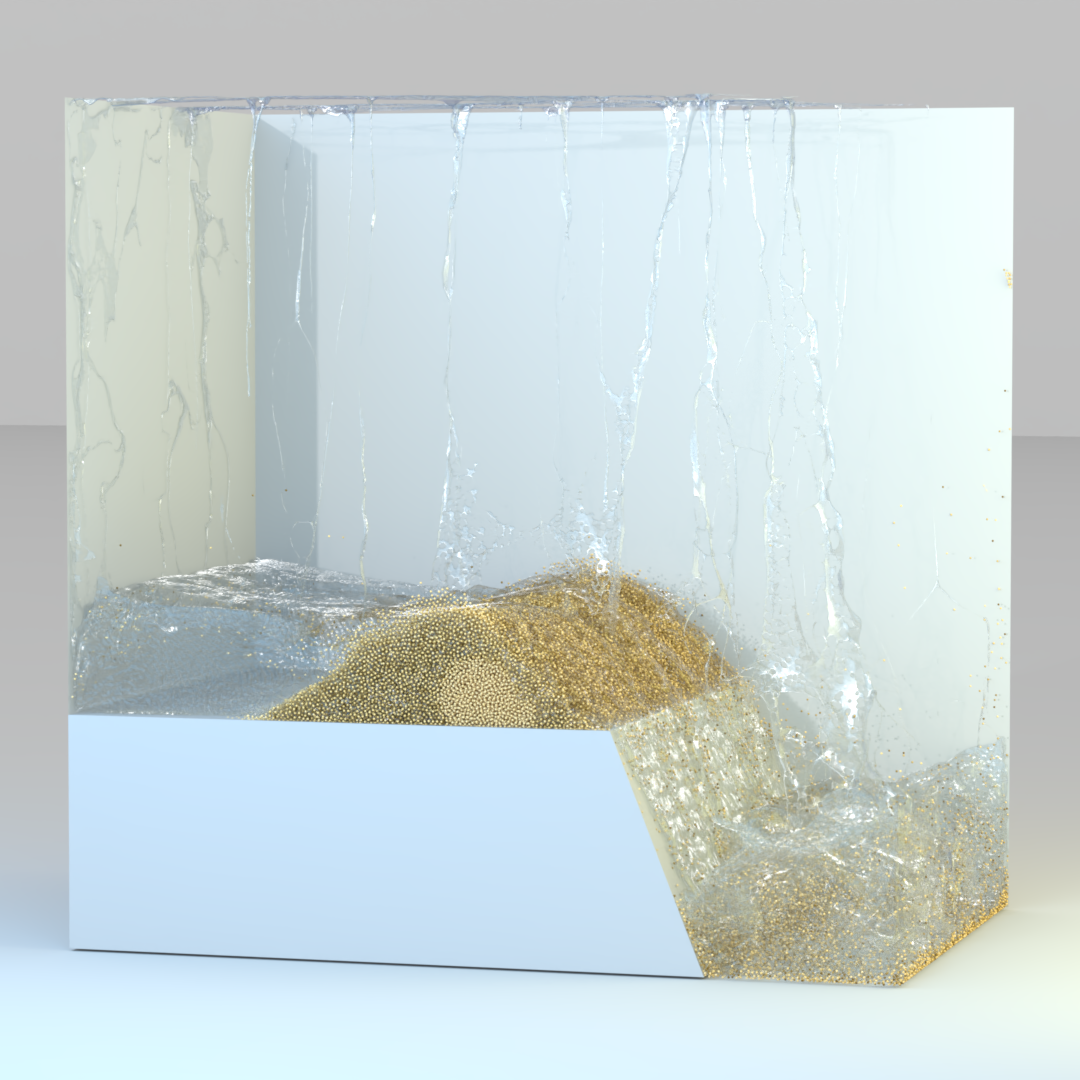}
\hfill
\includegraphics[width=\imagelength]{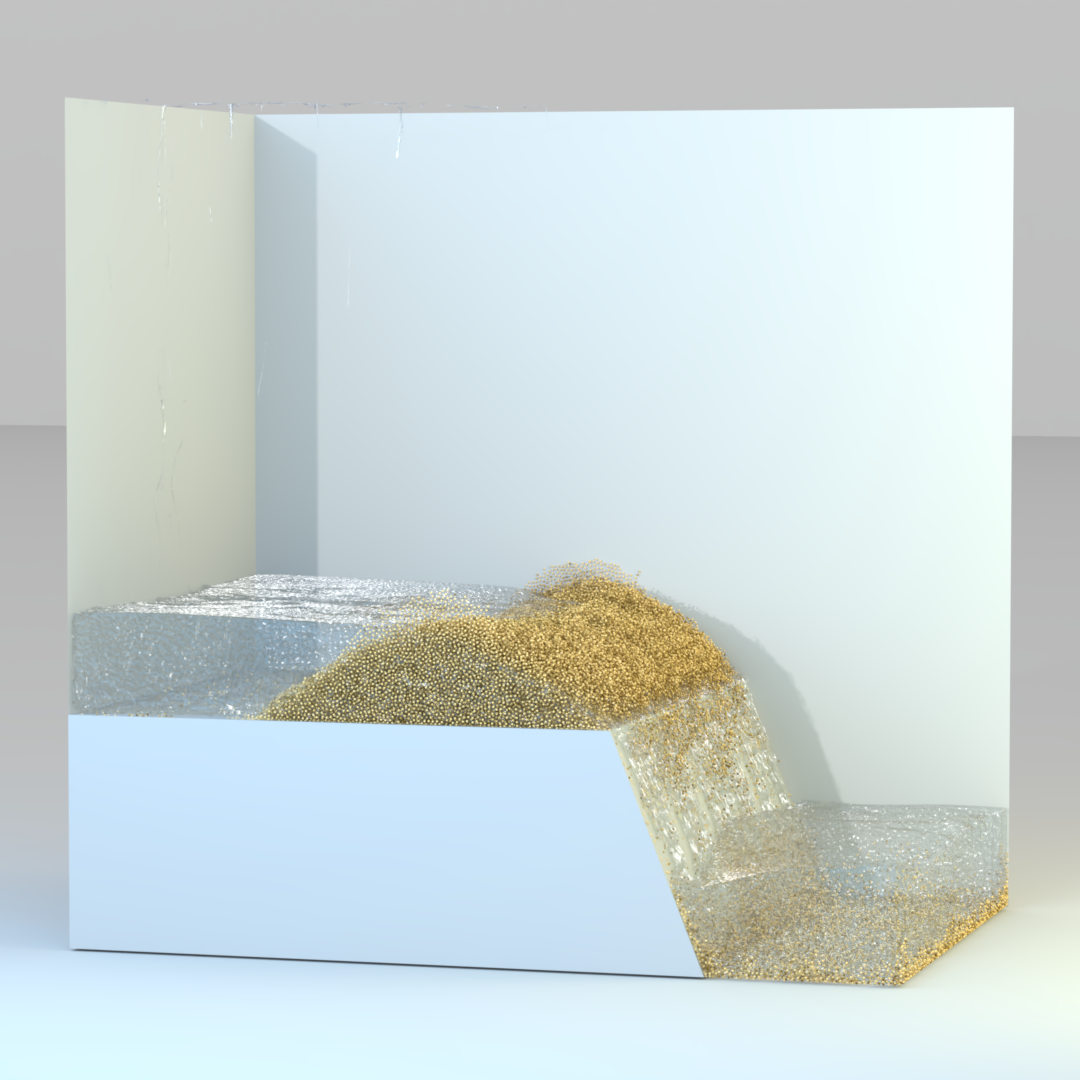}
\hfill
\includegraphics[width=\imagelength]{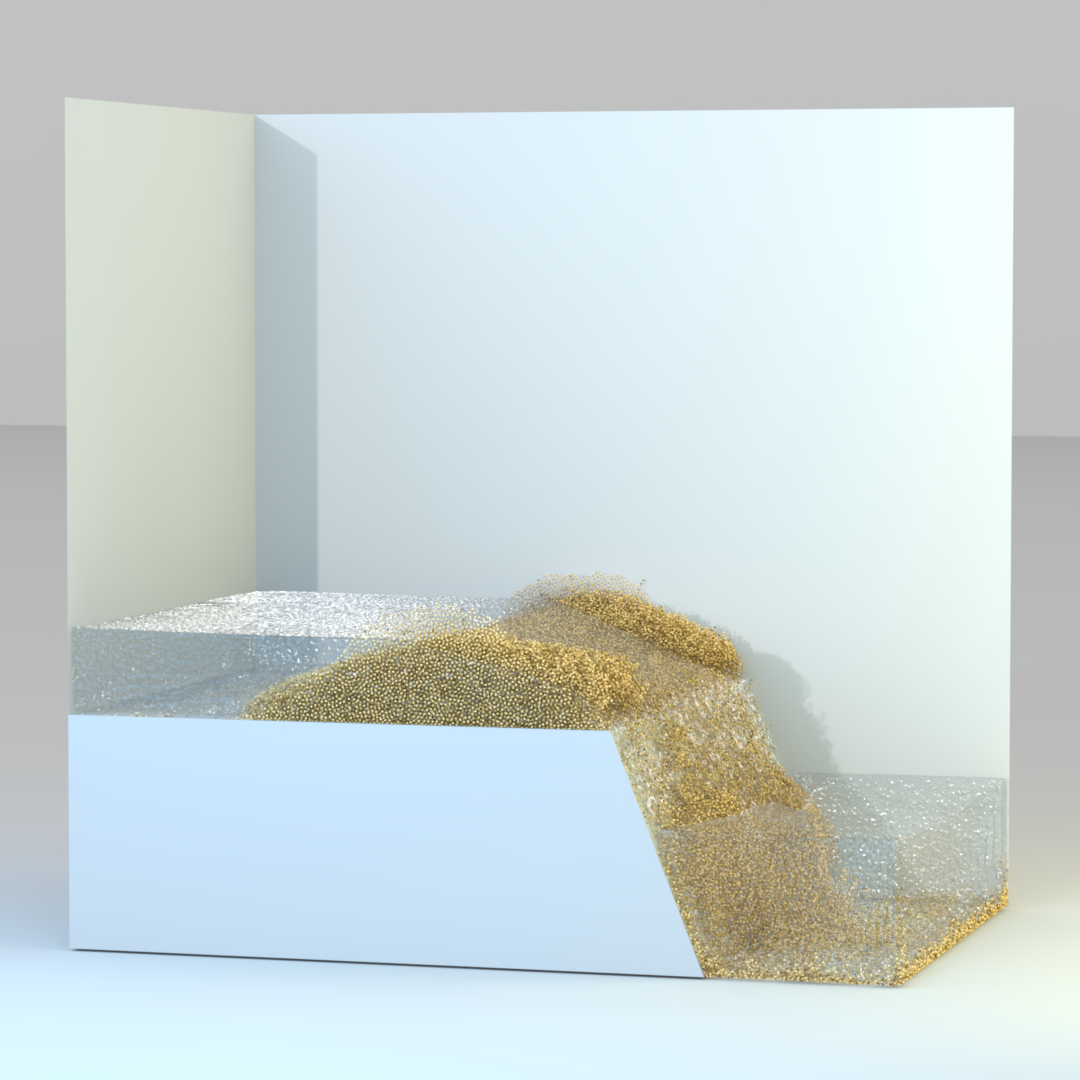}
\vspace{-2.2em}
\caption{Dam Breaking. In this figure we show the dam breaking simulation. The six figures show the state of $t = 0.5, 1, 1.5, 2, 4, 8\ \mathrm{s}$ after the water wall falls. 
}
\label{fig:dam}    
\end{figure*}

\paragraph{Cat Litter}
In Figure~\ref{fig:cat}, we simulate the behavior of wetting cat litter. Initially, there is a pile of dry cat litter in a box. After absorbing a drop of water, due to the concentration gradient force, the wetted cat litter granules aggregate into a clump. When scooping up a portion of the cat litter, the dry granules slip off the shovel due to a lack of internal stress, while the wetted granules maintain their aggregated shape and remain on the shovel. This test demonstrates the capability of the DEM particles in our algorithm to absorb water while preserving their shape stability under external forces.



\paragraph{Sandfall}
In Figure~\ref{fig:sandfall}, we pour $1.2$ M granules into the water, which shows the potential of our method in large-scale sand--water simulations. As granules continuously settle into the water, the liquid level elevates progressively. Upon the accumulated sand exceeding the liquid interface, most of the water is absorbed into the interstices. Subsequent deposition then forms a distinct dry layer at the top of the saturated substrate.

\paragraph{Dam Breaking}
Figure~\ref{fig:dam} shows the dam-breaking simulation. In our simulation, the sand dam relies on inter-granular friction to withstand the water's impact, generating reaction force that causes the fluid to rise and splash along the sand pile's surface, eroding the dam crest. Simultaneously, the sand pile transitions from dry to wet, and water saturation creates internal flow pathways. As the sand becomes increasingly saturated, the cohesion between granules weakens. The dam begins to behave more like a fluid and eventually collapses in its middle.
In contrast to the sand in SPH-DEM, we emphasize the granular characteristics of sand brought by DEM and the effect of the reaction force propagating into the fluid through the velocity projection on the fluid motion. In contrast to the MPM method in \cite{tampubolon2017multi, gao2018animating}, which relies on the deformation obtained on the mesh to obtain the stresses, the particles in our method resist the external forces by the interaction with each other, which is much tighter and the whole body is less prone to be washed away.

\section{Conclusion \& Future Work}
We have developed a novel coupling strategy between granules and fluids to simulate the movement of sand--water mixtures. Through a series of simulations and real-world experiments, GIC demonstrates a faithful reflection of physical phenomena in the real world. To the consideration of higher performance and broader application, we believe our method could be accelerated and extended as below.

\paragraph{Computational Efficiency}
Simulating realistic sand movement needs a large number of DEM particles, and due to the Rayleigh criterion, the DEM solution process requires a much smaller time step compared to the fluid simulation. These factors contribute to a relative high computational cost. To address this, we look forward to developing or adopting a stabler and more efficient DEM algorithm.
Additionally, in our cat litter experiment (Fig.~\ref{fig:cat}), the lifting motion of the shovel and the sand particles above it follows a predefined motion curve rather than being solved physically, as handling fast-moving rigid boundaries poses significant challenges. Using boundary particles would require extremely small time steps, drastically increasing computational costs despite the predominantly rigid motion. This observation highlights the need for exploring more efficient representations and optimization strategies for simulating fast-moving rigid boundaries within the DEM framework, improving both accuracy and computational efficiency.

\paragraph{Rotation and Angular Momentum}
In this work, our model only focuses on the translational motion of rigid spherical sand particles in the fluid, neglecting rotational effects. Although the velocity gradient in water would generate tangential viscous forces, exerting torque on the particles and leadig rotation, our current model does not account for such phenomena. While the IDP method could be integrated into the APIC framework \cite{kugelstadt2019implicit}, we opted for the FLIP method for fluid dynamics due to the absence of angular momentum information in our sand particles. Furthermore, the lack of proper transport equations for angular momentum in the sand--water mixture makes it inappropriate to apply an affine transformation to set the fluid’s target state without this foundational support. In the future, we aim to incorporate particle rotation and the resulting Magnus force into the mixture system to simulate more complex physical phenomena \cite{MagnusUeberDA}, enhancing the realism and accuracy of our model.
\paragraph{Interfacial Effects}
In our current algorithm, sand particles employ isotropic kernel functions to aggregate volume information onto the grid, which is interpreted as the spatial volume fraction of sand. Unlike PIC particles, which act as carriers of information for advection, DEM particles represent actual physical entities. While the Power PIC method achieves optimal spatial distribution by moving particles directly to their centroids on the grid, using non-uniform kernel functions for sand would implicitly assume non-spherical particles, necessitating additional schemes to account for their deformation and mass distribution. Furthermore, optimizing based on earth mover's distance implies a separated phase boundary condition, which is not applicable to sand--water mixtures, as sand is not water-repellent. This emphasizes the need to incorporate interfacial effects into the optimal transport process, enabling a more accurate representation of the complex interactions within sand--water mixtures.

\paragraph{Foam Simulation}
As demonstrated in the experiment of small ball (Fig.~\ref{fig:small}), DEM particles with low density can float on the water, showcasing the potential of extending our framework to simulate interactions between fluids and foam. By replacing the current physics model applied to DEM particles with one specialized to foam dynamics, we can explore a broader range of complex fluid-structure interactions. This extension would not only enhance the applicability of our method but also open doors to studying phenomena such as foam formation, stability, and its interaction with surrounding fluids, paving the way for innovative applications in both scientific and industrial domains.

\bibliographystyle{ACM-Reference-Format}
\bibliography{main}

\appendix
\section{Water--Sand Mixture in Kinetic Theory}
\label{sec:kinetic}

In this section, we review the key points of the kinetic theory description for sand--water mixtures and derive the mass and momentum conservation equations in the macroscopic perspective, expressed in terms of mass-weighted averages. For a more comprehensive introduction to the kinetic theory framework, readers may refer to \citet{hsu2003two, pitaevskii2012physical}.



In the kinetic theory description, we divide the system into three hierarchical levels: the microscopic thermal motion of fluid molecules, the mesoscopic translational motion of sediment particles, and the macroscopic evolution of the entire mixture. In the sand--water mixture system, the motion of granules is driven by the macroscopic flow of the fluid, while the space occupied by the fluid is variable. Our study primarily focuses on the granules, noting that the number of microscopic particles is excessively large compared to the granules. This relative number makes it impractical to directly apply the ensemble averaging method from classical statistical mechanics, which is designed for a fixed number of particles.
Instead, we adopt the concept of ``fluid seen'' proposed by \citet{peirano2002probabilistic} to characterize the state of the fluid with respect to the granules.


Based on a unified degree of freedom for both the fluid and granules, we establish a probability density function for the spatial configuration of the granular system. By performing ensemble averaging on each phase independently to derive their macroscopic statistical descriptions, we apply Favre averaging with respect to mass, decomposing the variables into time-averaged quantities and fluctuating components, thereby decoupling the correlations between statistical measures \cite{soo2018particulates}. This process enables the derivation of the ensemble-averaged forces acting on the granules, which encompass not only the drag force resulting from velocity differences but also a diffusion force originating from concentration gradients.

\subsection{Indicator Function and Ensemble Average}
\label{sec:ind}
For a given granule particle $i$ occupying space $\Omega_{\text{s}_i}$, the fluid parcel $i$ is defined as the smallest such parcel contact with this particle and influencing its motion, and its occupied space is denoted as $\Omega_{\text{f}_i}$.
The particle and this fluid parcel form a particle-fluid pair $\Omega^i = \Omega_{\ts_i} \cup \Omega_{\tf_i}, i = 1, \ldots, N$, satisfying $\Omega^i \cap \Omega^j = \emptyset$ for $i \ne j$.

For a specific state
\begin{equation}
    (\mathcal{C}^N, \mathcal{D}^N) \coloneqq \qty(\vb*{r}_{\tf_i}; \vb*{r}_{\ts_i} \mid i = 1, \ldots, N)
\end{equation}
in the configuration space of $N$ sediment particles,
define the probability density function $F = F(\mathcal{C}^N, \mathcal{D}^N; t)$, satisfying the normalization condition
\begin{equation}
    \iint \dd{C^N} \dd{D^N}\, F = 1,
\end{equation}
where $\dd{C^N} = \dd{\vb*{r}}_{\tf_1} \dd{\vb*{r}}_{\tf_2} \ldots \dd{\vb*{r}}_{\tf_N}$ and $\dd{D^N} = \dd{\vb*{r}}_{\ts_1} \dd{\vb*{r}}_{\ts_2} \ldots \dd{\vb*{r}}_{\ts_N}$.
To fully describe the entire system, the state in the configuration space is insufficient. For all possible particle distributions, a point is occupied by either a particle or the fluid, reflecting their respective properties. To indicate the distribution of each phase, we define the phase indicator function as
\begin{equation}
    \chi_{ki} = \chi_{ki}\qty(\vb*{x} \mid \mathcal{C}^N, \mathcal{D}^N) = \begin{cases}
        1, & \vb*{x} \in \Omega_k^i, \\
        0, & \mathrm{else}.
    \end{cases}
\end{equation}
Thus, the volume fraction of phase $k$ near a point can be written as
\begin{equation}
    \bar{\alpha}_k = \iint \dd{C^N}\dd{D^N} \sum_{j=1}^N \chi_{kj} F.
\end{equation}
The ensemble average of the physical quantity $\psi$ is
\begin{equation}
\label{eq:exp_shi}
\begin{aligned}
    \overline{\alpha_k\psi_k} &= \iint{\dd{C^N}\dd{D^N}\sum_{j=1}^N{\chi_{kj}\psi_{kj}}F} \\
    & = \int{\dd{K}^N \ev{\sum_{j=1}^N{\chi_{kj}\psi_{kj}}}F(\mathcal{K}^N; t)},
\end{aligned}
\end{equation}
where $\ev{\cdot}$ denotes the ensemble average over another term.

\subsection{Governing Equation}
Taking the time derivative of Eq. \eqref{eq:exp_shi}, we can obtain that
\begin{equation}
\label{eq:first_diff}
    \pdv{t}\qty(\overline{\alpha_k\psi_k}) = \overline{\alpha_k \pdv{\psi_k}{t}} + \overline{\psi_k \ev{\dot{\Gamma}_k}} + \int{\dd{K}^N \ev{\sum_{j=1}^N{\chi_{kj}\psi_{kj}}}\pdv{F}{t}},
\end{equation}
where
\begin{equation}
    \overline{\psi_k \ev{\dot{\Gamma}_k}} = \int{\dd{K^N}F\ev{\sum_{j = 1}^N \psi_{kj}\pdv{\chi_{ki}}{t}}},
\end{equation}
where $\pdv*{\chi_{ki}}{t}$ denotes the time variation of the indicator function, reflecting phase transitions in the two-phase flow. Using the normalization property of the distribution function, the corresponding conservation laws is
\begin{equation}
    \qty(\pdv{t} + \sum_{j=1}^N \vb*{v}_{kj}\vdot \grad) F(\mathcal{K}^N; t) = 0.
\end{equation}
By applying integration by parts, the third term in Eq. \eqref{eq:first_diff} could be rewritten as:
\begin{equation}
    \begin{aligned}
        \int{\dd{K}^N \ev{\sum_{j=1}^N{\chi_{kj}\psi_{kj}}}\pdv{F}{t}} & = -\int{\dd{K^N}F \ev{\sum_{j=1}^N{\chi_{kj}\psi_{kj}}} \sum_{j=1}^N{\div\vb*{v}_{kj}}} \\
        & {} + \div\int{\dd{K^N}F\sum_{j=1}^N\qty(\ev{\chi_{kj}\psi_{kj}}\vb*{v}_{kj})} \\
        & {} - \int{\dd{K^N}F\sum_{j=1}^N\qty(\vb*{v}_{kj}\vdot\ev{\chi_{kj}\grad\psi_{kj}})}.
    \end{aligned}
\end{equation}
Since the sum of the velocity divergence over the entire space is $0$, the first term in the integral is eliminated. Substituting this result into Eq. \eqref{eq:first_diff}, the governing equation for the physical quantity $\psi$ is
\begin{equation}
\label{eq:second_diff}
    \pdv{t}\qty(\overline{\alpha_k\psi_k}) + \div(\overline{\alpha_k\psi_k\vb*{v}_k}) = \overline{\alpha_k \frac{\mathrm{D}_k{\psi_k}}{\mathrm{D}{t}}} + \overline{\psi_k \ev{\dot{\Gamma}_k}}.
\end{equation}

Substituting $\psi_k = \rho_k$ into Eq. \eqref{eq:second_diff} yields the mass conservation equation:
\begin{equation}
\label{eq:Amass}
    \pdv{t}\qty(\overline{\alpha_k\rho_k}) + \div(\overline{\alpha_k\rho_k\vb*{v}_k}) = \overline{\rho_k \ev{\dot{\Gamma}_k}}.
\end{equation}
The constancy of the density of the fluid or sediment during the mesoscopic evolution process is adopted. 
Substituting $\psi_k=\rho_k\vb*{v}_k$, leads to the momentum conservation equation:
\begin{equation}
\label{eq:Amomentum}
    \pdv{t}\qty(\overline{\alpha_k\rho_k\vb*{v}_k}) + \div(\overline{\alpha_k\rho_k\vb*{v}_k\vb*{v}_k}) = \overline{\alpha_k \frac{\mathrm{D}_k{(\rho_k\vb*{v}_k)}}{\mathrm{D}{t}}} + \overline{\rho_k\vb*{v}_k \ev{\dot{\Gamma}_k}}.
\end{equation}
For this equation, the phase transition term corresponds to the momentum exchange between fluid particles and sand particles at the mesoscopic level during the absorption. 

For Eq. \eqref{eq:Amomentum}, the fluid and sand particles should be considered separately. If the term corresponds to the fluid, the mesoscopic motion equation of the fluid element can be expressed as
\begin{equation}
    \rho_\tf\frac{\mathrm{D}_{\tf}\vb*{v}_{\tf i}}{\mathrm{D}t} = \rho_\tf \vb*{b} + \div \vb*{\sigma}_{\tf i},
\end{equation}
where $\vb*{b}$ is the external force, $\vb*{\sigma}_\tf$ is the stress tensor. Substituting them into Eq. \eqref{eq:Amomentum} and integrating yields
\begin{equation}
   \overline{\alpha_\tf \frac{\mathrm{D}_\tf{(\rho_\tf\vb*{v}_\tf)}}{\mathrm{D}{t}}} = \bar{\alpha}_\tf \rho_{\tf}\vb*{b} + \div(\overline{\alpha_\tf\vb*{\sigma}_{\tf}}) + \bar{\vb*{M}}_\tf, 
\end{equation}
where
\begin{equation}
    \bar{\vb*{M}}_\tf = \int{F \dd{S} \vb*{n}_\tf \vdot \ev{\vb*{\sigma}_\tf}}
\end{equation}
is the interphase force acting on the fluid from the granules. For the granules, they experience the force $\bar{\vb*{M}}_{\ts} = -\bar{\vb*{M}}_\tf$ exerted by the fluid and the stress tensor $\vb*{\sigma}_{\ts}$ resulting from inter-particle collisions. Thus, the equations of motion for both phases can be written in the same form as
\begin{equation}
\label{eq:third_diff}
        \pdv{t}\qty(\overline{\alpha_k\rho_k\vb*{v}_k}) + \div(\overline{\alpha_k\rho_k\vb*{v}_k\vb*{v}_k}) =
        \bar{\alpha}_k\rho_k\vb*{b} + \div(\overline{\alpha_k\vb*{\sigma}_k}) + \bar{\vb*{M}}_k
        \overline{\rho_k\vb*{v}_k \ev{\dot{\Gamma}_k}}.
\end{equation}

\subsection{Weighted Average of Mass}
For the correlation term $\overline{\alpha_k \rho_k \vb*{v}_k}$ in Equation \eqref{eq:third_diff}, the specific evolution of the velocity cannot be directly obtained through ensemble averaging. Since the volume fraction $\alpha_k$ of each phase may change over time, this mathematical property exhibits the characteristics of compressible flow. To address the issue of density variations, the mass-weighted averaging method, known as Favre averaging, is adopted for compressible flow \cite{soo2018particulates, hsu2004toward}. This approach decomposes the physical quantity $\psi_k$ into a mass-weighted average $\tilde{\psi}_k$ and a fluctuating component \(\psi_k'\).
\begin{equation}
\label{eq:decom}
    \psi_k = \tilde{\psi}_k + \psi_k',
\end{equation}
where the mass-weighted average is defined as:
\begin{equation}
    \tilde{\psi}_k = \frac{\overline{\alpha_k\rho_k\psi_k}}{\bar{\alpha}_k\rho_k}.
\end{equation}

Applying the decomposition strategy of \eqref{eq:decom} to Eq. \eqref{eq:Amass} and \eqref{eq:third_diff}, it yields
\begin{equation}
    \pdv{t}\qty(\bar{\alpha}_k\rho_k) + \div(\bar{\alpha}_k \rho_k \tilde{\vb*{v}}_k) = 0,
\end{equation}
\begin{equation}
    \pdv{t}\qty(\bar{\alpha}_k\rho_k\tilde{\vb*{v}}_k) + \div(\bar{\alpha}_k \rho_k \tilde{\vb*{v}}_k\tilde{\vb*{v}}_k) = \bar{\alpha}_k\rho_k\vb*{b} + \tilde{\vb*{M}}_k + \div(\bar{\alpha}_k \tilde{\vb*{\sigma}}_k).
\end{equation}

The interphase force, acting on the particles due to the surrounding fluid flow, can be expressed as the dominant drag force
\begin{equation}
    \vb*{M}_{\ts} = \overline{\alpha_{\ts}\rho_{\ts} \frac{\vb*{v}_{\tf} - \vb*{v}_{\ts}}{\tau_{\ts}}} = \frac{\rho_{\ts}}{\tau_{\ts}}\qty(\bar{\alpha}_{\ts} \bar{\vb*{v}}_{\tf} + \overline{\alpha_{\ts}' \vb*{v}_{\tf}'} - \bar{\alpha}_{\ts}\vb*{v}_{\ts}),
\end{equation}
where $\tau_{\ts}$ is the velocity relaxation time of the particles, which is related to the particle radius $r$, the fluid viscosity coefficient $\mu$, the fluid density $\rho_{\tf}$, and the volume fraction $\bar{\alpha}_{\ts}$. 
For the correlation terms, the average velocity expression proposed by \citet{toorman2008vertical}
\begin{equation}
    \bar{\vb*{v}}_{\tf} = \tilde{\vb*{v}}_{\tf} - \frac{1}{\bar{\alpha}_{\tf}}\overline{\alpha_{\tf}' \vb*{v}_{\tf}'},
\end{equation}
is adopted, and the turbulence correlation term is simplified using a velocity gradient model
\begin{equation}
    \overline{\alpha_{\tf}'\vb*{v}_{\tf}'} = -\vb*{D}_{\ts} \grad{\bar{\alpha}_{\tf}}, \qq{} \overline{\alpha_{\ts}'\vb*{v}_{\tf}'} = -\vb*{D}_{\ts} \grad{\bar{\alpha}_{\ts}}
\end{equation}
where $\vb*{D}_{\ts}$ is the diffusion tensor related to the volume fraction, the relative velocity between the two phases and the average distance between particles.
Using ensemble averaging, the expression for the mean force acting on the particles is derived as
\begin{equation}
    \vb*{M}_{\ts} = \frac{\bar{\alpha}_{\ts} \rho_{\ts}}{\bar{\tau}_{\ts}}\qty(\vb*{v}_\tf - \vb*{v}_{\ts}) - \frac{\rho_{\ts} \vb*{D}_{\ts}}{\bar{\alpha}_\tf \bar{\tau}_{\ts}}\vdot \grad \bar{\alpha}_{\ts},
\end{equation}
where there are not only drag forces due to velocity differences, but also diffusion forces by concentration gradients.

\subsection{Remaining Interphase Forces}
In a flow field with a pressure gradient, particles experience a pressure gradient force:
\begin{equation}
    \vb*{F}_p = -\int{p\dd{\vb*{S}}} = -\int{\grad p \dd{V}} = -\frac{4}{3}\pi r^3 \grad p.
\end{equation}

When a particle accelerates in a fluid with relative acceleration, it induces the surrounding fluid to accelerate as well, creating an effect analogous to an increase in the particle's apparent mass. To quantify this phenomenon, consider the acceleration of a particle in a stationary, inviscid, and incompressible fluid. By adopting a spherical coordinate system with the particle's velocity direction as the polar axis, the increased pressure distribution on the particle's surface can be derived as
\begin{equation}
    \Delta p = -\frac{\rho_\tf r}{2}\dv{\vb*{v}_{\ts}}{t}\cos\theta.
\end{equation}
Through integration over the sphere, the added mass force can be determined as
\begin{equation}
    \vb*{F}_v = -\frac{2}{3}\pi r^3 \rho_\tf \dv{\vb*{v}_{\ts}}{t}.
\end{equation}

\section{Push-Out Neumann Boundary Condition for density projection}
\label{Push-Out BC}
In this section, the Push-Out Neumann Boundary Condition (BC) for density projection within our framework is derived. Signed distance fields (SDFs) are employed to represent boundaries. For particles inside the boundary, the required position change is expressed as
\begin{equation}
\label{eq:SDF}
    \delta x = -\frac{\grad d\qty(\vb*{x})}{\norm*{\grad d\qty(\vb*{x})}}d\qty(\vb*{x}).
\end{equation}
Consider a boundary cell in the positive grid direction relative to the fluid domain. By utilizing the ghost pressure $p_{\mathrm{ghost}}$ in the solid cells, the position change on a face of the MAC grid with index $i$ is 
\begin{equation}
\label{eq:pos change}
    \delta x_{h_i}=-\frac{\Delta t^2}{\rho_0}\frac{p_{\mathrm{ghost}}-p_\tf}{\Delta x},
\end{equation}
where $p_\tf$ denotes the pressure of the fluid cell. Next, Eq. \eqref{eq:ppel-2} is discretized using finite differences as
\begin{equation}
\label{eq:BCeq}
    A\qty(\qty(\sum_j\beta_j+\beta_{\mathrm{ghost}})p_\tf-\beta_{\mathrm{ghost}}p_{\mathrm{ghost}}-\sum\beta_j p_j)=1-\frac{\alpha_\tf^*}{\alpha_\tf'},
\end{equation}
where $A = \Delta t^2 / (\alpha_\tf'\Delta x^2)$, $\beta_j = (\alpha_j' + \alpha_\tf')/2$, $\beta_{\mathrm{ghost}} =  \alpha_\tf' /2 $. Substituting $p_{\mathrm{ghost}}$ with $\delta x_{h_i}$ from Eq. \ref{eq:pos change} into Eq. \eqref{eq:BCeq} yields
\begin{equation}
\label{eq:BCeq_1}
    A\qty(p_\tf\sum_j\beta_j-\sum\beta_j p_j)=1-\frac{\alpha_\tf^*}{\alpha_\tf'} - \frac{\beta_{\mathrm{ghost}}}{\alpha_\tf'}\frac{\delta x_{h_i}}{\Delta x}.
\end{equation}
If multiple particles are inside the boundary cell, the largest $\delta x_{h_i}$ is adopted to ensure all particles are pushed out of the boundary.

\end{document}